\clearpage

\documentclass[fleqn,12pt]{wlscirep}

\usepackage[utf8]{inputenc}
\usepackage[T1]{fontenc}
\usepackage{lineno}
\usepackage{amssymb}
\usepackage{longtable}
\usepackage{multirow}
\usepackage{multicol}

\graphicspath{ {./Figures/} }

\usepackage{amsmath, amsthm, amssymb}
\usepackage{array}
\usepackage{soul}
\linenumbers

\usepackage{subfiles}

\title{Neural activity in quarks language: Lattice Field Theory for a network of real neurons}

\author[1,*]{Giampiero Bardella}
\author[1,*]{Simone Franchini}
\author[2]{Liming Pan}
\author[1]{Riccardo Balzan}
\author[1]{Surabhi Ramawat}
\author[1]{Emiliano Brunamonti}
\author[1]{Pierpaolo Pani}
\author[1]{Stefano Ferraina}

\affil[1]{Sapienza University of Rome, 00185, Rome, Italy}
\affil[2]{University of Science and Technology of China, 230026, Hefei, China}
\affil[*]{To whom correspondence should be addressed;}
\affil[ \ ]{e-mail: giampiero.bardella@uniroma1.it, simone.franchini@yahoo.it}

\begin{abstract}
Brain-computer interfaces surged extraordinary developments in recent years, and a significant discrepancy now exists between the abundance of available data and the limited headway made in achieving a unified theoretical framework. This discrepancy becomes particularly pronounced when examining the collective neural activity at the micro- and meso-scale, where a coherent formalization that adequately describes neural interactions is still lacking. Here, we introduce a mathematical framework to analyze systems of natural neurons and interpret the related empirical observations in terms of lattice field theory, an established paradigm from theoretical particle physics and statistical mechanics. Our methods are tailored to interpret data from chronic neural interfaces, especially spike rasters from measurements of single neurons activity, and generalize the maximum entropy model for neural networks so that also the time evolution of the system is taken into account. This is obtained by bridging particle physics and neuroscience, paving the way to particle physics-inspired models of neocortex. 

\end{abstract}

\begin{document}
\nolinenumbers
\flushbottom
\maketitle

\thispagestyle{empty}


\clearpage

\section{Introduction}
\label{Introduction}

Integrating observations of neural activity into some coherent theoretical framework is still a challenging task due to the volume and diversity of experimental data. High-resolution recording techniques allow for simultaneous sampling of hundreds of neurons \cite{Pandarinath2018,Barack2021,Stavisky2021,Pang2023,Faskowitz2023,Gardner2022,Shi2023,Genkin2021,Pinotsis2023,Bullard2019,Z.Wei2020,Chandrasekaran2021,Pani2022,Stringer2019,Pachitariu2017}. Although current probes for \textit{in vivo} experiments usually record only a fraction of active neurons, the next generations promise to greatly increase this crucial parameter. Concerning the time scanning rate, interfaces performances have exceed the typical timescale of the neuron's activity long ago, and a gigantic amount of data has been accumulated and published, along with a variety of methods and theories. While some progress has been made in large-scale modeling, micro and meso-scale lag behind. For better or worse, the situation resembles the "Zoo of particle physics" prior to the introduction of the Standard Model. In this paper we introduce a lattice field theory (LFT) \cite{Wilson1974,Balian1974,Lee1983,Lee1987,Parisi1989,Wiese2009,Gupta2011,Zohar2015,ParottoQCD2018,FaccioliSFT,Magnifico2021,Buice2007,Buice2013,Fagerholm2021,Gosselin2022,Halverson2022,Tiberi2022,Gornitz1992,Deutsch2004,Singh2020,Franchini2023} that is tailored to interpret data from multisite brain-computer interfaces (BCIs) in a systematic and physically-grounded way, and that links the microscopic parameters to the experimental observations through well-known renormalization procedures. In short, LFTs discretize the space-time into a lattice grid and are commonly used in theoretical particle physics to facilitate numerical simulations and intractable calculations. Our starting point will be from a novel "kernel"\cite{Franchini2023} approach to LFTs, in part because of its simplicity and part because it allows a natural connection with the theory of spin glasses \cite{Franchini2023,Franchini2021,FranchiniREM2023,Concetti2018,Mezard1987,Mezard2009}, which have been proposed as model of neural activity \cite{Schneidman2006,Tkacik2009,Tkacik2013,Meshulam2021,Meshulam2023,Tang2008,Treves1991,Hermann2012} and for patterns storage in memory and learning \cite{Hopfield1982,Amit1985,Toulouse1986,Treves1988,Kwang2023}. We will assume that time evolution can be characterized by a discrete non relativistic process of interacting binary fields, or Qubits \cite{Gornitz1992,Deutsch2004,Singh2020}, that from a neuroscienence point of view can be interpreted as a field theoretic version of the Free Energy principle of the Bayesian Brain Theory (see for example Friston et al. \cite{Friston2010,Friston2019,Fagerholm2021}). We fully develop the formalism and the basic principles for binary raster diagrams (although the arguments can be readily extended to any Potts-like model with multi-spin interactions). The scope of this paper is to present the theory in full mathematical detail, so that it can be of reference in a wide range of settings from single neuron recordings to multi-layer perceptron's networks and quantum Touring machines.

\clearpage

\tableofcontents

\clearpage

\section{Main results}
\label{Results}

\subsection{Neural activity in terms of Lattice Field Theory}

\begin{linenomath}

It is widely accepted that the most basic computational units of the brain are the neurons. Neurons receive electrochemical inputs from other neurons through dendrites, which are then integrated into the cell body. When the integration reaches a threshold, the neuron generates electrical impulses called action potentials, or spikes. If such a threshold is not reached, no spike is generated. When recording neural activity, e.g., during a neurophysiology experiment, it is usual to collect the timing and occurrence of spikes of an arbitrary number $N$ of individual neurons and align them, in an arbitrary time window $T$, with events or stimuli specific to the chosen experimental paradigm. The matrix with $N$ rows (neurons) and $T$ columns (time) that encodes this information is called a spike raster. Calling $V$ the space and $S$ the time in which our system "lives,"  we note that $V$ is regularized by the intrinsic discretization of its units, i.e., the neurons are discrete objects, and that $S$ is regularized by the physiological existence of an absolute refractory period and fixed by the natural temporal ordering of the observed dynamical evolution. This implies that when studying an ensemble of neurons, we can represent space-time with a set of discrete points, or sites of a lattice. Formally, we define a spatial mapping onto the following ordered set of vertices (see section \ref{Fundamentals}),
\begin{equation}
    V:=\left\{ 1\leq i\leq N\right\},\ \ \ \ S:=\left\{ 1\leq\alpha\leq T\right\}
\end{equation}
where the time window is regularized into sub-intervals $\alpha$ according to a hypothetical "clock time" $\tau$, corresponding to the minimum time between two computational operations of the neuron. Considering that, after a spike, a neuron enters a refractory period during which it is temporarily unable to generate another one, as $\tau$ it would be natural to consider: charge time + discharge time + absolute refractory period + relative refractory period. However, since $\tau$ may vary significantly depending on the experimental setting, for an accurate digitalization of the signal it is convenient to consider its smallest possible value, i.e., the typical duration of a spike: $\tau\approx1ms$.
Within a $\tau$, the neuronal computational unit $i$ can be either silent or active, which can be represented by a binary variable. The raster (or kernel) $\Omega$  can be explicitly written as:
\begin{equation}
\Omega:= \{ \varphi_{i}^{\alpha}\in\{0,1\} :i\in V, \alpha\in S\}
\label{eq:2}
\end{equation}
in which $\varphi_{i}^{\alpha}$ is the binary variable representing the activity of the $i$-th neuron at time $\alpha$. After these simple observations, one recognizes that $\Omega$ naturally provides all the necessary information to describe the observed neural dynamic. From now on we will refer to $\Omega$ as the (neural) activity kernel, or simply, kernel \cite{Franchini2023}. The neural dynamics is expected to follow some causal evolution influenced by the prior states, i.e., a dynamical process with memory. As already argued by several authors \cite{Buice2007,Buice2013,Qiu2014,Fagerholm2021,Halverson2022,Tiberi2022}, it is reasonable to assume that such dynamics can be described by a quantum evolution, so that the formalism of quantum field theory \cite{Wilson1974,Lee1983,Lee1987,Parisi1989,Wiese2009,Gupta2011,Zohar2015,ParottoQCD2018,Buice2007,Buice2013,Halverson2022,Tiberi2022,Gornitz1992,Deutsch2004,Singh2020,Franchini2023} can be applied: this is an harmless assumption, since classical evolution can be always retrieved as a sub-case of quantum evolution. Then, let assume that the evolution of $\varphi_{i}^{\alpha}$ in $\alpha$ can be characterized by a discrete process of interacting binary fields, or Qubits \cite{Gornitz1992,Deutsch2004,Singh2020,Franchini2023}. We can model the time evolution of $\varphi_{i}^{\alpha}$ by considering its statistical mechanics counterpart on a lattice  \cite{Wilson1974,Lee1983,Lee1987,Parisi1989,Wiese2009,Gupta2011,Zohar2015,ParottoQCD2018,Buice2007,Buice2013,Halverson2022,Tiberi2022,Gornitz1992,Deutsch2004,Franchini2023}: to do so, we only need to formally define a few quantities, familiar to neuroscientists, that can be obtained from the binary kernel $\Omega$. Just like a system of particles, we can describe how the $N$ neurons, represented by $\Omega$, interact with their surroundings in the discrete lattice space-time with a single expression enclosing static and dynamic properties. In short, we postulate the existence of the lattice action 
\begin{equation}
\mathcal{A}:\{0,1\}^{VS}\rightarrow\mathbb{R},
\label{eq:The_Action}
\end{equation}
that allows to derive the Lagrangian description of the system and, through the principle of least action \cite{FeynmanQM,FaccioliSFT}, the corresponding statistical theory. Hence, let $\mathcal{O}$ be a test function of $\Omega$, we denote the ensemble average respect to $\mathcal{A}$ with angle brackets, and formally define it as follows:
\begin{equation}
\langle \mathcal{O}(\Omega) \rangle := \sum_{\ \Omega\in\{0,1\}^{VS}} \mathcal{O}(\Omega)\ \ \frac{\exp [-\lambda \mathcal{A}\ (\Omega)]}{\sum_{\ \Omega'\in\{0,1\}^{VS}} \ \exp [-\lambda \mathcal{A}\left(\Omega'\right)]}.
\label{eq:The_Average}
\end{equation}
The classical (non-quantum) limit of the theory is obtained taking the limit of infinite $\lambda$, corresponding to a zero Planck constant, or the zero temperature limit of canonical statistical mechanics \cite{Huang2003}. 
\end{linenomath}
Let introduce $\Phi$ the space correlation matrix, and $\Pi$ the time correlation matrix (joint-spike matrix, JS),
\begin{equation}\begin{split}
&\Phi:=\{\phi_{ij} \ \in\left[0,1\right]:\ \ i,j\in V\},\ \ \ \phi_{ij} \ :=\frac{1}{T}\sum_{\alpha\in S}\varphi_{i}^{\alpha}\,\varphi_{j}^{\alpha}\\
&\Pi:=\{p^{\alpha\beta}\in\left[0,1\right]:\, \alpha,\beta \in S\},\ \ \ p^{\alpha\beta}:=\frac{1}{N}\sum_{i\in V}\varphi_{i}^{\alpha}\varphi_{i}^{\beta}
\end{split}\end{equation} 
Indicating the transpose operation with the symbol $\dagger$, these are straightforwardly obtained from the kernel through the relations 
\begin{equation}
\Omega\,\Omega{}^{\dagger}/T=\Phi,\ \ \ \Omega{}^{\dagger}\Omega/N=\Pi,    
\end{equation}
which are often used automatically (and unconsciously) to calculate spatial and temporal correlations between experimental data. Combining these quantities, we can write a simplified action
\begin{equation}
\mathcal{A}\left(\Omega|\,A,B,I\right):=T\sum_{i\in V}\sum_{j\in V}A_{ij}\,\phi_{ij}+N\sum_{\alpha\in S}\sum_{\beta\in S}B^{\alpha\beta}p^{\alpha\beta}+\sum_{i\in V}\sum_{\alpha\in S}I_{i}^{\alpha}\varphi_{i}^{\alpha}.
\label{eq:action}
\end{equation}
where the matrix $A$ of potential interactions and the matrix $B$ of kinetic interactions control the theory and $I$ is the input kernel that collects the external influences. The full derivation of eq. (\ref{eq:action}), omitted here for the sake of conciseness, can be found in section \ref{LFT}. This is our first main result: an explicit expression for the action of a network of real neurons that depends on easily accessible experimental quantities. Although commonly used to derive and comment on empirical results in neuroscience research, the empirical correlation matrices were, hitherto, not related to each other or ascribable to a precise physical meaning. The action $\mathcal{A}$ provides a recipe for interpreting them in a physically grounded way, setting them within a general theoretical framework that portrays the dynamics of a system in terms of kinetic and potential energies. This entails being able to put a plethora of experimental results under one theoretical hat, using a coherent physical theory. The ingredients of the recipe are the three observables $\Phi$, $\Pi$ and $\Omega$, that encode all the information about the system, the parameters of the theory $A$, $B$, that control the fluctuations, and the boundary conditions $I$. We will call these triplets the hypermatrix and the inverse hypermatrix respectively, since each group of observables can be arranged into a single matrix as in Figure \ref{fig:1}. Many properties of the inverse matrices can be inferred by simple self consistency conditions: for example causality imply that $B$ is upper triangular (see section \ref{lagrangianTHEO}). Our method represent a natural generalization of the maximum entropy principle proposed in the works of Schneidman, Tkacik and colleagues \cite{Schneidman2006,Tkacik2009,Tkacik2013,Meshulam2021,Tang2008} where an Ising model with variable couplings is used to fit \textit{ex-vivo} recordings of a salamander retina (see sections \ref{Results MAXENT}, \ref{Results SALAMAN} and \ref{MAGNETIC}). Moreover, it is remarkable that the Principal Component Analysis (PCA) and the maximum entropy principle can be linked in a natural way within the proposed LFT context. The PCA is probably the most used numerical method in many scientific fields and, in neuroscience, a large amount of data from decades of research is already available in this form to feed machine learning methods. Given the operator relations between the kernel and the correlation matrices, it can be shown (section \ref{Results PCA}) that both PCA in space domain and the maximum entropy principle are special cases of the proposed LFT with zero kinetic term ($B=0$), while the action associated to a PCA in the time domain is that of a purely kinetic LFT ($A=0$), ie., the dual of the maximum entropy principle in the so-called momentum space. Although in this paper we will stick to a strictly real-space analysis, we notice that there are also many powerful spectral methods that can be used to analyze rectangular arrays, like the singular value decomposition (SVD). For example,using the SVD, one finds that the spectrum of $\Phi$ and $\Pi$ is the same as that of a scaling factor. This last fact has its own physical importance and will be discussed elsewhere.

\subsection{Generalization of the maximum entropy principle}
\label{Results MAXENT}

We show that our LFT is a generalization of the maximum entropy principle as presented in the work of Schneidman, Tackcik and others\cite{Schneidman2006,Tkacik2009,Tkacik2013,Meshulam2021,Tang2008}. With some algebra and a global rescaling of the quantities (see section \ref{MAGNETIC}) we can write the action in the magnetic representation:
\begin{equation}
\mathcal{A}\left(M|A,B,h\right)=\sum_{i\in V}\sum_{\alpha\in S}\ h_{i}^{\alpha}\sigma_{i}^{\alpha}+\frac{T}{4}\sum_{i\in V}\sum_{j\in V}\ A_{ij}\,c_{ij}+\frac{N}{4}\sum_{\alpha\in S}\sum_{\beta\in S}\ B^{\alpha\beta}q^{\alpha\beta} .
\end{equation}
where $h$ is linearly related to $I$ (see section \ref{MAGNETIC}). In the spin case the hypermatrix will therefore consist of $M$, $C$ and $Q$. By expanding the definition of $c_{ij}$ we immediately note that in the limit $B\rightarrow0$ we have:
\begin{equation}
\mathcal{A}\left(M|\,A,0,h\right)=\sum_{\alpha\in S}\sum_{i\in V}\ h_{i}^{\alpha }\sigma_{i}^{\alpha}+ \frac{1}{4} \sum_{\alpha\in S}\sum_{i\in V}\sum_{j\in V}\ A_{ij}\,\sigma^{\alpha}_{i}\sigma^{\alpha}_{j} .
\end{equation} 
In this way, a replicated version \cite{Mezard1987} of the Hamiltonian (i.e., the total energy of the system) used in the works of Schneidman and Tkacik is obtained. Notice that the limit $T\rightarrow1$, corresponding to a single replica of the system, it is exactly equivalent to the max entropy model\cite{Schneidman2006,Tkacik2009,Tkacik2013}
\begin{equation}
\mathcal{A}\left(M|\,A,0,h\right)=\sum_{i\in V}\ h_{i}^{1 }\sigma_{i}^{1}+ \frac{1}{4} \sum_{i\in V}\sum_{j\in V}\ A_{ij}\,\sigma^{1}_{i}\sigma^{1}_{j} .
\end{equation} 
Thus, the max entropy principle is recovered as specific case of a field theory with zero kinetic energy. For what we have shown, the authors approximate the activity with a LFT at equilibrium with $B=0$, which corresponds to a field theory with zero kinetic energy.

\subsection{Scaling test for axonal connectivity}
\label{Results SALAMAN}

Let us now show a simple possible application to neural recordings, considering the case of the salamander retina dataset shown in the papers of Schneidman, Tkacik and colleagues \cite{Schneidman2006,Tkacik2009,Tkacik2013}. Remarkably, they where able to reconstruct the $A$ matrix for small groups of neurons in a salamander retina\, thus obtaining both the correlation matrix and its dual in the parameter space. In Figure 1f of their paper\cite{Tkacik2009} they show the distributions of the reconstructed couplings $\tilde{J}_{ij}$ for some values of $N$. The distributions of $\tilde{J}_{ij}$ are indeed approximately Gaussian, and would be very interesting to verify the scaling of the variance of such distributions at various $N$. Concerning the space couplings, for mean field models \cite{Franchini2023,Franchini2021,FranchiniREM2023} the pairwise interaction is the sum of two terms 
\begin{equation}
A_{ij}=\tilde{J}_{0}+\tilde{J}_{ij},
\end{equation}
and hence for the thermodynamic limit to exist it is necessary that the $h_{i}$ are of order $O\left(1\right)$ in the number of neurons, and that $\tilde{J}_{0}$ and $\tilde{J}_{ij}$ scale correctly. This depends on
the connectivity of the matrix of axonal adjacencies (axon matrix):
\begin{equation}
\Lambda=\left\{ \Lambda_{ij}\in\left\{ 0,1\right\} :\,i,j\in V\right\} .
\end{equation}
Average connectivity is defined as follows:
\begin{equation}
g\left(\Lambda\right):=\frac{1}{N}\sum_{i\in V}\sum_{j\in V}\Lambda_{ij}.
\end{equation}
Therefore, to normalize correctly, one must take
\begin{equation}
\tilde{J}_{0}=\frac{1}{g\left(\Lambda\right)}J_{0}\Lambda_{ij},\ \ \ \tilde{J}_{ij}=\frac{1}{\sqrt{g\left(\Lambda\right)}}J_{ij}\Lambda_{ij}
\end{equation}
with $J_{0}$ and $J_{ij}$ Gaussian variables of unit variance and
zero mean. For fully connected models we have $g\left(1\right)=N$, and
then 
\begin{equation}
\tilde{J}_{0}=\frac{1}{N}J_{0},\ \ \ \tilde{J}_{ij}=\frac{1}{\sqrt{N}}J_{ij}
\end{equation}
For models with large connectivity but sub-linear in the number of neurons, one can consider $g\left(\Lambda\right)=N^{\alpha}$ with
$0<\alpha<1$, 
\begin{equation}
\tilde{J}_{0}=\frac{1}{N^{\alpha}}J_{0}\Lambda_{ij},\ \ \ \tilde{J}_{ij}=\frac{1}{\sqrt{N^{\alpha}}}J_{ij}\Lambda_{ij}
\end{equation}
whereas for finite-dimensional models we have that $g\left(\Lambda\right)=O\left(1\right)$. From preliminary analysis (we extracted data from Figure 1f of Tkacik et al. 2009\cite{Tkacik2009} with G3data and performed a Gaussian fit to find the variances, Figure 6) we confirm that the couplings are approximately Gaussian, but the scaling exponent appears to be $\alpha=1/2$ and not $\alpha=1$ as for the Sherrington-Kirkpatrick model\cite{Franchini2023,Franchini2021,FranchiniREM2023,Mezard1987,Charbonneau2023}. This would be interesting, since the system would admit a thermodynamic limit and still sufficient connectivity to manage the data with a mean-field theory\cite{Franchini2023,Franchini2021,FranchiniREM2023,Mezard1987,Charbonneau2023}.  Also, it would be very interesting to see the full hypermatrix to which the covariance matrix in Figure 1 of Tkacik et al. 2009\cite{Tkacik2009} belongs, and even more interesting would be to fit such hypermatrix with the field theory presented in this paper.

\subsection{Relation with the principal component analysis}
\label{Results PCA}

Here we show that the PCA can also be interpreted as a special case of our LFT. In particular, the PCA can be understood as projecting the data into a subspace such that the projected data has a minimum discrepancy with the original one.  In the following, we explicitly consider only the projection into the spatial domain, but the temporal projection is similar. Then, let $V'$ be a subset of $V$ with size $n<N$, let 
\begin{equation}
  X:=\{ x_{i}^k \in \mathbb{R}: i\in V, \, k\in V'\},
\end{equation}
be a real valued kernel with $N$ rows and $n$ columns, and let introduce the set of kernels with orthonormal columns (in this paragraph we denote by $\mathbb{I}$ the identity matrix)
\begin{equation}
  \mathcal{T}:=\{ X\in \mathbb{R}^{VV'}: X^{\dagger}X = \mathbb{I} \}.
\end{equation}
The columns span a $n$-dimensional subspace of $\mathbb{R}^N$, and the projection of $\Omega$ into this subspace is $XX^{\dagger}\Omega$. The PCA aims to identify the subspace such that the discrepancy between the operators $XX^{\dagger}\Omega$ and $\Omega$ is as small as possible in the Frobenius norm. The objective function is 
\begin{equation}
 \mathcal{A}(\Omega|X):=
  \Vert XX^{\dagger}\Omega-\Omega\Vert_{F}^{2}=
  \mathrm{Tr}(\Omega^{\dagger}\Omega - X^{\dagger}\Omega \Omega^{\dagger}X) 
\end{equation}
where $\Vert \cdot \Vert_F$ denotes the Frobenius norm, and where we applied $\Vert D \Vert_F^2 = \mathrm{Tr}(DD^{\dagger})$, with $\mathrm{Tr} (\cdot)$ indicating the trace of the operator. This and the othonormal constraint constitute a constrained optimization problem: calling $Y$ the solution to this problem we write
\begin{equation}
\mathcal{A}(\Omega|Y)=
\min_{X\in \mathcal{T}}\
\mathcal{A}(\Omega|X)=
\min_{X\in \mathcal{T}}\
\mathrm{Tr}(\Omega^{\dagger}\Omega - X^{\dagger}\Omega \Omega^{\dagger} X).
\end{equation}
The minimum is found by choosing $Y$ as the $n$ largest eigenvectors of $\Omega \Omega^{\dagger}$ (or largest left-singular vectors of $\Omega$), see Goodfellow et al.~\cite{Goodfellow2016} for a proof. In the end, one finds
\begin{equation}
\mathcal{A}(\Omega|Y)=
- \mathrm{Tr}(Y^{\dagger}\Omega \Omega^{\dagger} Y)=
- T \sum_{i \in V} \sum_{j\in V} \phi_{ij} \sum_{k\in V'}  y_{i}^{k} y_{j}^{k}.
\end{equation}
We conclude that the action of PCA in space domain is that of a LFT with $B=0$ and 
\begin{equation}
A_{ij}= - \sum_{k\in V'}  y_{i}^{k} y_{j}^{k},
\end{equation}
and is therefore a special form of the maximum entropy principle described above.

\subsection{A model for cortical recordings}

\begin{figure}[h!]
\centering
\includegraphics[width=0.85\linewidth]{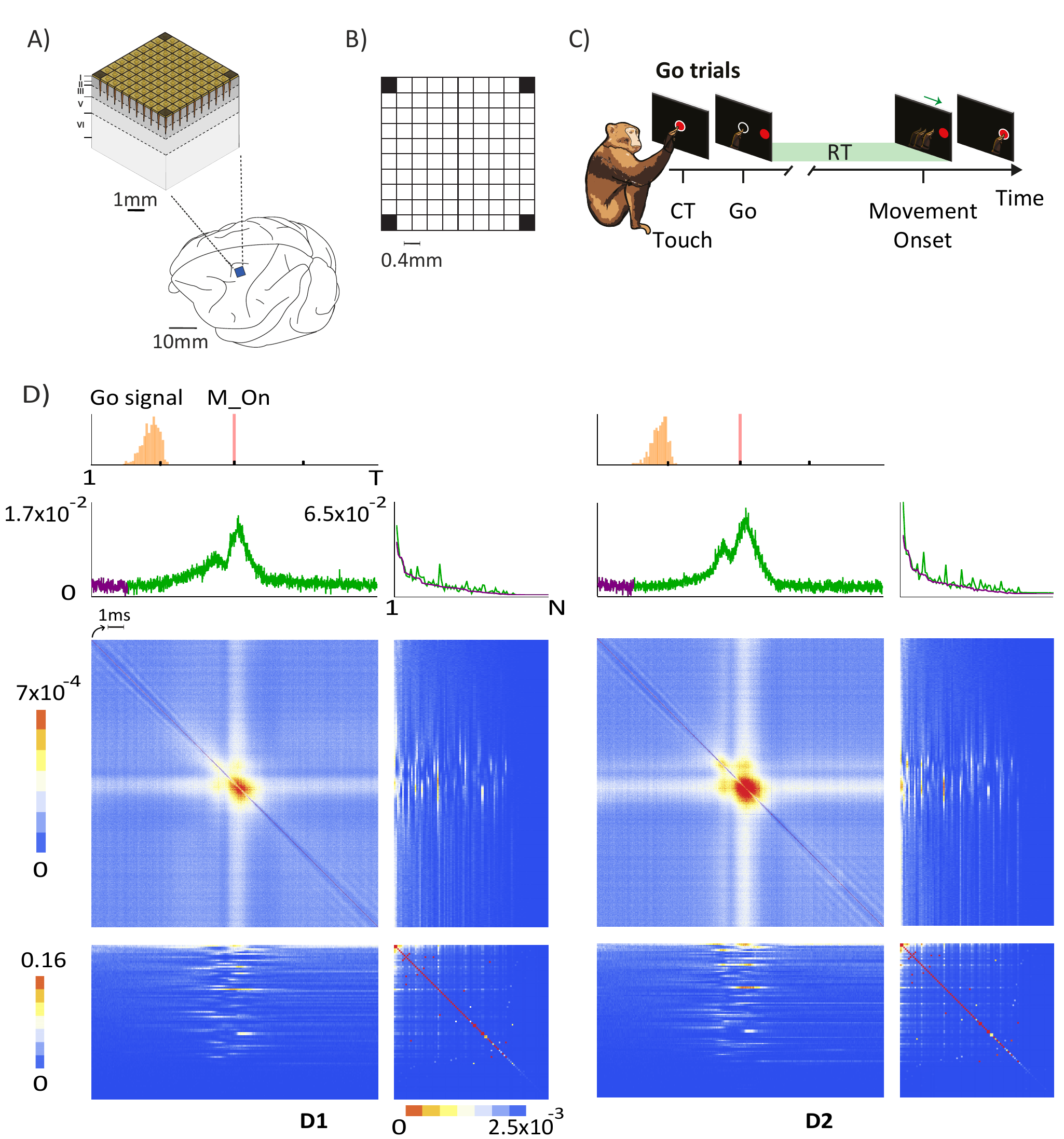}
\newline
\caption{
\textbf{Case study:} \textit{in vivo} recordings from the dorsal premotor cortex (PMd) of non-human primates during a behavioral task. \textbf{a) Cortical minitube sampling of the Utah array}: each electrode pitch is $\sim 40\mu m$ so that the listening volume of each electrode can be reasonably assumed of the order of the distance between the electrodes ($\sim 400\mu m$)\cite{Hill2014}. In the case of PMd, the Utah 96 samples activity from around the inner Baillager band\cite{Rapan2021,Opris2011} at around 1.5 mm penetration (see also Figure 7). 
\textbf{b) Decimated lattice} of the electrode kernel $\hat{\Omega}$ for Utah 96 interfaces.
\textbf{c) Behavioral task} that required visually guided arm movements toward a peripheral target (Go trials) that could appear in two opposite directions (D1 or D2). Monkeys had to reach and hold the peripheral target to get the reward. RT: reaction time; CT: central target; Go: Go signal appearance; M\_on: Movement onset. 
\textbf{d) Experimental hypermatrix} from the electrode kernel $\hat{\Omega}$ of eq. (\ref{eq:experikernel}) for D1 and D2. Neural activity is aligned [-1, +1]s around the M\_on to include the distributions of the stimuli (the Go signal, orange distribution and M\_on, magenta). Here the $I$ of eq. (\ref{eq:action}) shows the time markers for the stimuli presented during the task. Purple traces are the observables computed during a baseline period: the first 250 ms of the selected epoch. The neurons are sorted according to the activity in the first 250 ms of D1. Ticks are every 500 ms.
}
\label{fig:1}
\end{figure}

\begin{figure}[h!]
\centering
\includegraphics[width=0.85\linewidth]{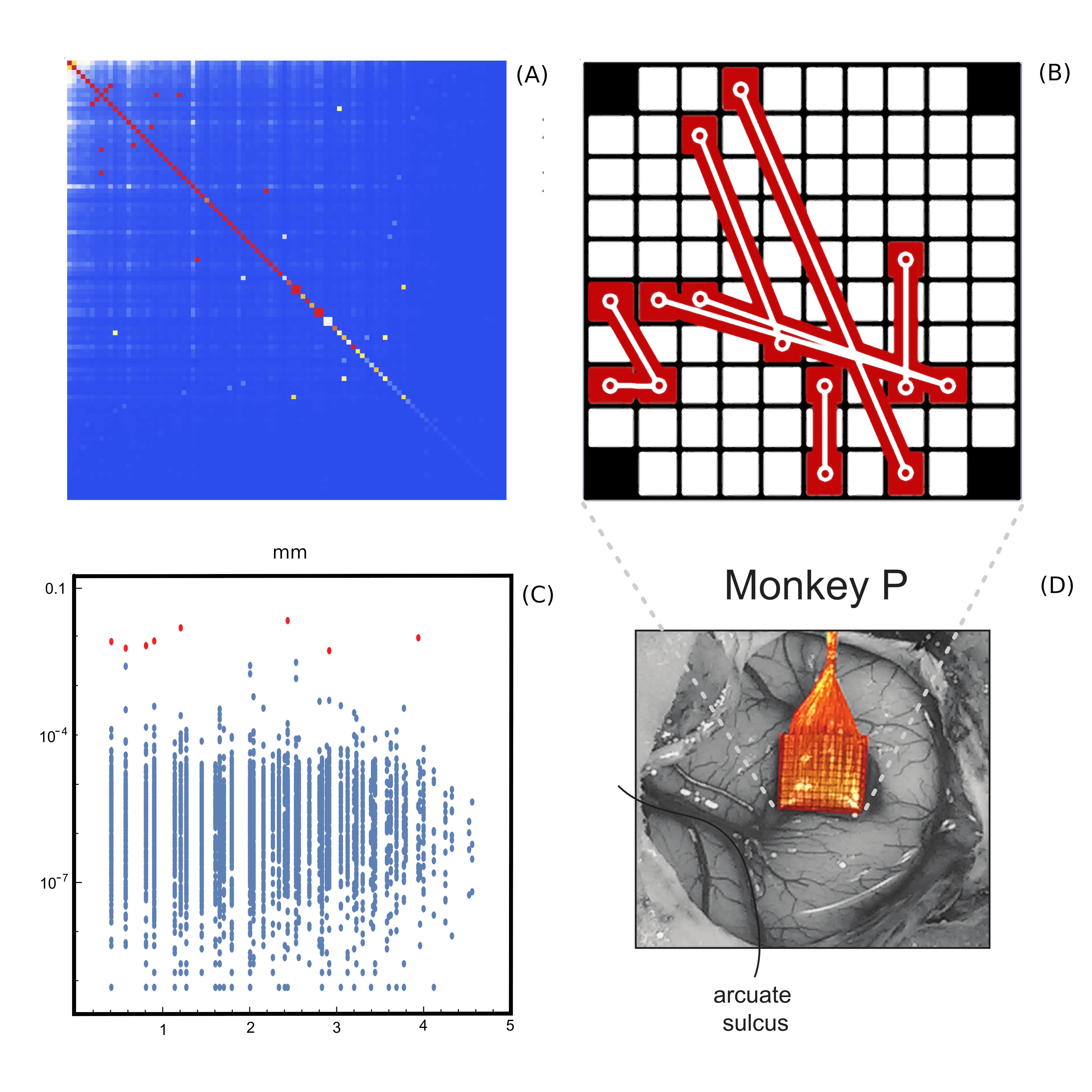}
\newline
\caption{
\textbf{a)} Trial-averaged correlation matrix D1+D2 (computed on the full $2s$ window) $\langle \hat{C}\rangle$, notice the eight most correlated red pairs in red. \textbf{b)} Space arrangement of the “channel” kernel $\hat{\Omega}$ for the Utah 96 BCIs. The lattice is charted by $x'y'$ (decimated lattice), the black corners are silent by default. In red we show the first eight most correlated pairs. \textbf{c)} Space covariance D1+D2 $\langle\delta \hat{C}\rangle$ vs euclidean distance (from lower Figure 7), the most correlated pairs are highlighted in red. The euclidean distance between the channels seems to be not very relevant at the $mm$ scale in this cortical area. \textbf{d)} Placement of the interface during surgery\cite{Bardella2020,Pani2022}.}
\label{fig:2}
\end{figure}

The LFT formalism makes it possible to compare cortical recordings of neural activity with a renormalized field theory. The most used interface to simultaneously record collective neural activity is the silicon-based multilectrode array Utah 96 (Blackrock Microsystems, Salt Lake City)\cite{Bullard2019,Chandrasekaran2021,Leber2017LongElectrodes}. To date, there are about 20 years of recordings of neural activity made with Utah 96, across different species and under hundreds of different experimental conditions, with thousands of kernels already available. The Utah array is a square grid with a 10 x 10 electrode arrangement with a total of 96 channels (the vertices of the square have no record). Due to its planar geometry, the length of its electrodes (around 1.5 mm penetration into the cortex) and their pitch ($40\mu$m) the Utah 96 is able to record from neurons belonging to horizontally separated cortical assemblies, sampled from the same superficial cortical layer $z$ (an example is given in Figure \ref{fig:1} panel a). We will refer to such assemblies as minitubes\cite{Mountcastle1997,Jones2000,Buxhoeveden2002,Hatsopoulos2010,Georgopoulos2010,Opris2011,Hill2014,Potjans2014,Markowitz2015,Cain2016,Hawkins2017,Chandrasekaran2017}. As a comparison, a multi-electrode array with a linear geometry, like a single shank with multiple contact points arranged in a vertical fashion (e.g., like Neuropixels\cite{Paulk2022}), would sample from neurons across various layers within a single minitube\cite{Georgopoulos2010,Opris2011}. Given that each electrode of the Utah array is designed to record approximately the activity of individual minitubes at a distance enough to avoid self-interaction terms, we can model any of its recording as a decimated minitube lattice, and the dynamic evolving around each electrode tip with an on/off field $\hat{\varphi}_{xy}^{\alpha}$ that identifies the state of the observed minitube (see section \ref{Renormalization}). Let us name $xyz$ the coordinates of a lattice such that $z$ represents the average height from the surface of the cortex at which a given layer is located. $V_{xyz}$ represents the volume occupied by (all) the neurons present there and $xy$ is the position of the minitube section in the horizontal plane. Each cortical layer $z$ has its kernel 
\begin{equation}
    \Omega_{z}:=\{\Omega_{xyz}^{\alpha}\in\left\{ 0,1\right\} ^{V_{xyz}}:\,xy\in\mathbb{L}_{2},\,\alpha\in S\},
\end{equation} 
where $\mathbb{L}_{2}$ is a two-dimensional lattice with an average lattice step around the diameter of the individual minitube. Calling $\mathbb{I}(\cdot)$ the indicator function, we can now define $\hat{\varphi}_{xy}^{\alpha}$ as: 
\begin{equation}
    \hat{\varphi}_{xy}^{\alpha}:=\mathbb{I}(\Omega_{xyz}^{\alpha}\neq0)
\end{equation}
which corresponds to assuming that the activation of any one of the neurons in the cell $V_{xyz}$ corresponds to the activation of the entire cell, and with high probability to the activation of the whole minitube. Next, to model the spacing between the probing contacts of the array we apply renormalization considering a decimated lattice $x'y'$ whose step is much greater than the diameter of the individual minitube. One way of renormalizing is the so called renormalization by decimation\cite{Wilson1983,Kadanoff2011,Efrati2014Real-spaceMechanics}, in which the details of the system at small scales are systematically simplified by integrating out most of the degrees of freedom (spins in magnetic systems, neurons in this context). In our case, it corresponds to leaving only those on the decimated lattice $x'y'$ at height $z$. This leads to the decimated activity kernel: 
\begin{equation}
\hat{\Omega}:=\{\hat{\varphi}_{x'y'}^{\alpha}\in\left\{ 0,1\right\} :\,x'y'\in\mathbb{L}'_{2},\,\alpha\in S\},\ \ \ \hat{\varphi}_{x'y'}^{\alpha}:=\mathbb{I}(\Omega_{x'y'z}^{\alpha}\neq0)
\label{eq:experikernel}
\end{equation}
Since $\hat{\Omega}$  describes the on/off activity recorded by each electrode tip, we call it the "electrode" kernel (the full derivation is in sections \ref{Renormalization} and \ref{decimated kernel}). $\hat{\Omega}$ yields the empirical hypermatrix shown in Figure \ref{fig:1} panel d. The space arrangement of $\hat{\Omega}$ for square multielectrode arrays like the Utah 96 is shown in Figure \ref{fig:1} panel b. Explaining how a cortical recording modeled in these terms is comparable to a renormalized field theory is our second main result. In section \ref{Renormalization} we show explicitly a very simple example of how to compute corrections for the effective theory in the semi-classical limit (near-zero Planck constant)\cite{Franchini2023}. It should be noted that more accurate renormalization schemes could be achieved by many other established methods\cite{Niemeijer1973,Niemeyer1974,Wilson1983,Kadanoff2011,Efrati2014Real-spaceMechanics,ParisiRG2001,Franchini2021,Tiberi2022,AngeliniRG2023}. In general, the exact renormalized theory for a generic cortical recording will depend on the details of the system, the recording interface, the experimental settings and other features that should be considered case by case.  Indeed, modeling and calculating the non-linear corrections for the effective theory would be one of the core aspects on which a transfer of expertise from nuclear physics and statistical mechanics to neuroscience could be crucial. In the case of Utah 96, the interface appears to be designed to take individual minitubes for each electrode at a distance enough to avoid self-interaction terms, so we can assume that, apart from systematic errors, sensor degradation, etc., the data can be identified with a decimated version of the kernel of columnar activities, i.e., with $\hat{\Omega}_{x'y'}$ defined before. It would be extremely interesting to reconstruct the couplings from an experimental hypermatrix, e.g.,of the motor control experiments presented in Pani et al.\cite{Pani2022} (see Figure \ref{fig:1} and \ref{fig:2}). However, this is a task with a significant computational burden: to reduce the number of computations required, one could eventually bin or decimate the kernel on a larger clock time. Anyway, we remark that even by looking at the hypermatrix alone, in particular at the kernel and the overlap matrix, it is already possible to appreciate most of the results of Pani et al.\cite{Pani2022} without resorting to numerical methods such as PCA (that can be however interpreted in this framework). In Pani et al.\cite{Pani2022} some modulation of the activity is observed between the Go and movement onset (M\_on), which is called "motor plan". We can also appreciate a stationary rhythmic activity before the Go signal (see Figure 6) and after the M\_on revealed by transverse waves in the joint spike matrix $\Pi$ that is not detected with standard methods (e.g., PCA), and that is suggestive of a time crystal \cite{Wilczek2012,Zhang2017}. A detailed discussion of the experiment shown in Figure \ref{fig:1} is in the section \ref{Neural recordings Methods}.

\subsection{Conclusions}

In conclusion, we showed that applying lattice methods from elementary particle theory to natural neurons should be possible and fruitful, but the understanding of this theoretical framework will still require substantial work. However, given the advanced development of LFTs\cite{Martinelli2023} and their vast range of applicability, knowledge exchange with neuroscience would be beneficial for the theoretical development of the latter in the near future (see Sections \ref{Micromodels} and \ref{Glassymove}), and for both in the long run (see Section \ref{Organoids}). The encounter opens exciting avenues for interdisciplinary research, facilitating connections between computational neuroscience and other fields of physics that utilize LFTs. Indeed, LFTs became of crucial importance far beyond their traditional realms, encompassing cellular automata\cite{Grinstein1985,Elze2014,Hooft2014}, number theory\cite{Gornitz1992} and computational modeling\cite{Fredkin1982,Capobianco2011,Mezard2009,Cranmer2023}. Thus, it is reasonable to think that they could do the same in neuroscience, once the appropriate common theoretical frame is established\cite{Buice2007,Buice2013,Fagerholm2021,Tiberi2022,Franchini2023}. With our manipulations we showed that determining the effective theory describing the dynamics of assemblies of neurons in the neocortex  is possible in terms of LFT, at least in our formalism\cite{Franchini2021}. Moreover, given its spatial symmetries\cite{Mountcastle1997,Jones2000,Buxhoeveden2002,Hatsopoulos2010,Georgopoulos2010,Opris2011,Hill2014,Potjans2014,Markowitz2015,Cain2016,Hawkins2017,Chandrasekaran2017}, it is possible that the topology of such a theory is either mean-field or two-dimensional, with the cortical layers behaving as interacting fields just as in elementary particle theory.  Rethinking neural interactions in this manner could significantly simplify the analytical construction of effective theories. This is because, whether in mean field\cite{Franchini2021,Tiberi2022,Franchini2023,FranchiniREM2023,ParisiRG2001,AngeliniRG2023} or dimension 2\cite{Niemeijer1973,Niemeyer1974,Kogut1975,Wilson1983,Kadanoff2011,Efrati2014Real-spaceMechanics}, established schemes for analyzing, simulating, renormalizing, and, in some cases, exactly solving such theories already exist. In truth, it is not possible to predict what a collision between particle physics and neuroscience might lead to, yet the arguments in this work strongly suggest that it would be worth finding out.

\clearpage

\clearpage

\clearpage

\section{Theoretical methods}
\label{TheoMethods}

\subsection{Fundamentals}
\label{Fundamentals}

Here we introduce some fundamental notations to describe a patch of cortical tissue as a dynamical system on lattice. In these first sections we will consider variables that are aimed to model the whole part of the nervous system involved in the neural computation, of which the actually observed neurons (e.g., during a neurophysiology experiment) are typically a sparse subset. This part is structured as follows: in the first sections we introduce the notation and the basic quantities, the observables and their physical significance, various notions of ergodicity and the fundamental one of stationarity. A discrete Lagrangian description of the system is then introduced in terms of LFT, relations to observables and a general statistical theory are established. A simple kernel “renormalization” scheme, based on Franchini 2023\cite{Franchini2023}, is then introduced to deal with the problem of relating the microscopic theory with marginals on a sparse subset of neurons. Finally, some applications to neuroscience are described, such as the possibility of constructing an effective theory by renormalizing by decimation \cite{Kadanoff2011,Efrati2014Real-spaceMechanics,Niemeijer1973,Niemeyer1974} a two-dimensional lattice of cortical columns. For the detailed derivation of the Lattice Field Theory (LFT) in kernel formalism\cite{Franchini2023} and its associated statistical theory please refer to section \ref{LFT} and \ref{STAT}.

\subsubsection{Map on vertex}

Let $N$ be the number of neurons involved in the task, these are arbitrarily mapped onto the ordered set of vertices 
\begin{equation}
V:=\left\{ 1\leq i\leq N\right\} .
\end{equation} 
Indexing $i\in V$ is defined short of a two-way map that shuffles the index
\begin{equation}
\theta:V\rightarrow V,\ \ \ \theta^{-1}:V\rightarrow V
\end{equation}
this map constitutes a parameter of the representation, and ideally should be chosen so as to highlight the function of the various neurons observed during the task, i.e. highlight any groups of neurons that belong to the same population, or computational structure.

\subsubsection{Identify the $\theta$ highlighting the space-time structures}

The neurons involved in the task (all of them, not just those actually observed) could be partitioned into further subgroups based on function, location, average activity, and times when this varies. For example, it is possible to sort neurons by average activity, or by the earliest time at which
they significantly change their initial state. In the case of multi-electrode array interfaces, it is also possible to define a unique sorting of the experimental rasters based on the position of the sensor channels. Given the sampling rate of neural interfaces, it is assumed that the temporal sampling of the data is much finer than any functional level of interest, while the spatial resolution could be severely limited. This issue will be addressed in subsequent sections; for this one we assume $\theta$ arbitrary unless otherwise stated.

\subsubsection{Multiscale analysis}

To study the system we will sometimes make use of the "multiscale" representation of the type described at the end of Section 3 of \cite{Franchini2023} (Definitions 7 and 8, Figures 3.1 and 3.2). A workable definition requires the introduction of some degree of theoretical description, but in general it can be formalized with a joint partition of V and S in the computational cells, if any, of the various structures. Typically, structures at multiple scales, both spatial and temporal, will be identified. The most general partition is of the type described in Section \ref{LFT}, in which a sequence L of nested partitions (refinements) of the kernel S are the events and sub-events into which the data taking can be categorized, e.g., in the setting described in \cite{Pani2022} we will have the single task level, which would be of the order of seconds, then the various sub-events (such as Go signal, Movement onset, etc.) on the 100 ms, and finally the single actual "computations", which might be isolated around 1-10 ms. Given the sampling rate of the most commonly used sensors, it is assumed that at the temporal level the data are scanned on a much finer scale than any functional level of interest, while the spatial resolution could be severely limited. This issue will be addressed in subsequent sections; for this one we assume $\theta$ arbitrary unless otherwise stated.

\subsubsection{Dynamical system}

In general, we assume that the vector 
\begin{equation}
\varphi_{V}\left(t\right):=\left\{ \varphi_{i}\left(t\right)\in\left\{ 0,1\right\} :\,i\in V\right\} 
\end{equation}
follows a causal evolution determined by the previous activity (dynamic process with memory) according to a hypothetical law $f$ 
\begin{equation}
\varphi_{V}\left(t\right)=f\left(\varphi_{V}\left(s\right):\,s<t\right),
\end{equation}
which in principle could be stochastic \cite{NelsonMechanics,Parisi1981}. We further assume that such a law can be described or approximated by a quantum time evolution\cite{Gornitz1992}, so that the formalism of statistical field theory\cite{Parisi1989} can be applied (notice that classical evolution is a subcase of quantum evolution).

\subsubsection{Time regularization}

Since the space is already regularized by the natural discretization of the computational units (neurons/minitubes), in order to switch to the all-lattice system, it remains to regularize the time window. The window is therefore quantized in $T$ sub-intervals of size $\tau$,
in turn mapped onto the ordered set of vertices  
\begin{equation}
S:=\left\{ 1\leq\alpha\leq T\right\} \end{equation}
preserving the temporal ordering. Notice that in contrast to $V$, the map between time intervals and $\alpha\in S$ is naturally fixed by the temporal ordering of the process. The preservation of the time ordering will be a key feature of our field theoretic approach, and is reminiscent of the time ordered product of Green functions that is used as starting point to derive the LSZ formula \cite{LSZ1955}.

\subsubsection{Clock time}

The time interval is quantized according to a hypothetical "clock"
time $\tau$, corresponding to the time between two fundamental computational operations by the neuron. 
As described in the main text it is convenient to consider $\tau=1 ms$. 

\subsubsection{Binary computation cell}

Within a clock time the computation unit can be possibly active or silent: this can be encoded by a binary variable,
\begin{equation}
\varphi_{i}^{\alpha}\in\left\{ 0,1\right\} 
\end{equation}
which is assumed to be the actual variable supporting the computation. The natural association is obviously with the active/silent state of the neuron during an interval equal to the clock time.
This leads to the definition of the (neural) activity kernel $\Omega$, i.e., a binary array
with $N$ rows and $T$ columns that encodes the entire activity of the observed neurons within the time window. Formally, the kernel is a function of the type: 
\begin{equation}
\Omega:VS\rightarrow\left\{ 0,1\right\} 
\end{equation}
We can introduce the symbol $\varphi_{i}^{\alpha}$ to represent the activity of the $i-$th neuron at time $\alpha$, in this way the kernel can be explicitly written as 
\begin{equation}
\Omega:=\left\{ \varphi_{i}^{\alpha}\in\left\{ 0,1\right\} :\,i\in V,\,\alpha\in S\right\} 
\end{equation}
The kernel is the general order parameter\cite{Franchini2023}, and is assumed to contain the entire information needed to describe the observed process. Given the presence of an absolute refractory period,
we expect that the regularized dynamical system 
\begin{equation}
\varphi_{V}^{\alpha}=f(\,\varphi_{V}^{\beta}:\,\beta<\alpha)
\end{equation}
is the one that naturally describes the evolution of the affected cortex sector, and not the possible continuous theory that would be obtained by taking $\tau\rightarrow0$.

\subsubsection{Kernel of the magnetizations}

To transform from the binary representation into the spin system we use the map $\sigma=2\,\varphi-1$ (which is equivalent to replacing the zeros by -1 ), that gives in the kernel of magnetizations:
\begin{equation}
M:VS\rightarrow\left\{ -1,1\right\} 
\end{equation}
explicitly, the magnetization kernel is
\begin{equation}
M:=\left\{ \sigma_{i}^{\alpha}\in\left\{ -1,1\right\} :\,i\in V,\,\alpha\in S\right\} 
\end{equation}
The two descriptions are mathematically equivalent, but note that some observables may have different meanings. For example, the correlation between binary spikes is 1 only when neurons fire together and 0 otherwise, while the product between two spins is equal to 1 if the variables are equal, and negative if opposite. In practice, the first variable is sensitive only to the event in which the two neurons fire simultaneously, putting on the same rank (same null value) the events in which only one of the neurons fires and the one in which they are silent together. The second, on the other hand, distinguishes only whether or not they do the same thing, regardless of whether they fire or not.

\subsection{Observables}
\label{Observables}

In the following we define the observables of interest that can be obtained from the kernel.

\subsubsection{Kernel offset}

The zero order observable is the global average, or "offset", of the kernel. This quantity is specific to the representation made with kernels and would be a kind of "ergodic" approximation of the activity in the space-time window that is considered. The offset is defined for both the activity and the magnetization kernels, respectively
\begin{equation}
\bar{\Omega}:=\frac{1}{T}\sum_{\alpha\in S}\frac{1}{N}\sum_{i\in V}\varphi_{i}^{\alpha},\ \ \ \bar{M}:=\frac{1}{T}\sum_{\alpha\in S}\frac{1}{N}\sum_{i\in V}\sigma_{i}^{\alpha}
\end{equation}
The two quantities are related by a linear relationship 
\begin{equation}
\bar{M}=2\bar{\Omega}-1
\end{equation}
From a mathematical point of view they are proportional to the Grand Sum (sum of all the elements of a matrix). Such a kernel corresponds to the thermodynamic limit of a gas on lattice
with particle density exactly equal to $\bar{\Omega}$. For magnetic systems, in highly connected ones this can occur due to the presence of a constant external field, or from a uniform fully connected interaction of the Curie-Weiss type. We don't know if there are other magnetic or gaseous systems that have this property and are substantially different from magnetization eigenstates.

\subsubsection{Row and column averages}

At this point we can move to the description of the observables of order one. We define the row averages, which would be the average activity of the state at time $\alpha$, and the column averages, that is the average firing rate of the $i-$th neuron in the time window considered. The average activity is 
\begin{equation}
\omega:=\left\{ \omega^{\alpha}\in\left[0,1\right]:\,\alpha\in S\right\} ,\ \ \ \omega^{\alpha}:=\frac{1}{N}\sum_{i\in V}\varphi_{i}^{\alpha},
\end{equation}
with $\omega^{\alpha}$ space average a time $\alpha$, while averages over the rows, is 
\begin{equation}
f:=\left\{ f_{i}\in\left[0,1\right]:\,i\in V\right\} ,\ \ \ f_{i}:=\frac{1}{T}\sum_{\alpha\in S}\varphi_{i}^{\alpha}.
\end{equation}
where $f_{i}$ is the time-averaged activity of the $i-$th neuron
in the time interval $S$. Notice that the average of the averages is in both cases equal to the offset 
\begin{equation}
\frac{1}{N}\sum_{i\in V}f_{i}=\frac{1}{T}\sum_{\alpha\in S}\omega^{\alpha}=\bar{\Omega}
\end{equation}
Similarly, starting from the kernel of magnetizations we obtain the vector of space averages over the columns, 
\begin{equation}
\mu:=\left\{ \mu^{\alpha}\in\left[0,1\right]:\,\alpha\in S\right\} ,\ \ \ \mu^{\alpha}:=\frac{1}{N}\sum_{i\in V}\sigma_{i}^{\alpha},
\end{equation}
the vector of time averages (on the rows) 
\begin{equation}
m:=\left\{ m_{i}\in\left[0,1\right]:\,i\in V\right\} ,\ \ \ m_{i}:=\frac{1}{T}\sum_{\alpha\in S}\sigma_{i}^{\alpha}.
\end{equation}
where $m_{i}$ is the time average of the $i-$th spin in the time interval $S$. Again, the average of the averages is equal to the
offset 
\begin{equation}
\frac{1}{N}\sum_{i\in V}m_{i}=\frac{1}{T}\sum_{\alpha\in S}\mu^{\alpha}=\bar{M}
\end{equation}
and between the averages made with spin and with spikes the same linear relationship holds: at this first level of description still no particular differences emerge between the two representations.

\subsubsection{Spectra of the averages (quantiles)}

We introduce the distributions (histograms) of the means
\begin{equation}
p_{f}\left(s\right):=\frac{1}{N}\sum_{i\in V}\delta\left(s-f_{i}\right),\ \ \ p_{\omega}\left(s\right):=\frac{1}{T}\sum_{\alpha\in S}\delta\left(s-\omega^{\alpha}\right)
\end{equation}
these two quantities are independent of the ordering of the index, and allow comparison of the statistics of row and column averages. These can be expressed as cumulants 
\begin{equation}
F_{f}\left(s\right):=\int_{0}^{s}du\,p_{f}\left(u\right),\ \ \ F_{\omega}\left(s\right):=\int_{0}^{s}du\,p_{\omega}\left(u\right)
\end{equation}
alternatively one can consider the "quantiles" $F_{f}^{-1}$ and $F_{\omega}^{-1}$ \cite{Steinbrecher2008,FranchiniREM2023} that are the inverse functions of the cumulants. In the case of quantiles, it is particularly easy to construct the form of the function, let use the notation $\left(i\right)$ and $\left(\alpha\right)$ for the maps that reorder the indices by increasing values (order statistics)
\begin{equation}
f_{\left(i\right)+1}\geq f_{\left(i\right)},\ \ \ \omega^{\left(\alpha\right)+1}\geq\omega^{\left(\alpha\right)},
\end{equation}
the quantiles of $f$ and $\omega$ are given by the following expressions:
\begin{equation}
F_{f}^{-1}\left(s\right)=f_{\left(i\right)}\,\mathbb{I}\left(s\in\left[\,\left(i\right)/T,\,\left(\left(i\right)-1\right)/T\,\right]\right),\ \ \ F_{\omega}^{-1}\left(s\right)=\omega^{\left(\alpha\right)}\mathbb{I}\left(s\in\left[\,\left(\alpha\right)/N,\,\left(\left(\alpha\right)-1\right)/N\,\right]\right).
\end{equation}
Magnetic versions of $f$ and $\omega$ are defined in the same way. Given the linear relationship between the averages of the two representations, one can use the same index as that used for f and $\omega$ and write directly 
\begin{equation}
F_{m}^{-1}\left(s\right)=m_{\left(i\right)}\,\mathbb{I}\left(s\in\left[\,\left(i\right)/T,\,\left(\left(i\right)-1\right)/T\,\right]\right),\ \ \ F_{\mu}^{-1}\left(s\right)=\mu^{\left(\alpha\right)}\mathbb{I}\left(s\in\left[\,\left(\alpha\right)/N,\,\left(\left(\alpha\right)-1\right)/N\,\right]\right)
\end{equation}
In what follows we may also refer to quantiles as "spectra" since for $N$ and $T$ finite are step functions whose support is naturally quantized in intervals of height $1/T$ for $f$, $1/N$ for $\omega$, $2/T$ for $m$ and $2/N$ for $\mu$. This is not very relevant in the thermodynamic limit, but it is certainly relevant in many experimental situations: the discussion will be taken up later after introducing the Wasserstein distance.

\subsubsection{Correlation matrices}

The variables of order two are the correlation matrices. It is possible to define a temporal correlation matrix, which in the case of the (neural) activity kernel would be the joint spikes matrix
\begin{equation}
\Phi:=\{\phi_{ij}\in\left[0,1\right]:\,i,j\in V\},\ \ \ \phi_{ij}:=\frac{1}{T}\sum_{\alpha\in S}\varphi_{i}^{\alpha}\,\varphi_{j}^{\alpha}
\end{equation}
and a spatial correlation matrix,
\begin{equation}
\Pi:=\{p^{\alpha\beta}\in\left[0,1\right]:\,\alpha,\beta\in S\},\ \ \ p^{\alpha\beta}:=\frac{1}{N}\sum_{i\in V}\varphi_{i}^{\alpha}\varphi_{i}^{\beta}
\end{equation}
In general, these matrices are obtained from the kernel through the relations
\begin{equation}
\Omega\,\Omega{}^{\dagger}/T=\Phi,\ \ \ \Omega{}^{\dagger}\Omega/N=\Pi
\end{equation}
Note, however, that these observables are not completely independent from the averages, and converge to the following matrices
\begin{equation}
\Phi_{0}:=\{f_{i}\,f_{j}\in\left[0,1\right]:\,i,j\in V\}, \ \ \ \ 
\Pi_{0}:=\{\omega^{\alpha}\omega^{\beta}\in\left[0,1\right]:\,\alpha,\beta\in S\}
\end{equation}
in the free-field approximation. We therefore define the connected correlation matrices, which describe the correlations between the fluctuations 
\begin{equation}
C^{*}:=\Phi-\Phi_{0},\ \ \ Q^{*}:=\Pi-\Pi_{0}
\end{equation}
these are zero in the free-field limit, and nontrivial in case the correlations between fluctuations are significant. Similarly to averages, we can also associate the correlations value distributions
\begin{equation}
p_{C^{*}}\left(s\right):=\frac{1}{N^{2}}\sum_{i\in V}\sum_{j\in V}\delta\left(s-c_{ij}^{*}\right),\ \ \ p_{Q^{*}}\left(s\right):=\frac{1}{T^{2}}\sum_{\alpha\in S}\sum_{\beta\in S}\delta(s-q_{*}^{\alpha\beta})
\end{equation}
which in the following we call fluctuation distributions.Again, we can define analogues in the magnetic description. We can associate the spin-spin correlation matrix typical of systems seen in statistical mechanics 
\begin{equation}
C:=\left\{ c_{ij}\in\left[1-,1\right]:\,i,j\in V\right\} ,\ \ \ c_{ij}:=\frac{1}{T}\sum_{\alpha\in S}\sigma_{i}^{\alpha}\,\sigma_{j}^{\alpha}
\end{equation}
and the overlap matrix commonly used in spin glass theory \cite{Franchini2023,Mezard1987}
\begin{equation}
Q:=\{ q^{\alpha\beta}\in\left[-1,1\right]:\ \alpha,\beta\in S\} ,\ \ \ q^{\alpha\beta}:=\frac{1}{N}\sum_{i\in V}\sigma_{i}^{\alpha}\sigma_{i}^{\beta}
\end{equation}
The relationship between kernels, correlations and overlap is still
the same
\begin{equation}
M\,M{}^{\dagger}/T=C,\ \ \ M{}^{\dagger}M/N=Q
\end{equation}
As before we can define the mean field matrices
\begin{equation}
C_{0}:=\left\{ m_{i}\,m_{j}\in\left[-1,1\right]:\,i,j\in V\right\}, \ \ \ \ 
Q_{0}:=\{ \mu^{\alpha}\mu^{\beta}\in\left[-1,1\right]: \ \alpha,\beta\in S \} 
\end{equation}
however, it is important to note that the matrices constructed by the magnetization kernel have a different meaning than those constructed by the binary kernel. To compare them, one should again consider the connected matrices 
\begin{equation}
C^{*}:=C-C_{0},\ \ \ Q^{*}:=Q-Q_{0}
\end{equation}
the latter contain only the correlations between fluctuations, and are the same for both spin and binary neurons minus a scaling factor. Note that the distribution of the levels of $Q^{*}$ is exactly the distribution of the overlaps mentioned in the Replica Symmetry Breaking (RSB) theory \cite{Mezard1987}.

\subsubsection{Differences between spin and lattice gas}

Although from a mathematical point of view the two descriptions are equivalent, some observables may mean different things. For example, in the case of activities the correlation between spikes $\varphi_{i}^{\alpha}\varphi_{j}^{\beta}$ is worth one only when the neurons fire together and zero in any other case, while the product between two spins $\sigma_{i}^{\alpha}\sigma_{j}^{\beta}$ is equal to one if the variables are equal and negative if opposite. In practice, the first variable is sensitive only to the event when the two neurons fire at the same time; the matrix of values is
\begin{equation}
\varphi_{i}^{\alpha}\varphi_{j}^{\beta}=\begin{cases}
\begin{array}{c}
1\\
0\\
0\\
0
\end{array}\ \ \begin{array}{c}
\varphi_{i}^{\alpha}=1,\ \varphi_{j}^{\beta}=1\\
\varphi_{i}^{\alpha}=1,\ \varphi_{j}^{\beta}=0\\
\varphi_{i}^{\alpha}=0,\ \varphi_{j}^{\beta}=1\\
\varphi_{i}^{\alpha}=0,\ \varphi_{j}^{\beta}=0
\end{array}\end{cases}
\end{equation}
putting on the same level (same null value) events in which only one neuron fires and when they are silent together. The second, on the other hand, distinguishes only whether or not they do the same thing, regardless of whether or not they fire
\begin{equation}
\sigma_{i}^{\alpha}\sigma_{j}^{\beta}=\begin{cases}
\begin{array}{c}
1\\
-1\\
-1\\
1
\end{array}\ \ \begin{array}{c}
\sigma_{i}^{\alpha}=+1,\ \sigma_{j}^{\beta}=+1\\
\sigma_{i}^{\alpha}=+1,\ \sigma_{j}^{\beta}=-1\\
\sigma_{i}^{\alpha}=-1,\ \sigma_{j}^{\beta}=+1\\
\sigma_{i}^{\alpha}=-1,\ \sigma_{j}^{\beta}=-1
\end{array}\end{cases}
\end{equation}
Notice that the relationship between the two observables is
\begin{equation}
\sigma_{i}^{\alpha}\sigma_{j}^{\beta}=4\varphi_{i}^{\alpha}\varphi_{j}^{\beta}-2(\varphi_{i}^{\alpha}+\varphi_{j}^{\beta})+1
\end{equation}
and that this in turn identifies a new observable
\begin{equation}
\varphi_{i}^{\alpha}+\varphi_{j}^{\beta}=\begin{cases}
\begin{array}{c}
2\\
1\\
1\\
0
\end{array}\ \ \begin{array}{c}
\varphi_{i}^{\alpha}=1,\ \varphi_{j}^{\beta}=1\\
\varphi_{i}^{\alpha}=1,\ \varphi_{j}^{\beta}=0\\
\varphi_{i}^{\alpha}=0,\ \varphi_{j}^{\beta}=1\\
\varphi_{i}^{\alpha}=0,\ \varphi_{j}^{\beta}=0
\end{array}\end{cases}
\end{equation}
Linearly related to its corresponding one calculated with the magnetizations 
\begin{equation}
\sigma_{i}^{\alpha}+\sigma_{j}^{\beta}=2(\varphi_{i}^{\alpha}+\varphi_{j}^{\beta})-1
\end{equation}
explicitly the sum of the spins worth 
\begin{equation}
\sigma_{i}^{\alpha}+\sigma_{j}^{\beta}=\begin{cases}
\begin{array}{c}
2\\
0\\
0\\
-2
\end{array}\ \ \begin{array}{c}
\sigma_{i}^{\alpha}=+1,\ \sigma_{j}^{\beta}=+1\\
\sigma_{i}^{\alpha}=+1,\ \sigma_{j}^{\beta}=-1\\
\sigma_{i}^{\alpha}=-1,\ \sigma_{j}^{\beta}=+1\\
\sigma_{i}^{\alpha}=-1,\ \sigma_{j}^{\beta}=-1
\end{array}\end{cases}
\end{equation}
This variable allows us to distinguish three cases (both silent, both active, one only active). Its intrinsic meaning is not yet clear, assuming it has any, however if we take an exponential transformation of the variables
\begin{equation}
z_{i}^{\alpha}:=\exp\lambda\sigma_{i}^{\alpha},\ \ \ \gamma_{i}^{\alpha}:=\exp2\lambda\varphi_{i}^{\alpha}
\end{equation}
we find that the equivalent relationship is with the products 
\begin{equation}
z_{i}^{\alpha}z_{j}^{\beta}=\gamma_{i}^{\alpha}\gamma_{j}^{\beta}\exp\left(-2\lambda\right).
\end{equation}
Let us now consider the difference variable, which we will later interpret as a generalized form of impulse to construct the kinetic term of the Lagrangian 
\begin{equation}
\varphi_{i}^{\alpha}-\varphi_{j}^{\beta}=\begin{cases}
\begin{array}{c}
0\\
1\\
-1\\
0
\end{array}\ \ \begin{array}{c}
\varphi_{i}^{\alpha}=1,\ \varphi_{j}^{\beta}=1\\
\varphi_{i}^{\alpha}=1,\ \varphi_{j}^{\beta}=0\\
\varphi_{i}^{\alpha}=0,\ \varphi_{j}^{\beta}=1\\
\varphi_{i}^{\alpha}=0,\ \varphi_{j}^{\beta}=0
\end{array}\end{cases}
\end{equation}
the link with the spin counterpart is 
\begin{equation}
\sigma_{i}^{\alpha}-\sigma_{j}^{\beta}=2(\varphi_{i}^{\alpha}-\varphi_{j}^{\beta}),
\end{equation}
the difference between spin is the vector
\begin{equation}
\sigma_{i}^{\alpha}-\sigma_{j}^{\beta}=\begin{cases}
\begin{array}{c}
0\\
2\\
-2\\
0
\end{array}\ \ \begin{array}{c}
\sigma_{i}^{\alpha}=+1,\ \sigma_{j}^{\beta}=+1\\
\sigma_{i}^{\alpha}=+1,\ \sigma_{j}^{\beta}=-1\\
\sigma_{i}^{\alpha}=-1,\ \sigma_{j}^{\beta}=+1\\
\sigma_{i}^{\alpha}=-1,\ \sigma_{j}^{\beta}=-1
\end{array}\end{cases}
\end{equation}
Notice that the value of the modulus for sums 
\begin{equation}
|\sigma_{i}^{\alpha}+\sigma_{j}^{\beta}|=\begin{cases}
\begin{array}{c}
2\\
0\\
0\\
2
\end{array}\ \ \begin{array}{c}
\sigma_{i}^{\alpha}=+1,\ \sigma_{j}^{\beta}=+1\\
\sigma_{i}^{\alpha}=+1,\ \sigma_{j}^{\beta}=-1\\
\sigma_{i}^{\alpha}=-1,\ \sigma_{j}^{\beta}=+1\\
\sigma_{i}^{\alpha}=-1,\ \sigma_{j}^{\beta}=-1
\end{array}\end{cases}
\end{equation}
is complementary to the modulus of differences
\begin{equation}
|\sigma_{i}^{\alpha}-\sigma_{j}^{\beta}|=\begin{cases}
\begin{array}{c}
0\\
2\\
2\\
0
\end{array}\ \ \begin{array}{c}
\sigma_{i}^{\alpha}=+1,\ \sigma_{j}^{\beta}=+1\\
\sigma_{i}^{\alpha}=+1,\ \sigma_{j}^{\beta}=-1\\
\sigma_{i}^{\alpha}=-1,\ \sigma_{j}^{\beta}=+1\\
\sigma_{i}^{\alpha}=-1,\ \sigma_{j}^{\beta}=-1
\end{array}\end{cases}
\end{equation}
it follows that the moduli of sum and difference satisfy the relationship
\begin{equation}
|\sigma_{i}^{\alpha}+\sigma_{j}^{\beta}|+|\sigma_{i}^{\alpha}-\sigma_{j}^{\beta}|=2
\end{equation}
and that the moduli of spin and neuron differences are directly proportional
\begin{equation}
|\sigma_{i}^{\alpha}-\sigma_{j}^{\beta}|=2|\varphi_{i}^{\alpha}-\varphi_{j}^{\beta}|
\end{equation}
One can therefore write the modulus of the sum of spins with a quantity proportional to the modulus of the difference of neurons (or spins) changed by sign. The formula for the product of the spins is therefore
\begin{equation}
\sigma_{i}^{\alpha}\sigma_{j}^{\beta}=1-|\sigma_{i}^{\alpha}-\sigma_{j}^{\beta}|=1-2|\varphi_{i}^{\alpha}-\varphi_{j}^{\beta}|
\end{equation}
while that for the sum of the spin is 
\begin{equation}
\sigma_{i}^{\alpha}+\sigma_{j}^{\beta}=2\sigma_{i}^{\alpha}\,(1-|\sigma_{i}^{\alpha}-\sigma_{j}^{\beta}|)=2\,(2\varphi_{i}^{\alpha}-1)\,(1-2|\varphi_{i}^{\alpha}-\varphi_{j}^{\beta}|)
\end{equation}
both proportional to the modulus of the difference between the neurons.

\subsection{Estimators for ergodicity breaking}
\label{ergodicity breaking estimators}

The idea of ergodicity is related to both space and time. With the term "ergodic,” we intend that the same information can be obtained by looking at a small portion of space for a large amount of time, or at a large portion of space for a small amount of time. From Yamamoto's book \cite{Yamamoto2004}: the idea of ergodicity arises if we have a single sample function of a stochastic process instead of the whole ensemble. A single sample function often provides little information about the statistics of the process. However, if the process is ergodic, that is, the time averages are equal to the ensemble averages, then all statistical information can be derived from a single sample function. When a process is ergodic, each sample function represents the entire process. Reflection should convince that the process must necessarily be stationary. Ergodicity thus implies stationarity. There are levels of ergodicity, just as there are levels (degrees) of stationarity. We will mostly consider two levels of ergodicity: ergodicity in mean and in correlation.

\subsubsection{Ergodicity in mean and correlation}

From Yamamoto book \cite{Yamamoto2004}, a process is said to be "ergodic on average" if the averages of rows and columns of the kernel are equal
\begin{equation}
\mu^{\alpha}=m_{i}
\end{equation}
Note that this implies averages $\mu^{\alpha}$ and $m_{i}$ constants in $\alpha$ and $i$, which by the definitions given earlier should necessarily be equal to the offset $\bar{M}$. Let $\chi_{i}^{k}$ the autocorrelation with period $k$ of the $i-$th neuron
\begin{equation}
\chi_{i}^{k}=\frac{1}{T}\sum_{\alpha\in S}\sigma_{i}^{\alpha}\sigma_{i}^{\alpha-k}
\end{equation}
and let $q^{\alpha,\alpha-k}$ the overlap between the states at the time $\alpha$ and $\alpha-k$
\begin{equation}
q^{\alpha,\alpha-k}=\frac{1}{N}\sum_{i\in V}\sigma_{i}^{\alpha}\sigma_{i}^{\alpha-k}
\end{equation}
A process is said to be "ergodic in correlation" if
\begin{equation}
\chi_{i}^{k}=q^{\alpha\alpha-k}
\end{equation}
from which it follows that the autocorrelation and overlap functions must be constant in $i$ and $\alpha$ respectively. Note that for a given $k$ the averages are equal
\begin{equation}
\frac{1}{N}\sum_{i\in V}\chi_{i}^{k}=\frac{1}{T}\sum_{\alpha\in S}q^{\alpha\alpha-k}
\end{equation}
and we can therefore introduce the period-averaged autocorrelation $k$ 
\begin{equation}
\Delta^{k}:=\frac{1}{T}\sum_{\alpha\in S}q^{\alpha\alpha-k}=\frac{1}{T}\sum_{\alpha\in S}\frac{1}{N}\sum_{i\in V}\sigma_{i}^{\alpha}\sigma_{i}^{\alpha-k}
\end{equation}
Which should highlight synchronous activity at a given time scale $k$ (if any). In a correlation-ergodic process this function describes the system completely. We should specify whether one should consider simple or connected correlations: if connected correlations are negligible, one has
\begin{equation}
\Delta_{0}^{k}:=\frac{1}{T}\sum_{\alpha\in S}\mu^{\alpha}\mu^{\alpha-k}
\end{equation}
which can be deduced from the averaged kernel, and should highlight the time scales at which first-order variables (ie, averages) are correlated. Therefore, to study true second-order ergodicity, we need to look at the connected autocorrelation 
\begin{equation}
\Delta_{*}^{k}:=\frac{1}{T}\sum_{\alpha\in S}\left(q^{\alpha\alpha-k}-\mu^{\alpha}\mu^{\alpha-k}\right)
\end{equation}
which is also independent from the choice of representation for the computational cell (either spin or lattice gas). This quantity must necessarily be calculated by averaging the matrices, and should track the correlation scales of the fluctuations in the considered time window. Note that for each of these observables it is also possible to calculate the error with a simple propagation: for a window of size T the error diverges as k approaches T , because of the reduction in the number of values over which the average is averaged.

\subsubsection{Stationarity}

Again, from Yamamoto's book \cite{Yamamoto2004}: a kernel is said to be stationary in a weak sense if the mean value of the columns $\mu^{\alpha}$ is a constant and the overlap $q^{\alpha\alpha-k}$ between the states at times $\alpha$ e $\alpha-k$ depends only on $k$ (and not $\alpha$). If a process is stationary in the weak sense, the autocorrelation function and the power spectral density function form a pair of Fourier transforms (Wiener-Khinchine theorem). Therefore, if we know or can measure the autocorrelation function, we can find the power spectrum density function, that is, which frequencies contain how much power in the signal. Ergodicity in correlation implies stationarity, but stationarity does not imply ergodicity in correlation.

\subsubsection{Ergodicity in distribution}

These notions of ergodicity can be relaxed, for example by considering the distance between the distributions of means $p_{m}$ and $p_{\mu}$. This form of ergodicity is less restrictive than the previous ones, since it does not require stationarity of the process: it is therefore some weaker form of ergodicity than those described by Yamamoto, and probably non-standard, although quite natural to consider in the context of kernels\cite{Franchini2023}. 

\subsubsection{Wasserstein metric}

An explicit definition of ergodicity in distributions requires defining a distance between distributions: a particularly interesting distance (essentially for its relevance in the context of optimal transport) is the Wasserstein metric of order k 
\begin{equation}
W_{k}\left(p_{m},\,p_{\mu}\right)^{k}:=\int_{0}^{1}ds\,\left|\,F_{m}^{-1}\left(s\right)-F_{\mu}^{-1}\left(s\right)\right|^{k}
\end{equation}
It can be shown that convergence with respect to distance of order $k$ is equivalent to the usual convergence in the weak sense plus the convergence of the first $k$ moments. For time windows with $N=T$ the formula becomes particularly simple, 
\begin{equation}
W_{k}\left(p_{m},\,p_{\mu}\right)^{k}:=\frac{1}{N}\sum_{i\in V}\left|\,m_{\left(i\right)}-\mu^{\left(i\right)}\right|^{k}
\end{equation}
for $N\neq T$ one must be careful to first establish an appropriate binning. The version that seems most interesting to us is that of order k=1, which is equivalent to weak convergence, and is also known as the "earth mover" distance, in that it establishes the optimal probability mass transport plan for transforming one distribution into the other (the distributions are thought of as two sand piles on the segment, and the cost of moving one unit of mass from one value to the other is taken proportional to the distance: in practice, it minimizes the work of transforming one pile into the other), we also have that the distance between cumulant is the same done with quantiles
\begin{equation}
\int_{0}^{1}ds\,\left|\,F_{m}^{-1}\left(s\right)-F_{\mu}^{-1}\left(s\right)\right|=\int_{-1}^{1}du\left|\,F_{m}\left(u\right)-F_{\mu}\left(u\right)\right|
\end{equation}
However, remember that all these quantities are ideally designed to study limit kernels, for $N$ and $T$ finite (and not even particularly large in our case) the number of support levels accessible for the two distributions could be very different depending on how many clock times are in the window (given that the number of observed neurons remains fixed). One must therefore carefully consider whether there are enough events in the time interval under consideration such that the spectrum of $m$ is reasonably comparable with that of $\mu$.

\subsection{Lattice Field Theory}
\label{LFT}

It has been proposed by many authors\cite{Lee1987,Hooft2014,Rovelli1995} that although QFTs are generally defined on a continuous support, it is perfectly possible to formulate physical theories directly in terms of difference equations, and still keep all the desirable symmetries and conservation laws of continuous formulations. This is an example of native lattice approach to quantum field theory, and its practical importance is growing with the available computational power and evolution of AI. Let introduce an index for the “mixed space” (or interval space)
\begin{equation}
\Lambda:=\{1\leq\ell\leq L\}
\end{equation}
that is a vertex set of $L=NT$ vertices marked by the index $\ell$, then let $M$ be a spin field on $\Lambda$ with components supported by $\left\{-1,1\right\}$. For now, we will represent the magnetization kernel as a spin vector on the mixed space-time lattice $\Lambda$, which collects the value of the field at all intervals.
\begin{equation}
M:=\{\sigma_{\ell}\in\Gamma:\ell\in\Lambda\}
\end{equation}
For the moment, $M$ is simply a vector that contains the field value for all the space-time points, or computational cells in our case. We will soon re-map $\Lambda$ into a multiplex lattice $VS$ where the proper time $\alpha$ (special dimension) is treated separately from the other dimensions (ordinary dimensions), that is, the “Kernel representation” \cite{Franchini2023}, but already at this point we can highlight another key idea of using an LFT to fit neural data. In contrast with what is usually done in max entropy approaches \cite{Schneidman2006}, in a LFT, the same neuron at two different times is considered like two different neurons, and it is the “action” function that ultimately correlates them in such a way that they look like the same neuron evolving in time. Then, let $\mathcal{O}$ be a test function of $M$. Following Feynman \cite{FeynmanQM} and many others, we postulate the existence of the analytic \textbf{action} function
\begin{equation}
\mathcal{A}:\mathbb{R}^{\Lambda}\rightarrow\mathbb{R}
\end{equation}
and that the averages can be computed from the Wick-rotated Gell-Mann-Low (WGL) formula\cite{Guerra1975}. The combined work of several authors showed that it is equivalent to the Gibbs \cite{Gibbs1902} average with $\mathcal{A}$ on behalf of the Hamiltonian, which corresponds to the principle of least action. The WGL formula is
\begin{equation}
\langle\mathcal{O}\left(X\right)\rangle=\sum_{X\in {\mathbb{R}}^{\Lambda}}\mathcal{O}\left(X\right)\,\frac{\exp\left[-\lambda\mathcal{A}(X)\right]}{\sum_{X'\in {\mathbb{R}^{\Lambda}}}\exp\left[-\lambda\mathcal{A}(X')\right]}
\end{equation}
and is suitable to describe any observable that depends on the field $X$. The continuous (or thermodynamic) limit of this theory is attained for $L\rightarrow\infty$, if it exists, while the zero temperature limit, $\lambda\rightarrow\infty$, correspond to the non-quantum limit of the theory. 

\subsubsection{Qubit Field Theory}

Binary quantum field theories where pioneered by C. F. von Weizsacker in the 50s with the “Ur” (Alternatives) theory \cite{Gornitz1992,Finkelstein2003}. The Ur theory is the earliest example of Qubit field theory \cite{Gornitz1992,Deutsch2004,Singh2020,Franchini2023} and is probably the simplest of all lattice quantum field theories (QFTs). Here we apply the Taylor theorem and other elementary mathematical methods to the Lee formulation of quantum mechanics \cite{Lee1983,Lee1987} in order to obtain a path integral formulation of the Ur theory via perturbative methods. We start from the assumption that the action is an analytic function of the field components, then Taylor's theorem can be applied to obtain a convergent perturbation theory\cite{Parisi1981}. Define the auxiliary functions
\begin{equation}
D_{\ell}(X):=\frac{\partial\mathcal{A}(X)}{\partial x_{\ell}},\ \ \ D_{\ell\ell'}\left(X\right):=\frac{\partial^{2}\mathcal{A}(X)}{\partial x_{\ell}\,\partial x_{\ell'}}\left[1-\mathbb{I}(\ell=\ell')\right]+\frac{1}{2}\frac{\partial^{2}\mathcal{A}(X)}{\partial x_{\ell}^{2}}\,\mathbb{I}(\ell=\ell')
\end{equation}
then, by Taylor's theorem, the action can be expanded around the null field, and this correspond to the one-, two- vertex interactions etc. 
\begin{equation}
\mathcal{A}(X)=\sum_{\ell\in\Lambda}D_{\ell}\left(0\right)x_{\ell}+\sum_{\ell\in\Lambda}\sum_{\ell'\in\Lambda}D_{\ell\ell'}\left(0\right)x_{\ell}x_{\ell'}+\ ...
\end{equation}
here we stop at second order to avoid complications, but a fourth order theory should be considered for accurate description of physical theories. To obtain a Ur theory we can take $|x_{\ell}|\rightarrow g$ with $g<1$ (a form of Ising limit \cite{IsingLimit}). Let introduce the tensors
\begin{equation}
F_{\ell}:=g\,D_{\ell}(0),\ \ \ F_{\ell\ell'}:=g^{2}\,D_{\ell\ell'}(0),\ \ ...
\end{equation}
By substituting into the series expansion before we obtain the first order perturbation theory of the Ur in magnetic representation 
\begin{equation}
\mathcal{A}(M)=\sum_{\ell\in\Lambda}F_{\ell}\,\sigma_{\ell}+\sum_{\ell\in\Lambda}\sum_{\ell'\in\Lambda}F_{\ell\ell'}\,\sigma_{\ell}\sigma_{\ell'}+\ ...\ 
\end{equation}
and the theory is controlled by the tensor sequence $F$. We can immediately recognize the Ising-like structure of the action, which can be related to the usual formulations of the Standard Model on lattice trough, for example, the Parotto mapping of QCD \cite{ParottoQCD2018}. In general, the statistical method \cite{Franchini2023,Parisi1989,Mezard1987} allows the problem of finding the quantum (thus also classical) time evolution of a system of interacting binary fields \cite{Gornitz1992,Deutsch2004,Singh2020,Franchini2023} to be transformed into a problem of classical statistical mechanics on lattice \cite{Lee1987,Parisi1989,Wiese2009,Gupta2011}, which can then be studied through canonical theory \cite{Gibbs1902,Huang2003}, renormalization \cite{Kadanoff2011,Efrati2014Real-spaceMechanics,Wilson1974,Wilson1983,Niemeijer1973, Niemeyer1974}, and other powerful mathematical methods \cite{Franchini2023,Mezard1987}. 

\subsubsection{Neural LFT}

In the following section we provide the complete derivation of the expressions of the action $\mathcal{A}$, both in the binary case and in the spin representation. Therefore, for the sake of completeness, some expressions and definitions given in the main text will be included again. For any physical (then limited) region of our analogue space-time, a map $\Theta$ exists that connects the kernel with the mixed space and vice versa. Let introduce a “grand map”
\begin{equation}
\Theta:VS\rightarrow\Lambda,\ \ \ \Theta^{-1}:\Lambda\rightarrow VS
\end{equation}
that establishes a biunivocal relation between the points of the mixed space $\Lambda$ and those of the observed space time VS. This map always exists for any physical (finite) discrete observable and is another free parameter of the theory, that can be tuned to highlight the space-time structures. Then, we relabel the points according to a double index as in \cite{Franchini2023}. Assuming that the neuron's computation is supported by the the field
\begin{equation}
\psi_{i}^{\alpha}:=\varphi_{i}^{\alpha}-\bar{\Omega}, 
\end{equation} 
where $\bar{\Omega}$ is the global offset for the neuron \textit{in vivo}, which could possibly be zero. By Taylor's theorem, the action of a lattice field theory $\mathcal{A}(\Omega|\,F)$ can be described by an expansion of the kind:
\begin{multline}
\mathcal{A}(\Omega|\,F)=\sum_{i\in V}\sum_{\alpha\in S}F_{i}^{\alpha}\psi_{i}^{\alpha}+\sum_{i\in V}\sum_{j\in V}\sum_{\alpha\in S}\sum_{\beta\in S}F_{ij}^{\alpha\beta}\psi_{i}^{\alpha}\psi_{j}^{\beta}+\\
+\sum_{i\in V}\sum_{j\in V}\sum_{h\in V}\sum_{\alpha\in S}\sum_{\beta\in S}\sum_{\gamma\in S}F_{ijh}^{\alpha\beta\gamma}\psi_{i}^{\alpha}\psi_{j}^{\beta}\psi_{h}^{\gamma}+\sum_{i\in V}\sum_{j\in V}\sum_{h\in V}\sum_{k\in V}\sum_{\alpha\in S}\sum_{\beta\in S}\sum_{\gamma\in S}\sum_{\delta\in S}F_{ijhk}^{\alpha\beta\gamma\delta}\psi_{i}^{\alpha}\psi_{j}^{\beta}\psi_{h}^{\gamma}\psi_{k}^{\delta}+\,...\label{eq:cvcg}
\end{multline}
Each term represent one-, two-, three-, and four-vertex interactions etc., while the tensors sequence $F$ collects the parameters of the theory. However, if we want to consider the same non-relativistic approximation used by Schneidman and colleagues\cite{Schneidman2006}, interactions with more than two vertices can be neglected, and also two-vertex interactions with four different indices.
Therefore, the proposed action reduces to:
\begin{equation}
\mathcal{A}(\Omega|\,A,B)=\sum_{i\in V}\sum_{j\in V}A_{ij}\,\sum_{\alpha\in S}\psi_{i}^{\alpha}\psi_{j}^{\alpha}+\sum_{\alpha\in S}\sum_{\beta\in S}B^{\alpha\beta}\sum_{i\in V}\psi_{i}^{\alpha}\psi_{i}^{\beta}
\end{equation}
The action depends explicitly on the correlation and overlap matrices, and is controlled by the matrix of potential interactions $A$ and by the matrix of kinetic interactions $B$. We can switch to the binary representation $\varphi$ through the transformation
\begin{equation}
\psi_{i}^{\alpha}\psi_{j}^{\beta}=\varphi_{i}^{\alpha}\varphi_{j}^{\beta}-\bar{\Omega}\,(\varphi_{j}^{\beta}+\varphi_{i}^{\alpha})+\bar{\Omega}^{2}
\end{equation}
doing the algebra one finds that the structure of the theory is the same: 
\begin{multline}
\sum_{i\in V}\sum_{j\in V}A_{ij}\,\sum_{\alpha\in S}\psi_{i}^{\alpha}\psi_{i}^{\beta}+\sum_{\alpha\in S}\sum_{\beta\in S}B^{\alpha\beta}\sum_{i\in V}\psi_{i}^{\alpha}\psi_{i}^{\beta}=\\
=\sum_{i\in V}\sum_{j\in V}A_{ij}\,\sum_{\alpha\in S}\varphi_{i}^{\alpha}\varphi_{j}^{\alpha}+\sum_{\alpha\in S}\sum_{\beta\in S}B^{\alpha\beta}\sum_{i\in V}\varphi_{i}^{\alpha}\varphi_{i}^{\beta}+\\
+\bar{\Omega}\sum_{\alpha\in S}\sum_{i\in V}\varphi_{i}^{\alpha}\sum_{j\in V}(A_{ij}-A_{ji})
+\bar{\Omega}\sum_{\alpha\in S}\sum_{i\in V}\varphi_{i}^{\alpha}\sum_{\beta\in S}(B^{\alpha\beta}-B^{\beta\alpha})\ \mathrm{+\ const.}\label{eq:gfghdhd}
\end{multline}
because the connected part is identical. If the global offset $\bar{\Omega}$ is not zero one must take into account the appearance of additional currents 
\begin{equation}
I_{i}^{\,0}:=\bar{\Omega}\sum_{j\in V}(A_{ij}-A_{ji}),\ \ \ I_{0}^{\alpha}:=\bar{\Omega}\sum_{\beta\in S}(B^{\alpha\beta}-B^{\beta\alpha})
\end{equation}
which, however, transform linearly. Ultimately, the action in the binary form is:
\begin{multline}
\mathcal{A}\left(\Omega|\,A,B\right)=\sum_{i\in V}I_{i}^{\,0}\sum_{\alpha\in S}\varphi_{i}^{\alpha}+\sum_{\alpha\in S}I_{0}^{\alpha}\sum_{i\in V}\varphi_{i}^{\alpha}
+\sum_{i\in V}\sum_{j\in V}A_{ij}\,\sum_{\alpha\in S}\varphi_{i}^{\alpha}\varphi_{j}^{\alpha}+\sum_{\alpha\in S}\sum_{\beta\in S}B^{\alpha\beta}\sum_{i\in V}\varphi_{i}^{\alpha}\varphi_{i}^{\beta}\label{eq:cuhgkv}
\end{multline}
Notice that by normalizing the sums the action can be rewritten using the first-order observables $f$, $\omega$, which are obtained from the kernel $\Omega$ through linear transformations, and with those of second order, the matrices $\Phi$ and $\Pi$, which are nonlinear observables: 
\begin{equation}
\mathcal{A}\left(\Omega|\,A,B\right)=T\sum_{i\in V}I_{i}^{\,0}f_{i}+N\sum_{\alpha\in S}I_{0}^{\alpha}\,\omega^{\alpha}+T\sum_{i\in V}\sum_{j\in V}A_{ij}\,\phi_{ij}+N\sum_{\alpha\in S}\sum_{\beta\in S}B^{\alpha\beta}p^{\alpha\beta},
\label{eq:hypermatrix_teo}
\end{equation} 
that is eq. (\ref{eq:action}) of the main text. For a single realization of the process the kernel is binary and the matrices can be obtained from the kernel (in this case the hypermatrix is a redundant representation). However, as we shall see, this does not apply in general to the averaged hypermatrix, where the correlations contain information about the ensemble fluctuations.

\subsubsection{Magnetic representation}
\label{MAGNETIC}

To switch to the magnetic representation we apply the usual bit-spin transformation $\sigma=2\,\varphi-1$:
\begin{multline}
\mathcal{A}\left(M|\,A,B\right)=\frac{1}{2}\sum_{i\in V}I_{i}^{0}\sum_{\alpha\in S}\sigma_{i}^{\alpha}+\frac{1}{2}\sum_{\alpha\in S}I_{0}^{\alpha}\sum_{i\in V}\sigma_{i}^{\alpha}+\\
+\frac{1}{4}\sum_{i\in V}\sum_{j\in V}(A_{ij}+A_{ij})\sum_{\alpha\in S}\sigma_{i}^{\alpha}+\frac{1}{4}\sum_{\alpha\in S}\sum_{\beta\in S}(B^{\alpha\beta}+B^{\beta\alpha})\sum_{i\in V}\sigma_{i}^{\alpha}+\\
+\frac{1}{4}\sum_{i\in V}\sum_{j\in V}A_{ij}\,\sum_{\alpha\in S}\sigma_{i}^{\alpha}\sigma_{j}^{\alpha}+\frac{1}{4}\sum_{\alpha\in S}\sum_{\beta\in S}B^{\alpha\beta}\sum_{i\in V}\sigma_{i}^{\alpha}\sigma_{i}^{\beta}\label{eq:cuhgkv-2}
\end{multline}
The structure of the interaction is identical except for a global rescaling 
\begin{equation}
\tilde{A}_{ij}=\frac{1}{4}A_{ij},\ \ \ \tilde{B}^{\alpha\beta}:=\frac{1}{4}B^{\alpha\beta},
\end{equation}
And an adjustment of currents with an additional term
\begin{equation}
\tilde{I}_{i}^{0}:=2\bar{\Omega}\sum_{j\in V}(\tilde{A}_{ij}-\tilde{A}_{ij})+\sum_{j\in V}(\tilde{A}_{ij}+\tilde{A}_{ji}),\ \ \
\tilde{I}_{0}^{\alpha}:=2\bar{\Omega}\sum_{\beta\in S}(\tilde{B}^{\alpha\beta}-\tilde{B}^{\beta\alpha})+\sum_{\beta\in S}(\tilde{B}^{\alpha\beta}+\tilde{B}^{\beta\alpha}).\label{eq:dvdvrg}
\end{equation}
The action in the magnetic representation is therefore
\begin{equation}
\mathcal{A}\,(M|\,\tilde{A},\tilde{B})=T\sum_{i\in V}\tilde{I}_{i}^{\,0}m_{i}+N\sum_{\alpha\in S}\tilde{I}_{0}^{\alpha}\mu^{\alpha}+T\sum_{i\in V}\sum_{j\in V}\tilde{A}_{ij}\,c_{ij}+N\sum_{\alpha\in S}\sum_{\beta\in S}\tilde{B}^{\alpha\beta}q^{\alpha\beta}
\end{equation}
In this case the hypermatrix will consists of $M$, $C$ and $Q$. We can recover the Ising Hamiltonian used in Schneidman et al.\cite{Schneidman2006} in the limit $T\rightarrow1$ and $B\rightarrow0$: 
\begin{equation}
\mathcal{A}\,(M|\,\tilde{A},0)=\sum_{i\in V}\sum_{j\in V}\tilde{A}_{ij}\,\sigma^{1}_{i} \sigma^{1}_{j}
\end{equation}
Thus, the max entropy principle is recovered as specific case of a field theory with zero kinetic energy. Since the theories are equivalent in the coming manipulations we will mainly use the spin representation.

\subsubsection{Classical Lagrange mechanics}

Let briefly recall the fundamental assumptions of the Lagrangian mechanics in continuous time: given the semi-compact (i.e. still not discretized in the time variable) real valued kernel 
\begin{equation}
X:=\left\{ x_{V}\left(t\right)\in\mathbb{R}^{N}:\,t\in\left[0,T\right]\right\} ,\ \ \ x_{V}\left(t\right):=\left\{ x_{i}\left(t\right)\in\mathbb{R}:\,i\in V\right\} 
\end{equation}
we assume the existence of the "Lagrangian" function, canonically interpreted \cite{Lagrangemechanics} as the difference between the kinetic and the potential energy of the system at given time. Then, the classical action functional is defined from the Lagrangian  by integrating over the time interval
\begin{equation}
\mathcal{A}\left(X\right):=\int_{0}^{T}dt\ \mathcal{L}\left[\,t,\,x_{V}\left(t\right),\,\partial_{t} x_{V}\left(t\right)\right],
\end{equation}
we denoted the derivative respect to time with $\partial_{t}:=d/dt$. Then, the associated evolution is a stationary point of the action, usually a minimum
\begin{equation}
Y:=\arg \ \min_{X} \  \mathcal{A}\left(X\right).
\end{equation}
The  equations of motion are also obtained from the Lagrangian, through the celebrated Euler-Lagrange equations
\begin{equation}
\frac{\partial\mathcal{L}}{\partial y_{i}}=\partial_{t}\left[\frac{\partial\mathcal{L}}{\partial\left(\partial_{t}y_{i}\right)}\right]
\end{equation}
that provide the stationary points of the action. On the other hand, determining weather a certain set of equations of motion
\begin{equation}
\Gamma_{V}\left(t,\,y_{V},\,\partial_{t}y_{V},\,\partial^2_{t}y_{V}\right)=0
\end{equation}
admit a Lagrangian descrition is known as the inverse problem of Lagrangian mechanics. Many authors contributed to this topic and the necessary conditions, called Helmoltz conditions, 
\begin{equation}
\frac{\partial\Gamma_{i}}{\partial(\partial^2_{t}y_{j})}=\frac{\partial\Gamma_{j}}{\partial(\partial^2_{t}y_{i})},
\end{equation}
\begin{equation}
\frac{\partial\Gamma_{i}}{\partial y_{j}}-\frac{\partial\Gamma_{j}}{\partial y_{i}}=\frac{1}{2}
\partial_{t}\left[\frac{\partial\Gamma_{i}}{\partial(\partial_{t}y_{j})}-\frac{\partial\Gamma_{j}}{\partial(\partial_{t}y_{i})}\right],
\end{equation}
\begin{equation}
\frac{\partial\Gamma_{i}}{\partial(\partial_{t}y_{j})}+\frac{\partial\Gamma_{j}}{\partial(\partial_{t}y_{i})}=2
\partial_{t}\left[\frac{\partial\Gamma_{j}}{\partial(\partial^2_{t}y_{i})}\right],
\end{equation}
have been worked out in many general contexts \cite{Helmolzcond1,Sarlet1982}. The problems dates back to Jacobi and has been attacked by many authors \cite{Douglas1939,Sarlet1982}. The lattice case has been studied by Crăciun and Opris \cite{Craciun1996}, Bourdin and Cresson \cite{Bourdin2013} and more recently by Gubbiotti \cite{Gubbiotti1,Gubbiotti2}. Although this has been indirectly shown already for many important models of neural network\cite{Fagerholm2021}, a systematic test of these conditions applied to the many equations of motions presented so far would be an important test for the quantum field theoretic description, which includes the free energy principle of Friston et al. \cite{Friston2010,Friston2019,Fagerholm2021}

\subsubsection{Lagrangian description of LFT} 
\label{lagrangianTHEO}

Assuming that the process evolves causally, it follows that the kinetic matrix $B$ must be upper triangular, that is, the state of the system at instant $\alpha$ depends only on the states realized in the previous $\beta\leq\alpha-1$. Therefore, we can define the sequence of time windows
\begin{equation}
\mathcal{S}:=\left\{ S_{\alpha}\subset S:\alpha\in S\right\} ,\ \ \ S_{\alpha}:=\left\{ 1\leq\beta\leq\alpha\right\}
\end{equation}
In this way it is possible to define the Lagrangian of the system
\begin{multline}
\mathcal{L}(\sigma_{V}^{S_{\alpha}}|\,A,B):=-\sum_{i\in V}\sum_{j\in V}A_{ij}\,\sigma_{i}^{\alpha}\sigma_{j}^{\alpha}-\sum_{\beta\in S_{\alpha-1}}B^{\alpha\beta}\sum_{i\in V}\sigma_{i}^{\alpha}\sigma_{i}^{\beta}=\\
=-\sum_{i\in V}\sum_{j\in V}A_{ij}\,\sigma_{i}^{\alpha}\sigma_{j}^{\alpha}-N\sum_{\beta\in S_{\alpha-1}}B^{\alpha\beta}q^{\alpha\beta}\label{eq:ssssss}
\end{multline}
where in the second line the definition of overlap $q^{\alpha\beta}$ is applied 
\begin{equation}
\mathcal{A}\left(M|\,A,B\right)=\sum_{\alpha\in S}\mathcal{L}(\sigma_{V}^{S_{\alpha}}|\,A,B).
\end{equation}
We can isolate the potential term from the kinetic term (which depends on the overlap)
\begin{equation}
H(\sigma_{V}^{\alpha}|\,A):=\sum_{i\in V}\sum_{j\in V}A_{ij}\,\sigma_{i}^{\alpha}\sigma_{j}^{\alpha},\ \ \ \ K(\,q^{\alpha S_{\alpha-1}}|\,B\,):=-N\sum_{\beta\in S_{\alpha-1}}B^{\alpha\beta}q^{\alpha\beta}.
\end{equation}
Thus, we can rewrite the Lagrangian in the canonical form\cite{Lagrangemechanics}
\begin{equation}
\mathcal{L}(\sigma_{V}^{S_{\alpha}}|\,A,B)=-H(\sigma_{V}^{\alpha}|\,A)+K(\,q^{\alpha S_{\alpha-1}}|\,B\,),
\end{equation}
where $q^{\alpha S_{\alpha-1}}$ is the $\alpha-$th row of the matrix of overlaps up to the time $\alpha-1$
\begin{equation}
q^{\alpha S_{\alpha-1}}:=\{q^{\alpha\beta}\in Q:\,\beta\in S_{\alpha-1}\}
\end{equation}
and this is enough to set the dynamics of the system. That the overlap-dependent term can be truly interpreted as a kinetic term is deduced by comparing with a simple Lagrangian system (see the work from Lee \cite{Lee1987} for an overview). We introduce the pulse (or "momentum") kernel
\begin{equation}
\partial M:=\left\{ \partial \sigma_{i}^{\alpha}\in\left\{ -2,\,0,\,2\right\} :\,\alpha\in S_{\alpha}\right\} ,\ \ \ 
\partial\sigma_{i}^{\alpha}:=\sigma_{i}^{\alpha}-\sigma_{i}^{\alpha-1}
\end{equation}
The lagrangian of the Markovian scalar field (without memory) is
\begin{equation}
\mathcal{L}(\sigma_{V}^{\alpha},\partial\sigma_{V}^{\alpha}|\,A,B_{0}):=-H(\sigma_{V}^{\alpha}|\,A)+\frac{1}{2}B_{0}\left\Vert \partial\sigma_{V}^{\alpha}\right\Vert _{2}^{2}
\end{equation}
with a few algebraic steps (e.g., Babylonian trick\cite{Kistler2021}) it can be shown that
\begin{equation}
\left\Vert \partial\sigma_{V}^{\alpha}\right\Vert _{2}^{2}=2N\,(1-q^{\alpha\alpha-1})
\end{equation}
therefore the Lagrangian can be rewritten as.
\begin{equation}
\mathcal{L}(\sigma_{V}^{\alpha},\partial\sigma_{V}^{\alpha}|\,A,B_{0})=-H(\sigma_{V}^{\alpha}|\,A)+B_{0}N\,(1-q^{\alpha\alpha-1}).
\end{equation}
In the case of our action, taking
\begin{equation}
B^{\alpha\beta}=-B_{0}\,\mathbb{I}\left(\alpha-1=\beta\right)
\end{equation}
where $\mathbb{I}(\cdot)$ is the indicator function. The associated Lagrangian becomes
\begin{equation}
\mathcal{L}(\sigma_{V}^{S_{\alpha}}|\,A,B)=-H(\sigma_{V}^{\alpha}|\,A)-B_{0}N\,q^{\alpha\alpha-1}
\end{equation}
and one can see immediately that the difference between the two Lagrangians
is
\begin{equation}
\mathcal{L}(\sigma_{V}^{S_{\alpha}}|\,A,B)-\mathcal{L}(\sigma_{V}^{\alpha},\,\partial\sigma_{V}^{\alpha}|\,A,B_{0})=-B_{0}N
\end{equation}
i.e. a constant that is irrelevant to the determination of dynamics. Moreover, the sign of the kinetic term is reversed with respect to that of the overlap term. It follows that the free field system is a sub-case of the general action described at the beginning, whose overlap term can be reduced to the kinetic term of the free Lagrangian.

\subsection{Statistical Field Theory}
\label{STAT}

So far, our theory is equivalent to a binary quantum field theory on lattice, i.e. \textbf{Qubit} field theory\cite{Gornitz1992,Deutsch2004,Singh2020,Franchini2023}, since the same results can be deduced by applying the \textbf{Wick rotation} (i.e., a rotation $i \rightarrow -1$  of the time units into the imaginary plane) \cite{WickRotation,Wickrotation1975,FaccioliSFT}  to a system of non-relativistic quantum oscillators  \cite{DiCastro1969,Wilson1974,Lee1983,Lee1987,Parisi1989,Wiese2009,Gupta2011,ParottoQCD2018,Buice2007,Buice2013,Fagerholm2021,Halverson2022}.
In general, the evolution of a Lagrangian system is determined by the principle of stationary action, which means that the kernel that satisfy it is not necessarily a
minimum of the action: it can also be a maximum, or a saddle point. Following Feynman\cite{FeynmanQM,FaccioliSFT}, for the quantum evolution we consider a Gibbs principle\cite{Huang2003} applied to the action, which is equivalent to the \textbf{principle of least action}. We define the action's partition function:
\begin{equation}
G\left(A,B\right)=\sum_{M\in\left\{ -1,1\right\} ^{VS}}\exp\left[-\lambda\mathcal{A}\left(M|\,A,B\right)\right]
\end{equation}
where we interpret the action as a Hamiltonian and look for its minimum. Here $\lambda$ is the inverse Planck constant, and plays the role of a temperature. The classical limit is recovered for $\lambda\rightarrow \infty$.
We also define the free action 
\begin{equation}
\Psi\left(A,B\right):=-\frac{1}{\lambda}\log G\left(A,B\right)
\end{equation}
which would be the analogue of the free energy. We then apply the steps to obtain the Gibbs principle \cite{Franchini2023}: first we manipulate the partition function, multiplying and dividing by a test measure to obtain the flat functional
\begin{equation}
\sum_{M\in\left\{ -1,1\right\} ^{VS}}\exp\left[-\lambda\mathcal{A}\left(M|\,A,B\right)\right]=\langle\exp\left[-\lambda\mathcal{A}\left(M|\,A,B\right)-\log\zeta\left(M\right)\right]\rangle_{\zeta}.
\end{equation}
Then we apply Jensen's inequality to the average versus the test measure
\begin{multline}
\langle\,\exp\left[-\lambda\mathcal{A}\left(M|\,A,B\right)-\log\zeta\left(M\right)\right]\rangle_{\zeta}\geq\\
\geq\exp\left[-\lambda\langle\mathcal{A}\left(M|\,A,B\right)\rangle_{\zeta}-\langle\,\log\zeta\left(M\right)\rangle_{\zeta}\right]=\exp\left[-\lambda\mathcal{F}\left(\zeta|A,B\right)\right]\label{eq:dss}
\end{multline}
so as to obtain the \textbf{free action} functional
\begin{equation}
\mathcal{F}\left(\zeta|\,A,B\right):=\langle\mathcal{A}\left(M|\,A,B\right)\rangle_{\zeta}+\frac{1}{\lambda}\langle\,\log\zeta\left(\sigma\right)\rangle_{\zeta}
\end{equation}
This functional is greater or equal to the free action for any test measure
\begin{equation}
\Psi\left(A,B\right)\leq\mathcal{F}\left(\zeta|\,A,B\right),\ \ \ \forall\zeta\in\mathcal{P}(\left\{ -1,1\right\} ^{VS})
\end{equation}
and one can see that the minimum is actually reached by the Gibbs measure
\begin{equation}
\eta\left(M|\,A,B\right):=\frac{1}{G\left(A,B\right)}\exp\left[-\lambda\mathcal{A}\left(M|\,A,B\right)\right]
\end{equation}
It can be verified that the measure satisfies the relationship with the free action
\begin{equation}
\mathcal{F}\left(\eta|\,A,B\right):=\inf_{\zeta\in\mathcal{P}\left(\left\{ -1,1\right\} ^{VS\,}\right)}\mathcal{F}\left(\zeta|\,A,B\right)=\Psi\left(A,B\right)
\end{equation}
If the system is assumed to be classical (i.e., non-quantum) the dynamics is obtained in the zero temperature limit. However, it could also have an intrinsic minimum temperature (equivalent to non zero Planck's constant).

\subsubsection{Description in terms of the correlation matrices}

Notice that in experiments where many neurons are recorded for long time, it may become useful to express the variational problem purely in terms of the correlation matrices, that are continuously supported. Recalling the definitions
\begin{equation} 
C\left(M\right):=M\,\,M^{\dagger}/T,\ \ \ Q\left(M\right):=M^{\dagger}M\,/N,
\end{equation}
and introducing the differential operators
\begin{equation}
dC:=\prod_{i\in V}\prod_{j\in V}dc_{ij},\ \ \ dQ:=\prod_{\alpha\in S}\prod_{\beta\in S}dq^{\alpha\beta},
\end{equation}
we can define the probability distribution of the pairs of correlation matrices that can be obtained from the same kernel, with a little abuse of notation
\begin{equation}
\mathcal{P}(C,Q):=\frac{1}{2^{\,NT}}\sum_{\Omega\in\left\{ 0,1\right\} ^{VS}}\,\prod_{i\in V}\prod_{j\in V}\delta[\,c_{ij}-c_{ij}\left(M\right)]\,\prod_{\alpha\in S}\prod_{\beta\in S}\delta[\,q^{\alpha\beta}-q^{\alpha\beta}\left(M\right)]
\end{equation}
Here, the product of delta functions is still not well defined in a mathematical sense, and there are many ways in which we could rigorously represent this distribution. For example, one could use the Fourier representation of the delta and implement the fact that by Singular Value Decomposition (SVD), the eigenvalues of the two matrices $C$ and $Q$ are the same, and that equating the spectra would be enough to characterize the distribution in the termodynamic limit. For this paper, we are not going to specify the exact form and assume only that the definition is such that the entropy density of $\mathcal{P}$ exists also for infinite $N$ and $T$, and that
\begin{equation}
\int_{C\in\left[-1,1\right]^{N^{2}}}dC\int_{Q\in\left[-1,1\right]^{T^{2}}}dQ\ \mathcal{P}(C,Q)=1
\end{equation}
Then, we could try to define a version of the free action that only depends on the correlation matrices. Let us introduce a transformed action functional
\begin{equation}
\mathcal{E}\left(C,Q|\,A,B\right):=NT\log2+\log\mathcal{P}(C,Q)+\mathcal{A}\left(C,Q|\,A,B\right),  
\end{equation}
the old action function $\mathcal{A}$ is as before
\begin{equation}
\mathcal{A}\left(C,Q|\,A,B\right) :=T\sum_{i\in V}\sum_{j\in V}A_{ij}c_{ij}+N\sum_{i\in V}\sum_{j\in V}B^{\alpha\beta}q^{\alpha\beta},
\end{equation}
using this new functional we can redefine the Gibbs measure
\begin{equation}
\eta\left(C,Q|\,A,B\right):=\frac{1}{G\left(A,B\right)}\exp\left[-\lambda\mathcal{E}\left(C,Q|\,A,B\right)\right]
\end{equation}
and the partition function
\begin{equation}
G\left(A,B\right)=\int_{C\in\left[-1,1\right]^{N^{2}}}dC\ \int_{Q\in\left[-1,1\right]^{T^{2}}}dQ\ \exp\left[-\lambda\mathcal{E}\left(C,Q|\,A,B\right)\right].
\end{equation}
Notice that the spectral properties of these matrices allows to connect with many other interesting settings: for example we showed that the PCA is a special form of LFT inference, with $I=0$ and $B=0$. Since in the applicative sections of this paper we will not deal with the large number of neurons (and clock times) at which this approximation is justified we don't discuss it further. We hope to address the matter in a dedicated work.

\subsubsection{Connection with Replica Theory}

The theory in the Lagrangian form allows to establish a connection with the replica theory \cite{Mezard1987,Charbonneau2023}
\begin{equation}
G\left(A,B\right)=\sum_{M\in\left\{ -1,1\right\} ^{VS}}\prod_{\alpha\in S}\exp\,[-\lambda\mathcal{L}(\sigma_{V}^{S_{\alpha}}|\,A,B)]
\end{equation}
and since the sum over the kernels is equivalent to a sum over $T$ replicas
of the system $\sigma_{V}$
\begin{equation}
\sum_{M\in\left\{ -1,1\right\} ^{VS}}=\sum_{\sigma_{V}^{1}\in\left\{ -1,1\right\} ^{V}}\sum_{\sigma_{V}^{2}\in\left\{ -1,1\right\} ^{V}}\,...\,\sum_{\sigma_{V}^{T}\in\left\{ -1,1\right\} ^{V}}
\end{equation}
the replicated system is obtained in the limit $B\rightarrow0$ (no kinetic term) 
\begin{equation}
\mathcal{A}\left(M|\,A,0\right)=-\sum_{\alpha\in S}H(\sigma_{V}^{\alpha}|\,A)
\end{equation}
with simple steps we arrive at the following relations
\begin{multline}
G\left(A,0\right):=\sum_{\sigma\in\left\{ -1,1\right\} ^{VS}}\prod_{\alpha\in S}\,\exp\,[-\lambda H(\sigma_{V}^{\alpha}|\,A)]=\\
=\sum_{\sigma_{V}^{1}\in\left\{ -1,1\right\} ^{V}}\exp\,[\,-\lambda H(\sigma_{V}^{1}|\,A)]\ ...\sum_{\sigma_{V}^{T}\in\left\{ -1,1\right\} ^{V}}\exp\,[\,-\lambda H(\sigma_{V}^{T}|\,A)]
=\prod_{\alpha\in S}Z\left(A\right)=Z\left(A\right)^{T}\label{eq:hhhhh}
\end{multline}
where $Z$ is the partition function associated with the Hamiltonian $H$ and
\begin{equation}
Z\left(A\right):=\sum_{\sigma_{V}^{1}\in\left\{ -1,1\right\} ^{V}}\exp\left[-\lambda H(\sigma_{V}^{1}|\,A)\right]
\end{equation}
As one can see, the partition function of the action converges to the partition function of the Hamiltonian replicated $T$ times. The interpretation of the replica trick\cite{Mezard1987}
\begin{equation}
\log Z\left(A\right)=\lim_{T\rightarrow0}\frac{1}{T}[Z\left(A\right)^{T}-1]=\lim_{T\rightarrow0}\frac{1}{T}\left[G\left(A,0\right)-1\right]
\end{equation}
is natural enough in this context: the formal limit $T\rightarrow0$ describes a situation in which the continuous limit of the theory ($\tau\rightarrow0$) is observed for an infinitesimal time.

\subsubsection{External input}

So far, we have only considered the evolution of an isolated system, but obviously in our case the input is crucial, so we must include it in the model. This can be done in a relatively simple way by introducing the Input kernel, which describes the input signal in the network
\begin{equation}
\mathcal{I}\left(M|\,I\right):=-\sum_{\alpha\in S}I_{V}^{\alpha}\cdot\sigma_{V}^{\alpha}
\end{equation}
which should be added to the action to obtain the description of the full system
\begin{equation}
\mathcal{A}\left(M|\,A,B,I\right):=\mathcal{A}\left(M|\,A,B\right)-\mathcal{I}\left(M|\,I\right)
\end{equation}
By introducing the input partition function
\begin{equation}
R\left(I\right):=\sum_{M\in\left\{ -1,1\right\} ^{VS}}\exp\,\left[-\lambda\mathcal{I}\left(M|\,I\right)\right]=\prod_{\alpha\in S}\prod_{i\in V}\,2\cosh\left(\lambda I_{i}^{\alpha}\right)
\end{equation}
and applying Gibbs principle we find the distribution of the input
\begin{equation}
\rho\left(M|\,I\right):=\frac{1}{R\left(I\right)}\exp\left[-\lambda\mathcal{I}\left(M|\,I\right)\right]=\frac{1}{R\left(I\right)}\prod_{\alpha\in S}\prod_{i\in V}\,\exp\left(-\lambda I_{i}^{\alpha}\sigma_{i}^{\alpha}\right)
\end{equation}
The partition function of the general action can be expressed in terms of the average of the isolated state respect to $\rho$ 
\begin{equation}
G\left(A,B,I\right)=R\left(I\right)\langle\,\exp\left[-\lambda\mathcal{A}\left(M|\,A,B\right)\right]\rangle_{\rho}
\end{equation}
Note also that the partition can also be expressed as the average
of the input over the measure of the isolated system 
\begin{equation}
G\left(A,B,I\right)=G\left(A,B\right)\langle\exp\left[-\lambda\mathcal{I}\left(M|\,I\right)\right]\rangle_{\eta}
\end{equation}
from which a relationship between partition functions and averages over states
\begin{equation}
R\left(I\right)\langle\,\exp\left[-\lambda\mathcal{A}\left(M|\,A,B\right)\right]\rangle_{\rho}=G\left(A,B\right)\langle\exp\left[-\lambda\mathcal{I}\left(M|\,I\right)\right]\rangle_{\eta}
\end{equation}
For example, the input kernel could model the signal arriving to the observed cortical area after a stimulus. In case of a motor task\cite{Bardella2020,Pani2022} could be the thalamic input arriving to the boundary neurons (in a topological sense) of the recorded cortical region following the Go stimulus, that is expected to be steady-state almost everywhere except around the time interval at which the motor plan is realized. If axonal and synaptic connections are reasonably stable then most of the observed variability could come from input noise from the rest of the network, or slightly different initial conditions, etc. To
include all possible effects one can introduce a "quenched" space-time noise term, i.e. a random field $\delta$ to be added to the input term $I$
\begin{equation}
\mathcal{I}\left(M|\,I,\,\delta\right):=-\sum_{\alpha\in S}I_{V}^{\alpha}\cdot\sigma_{V}^{\alpha}-\sum_{\alpha\in S}\delta_{V}^{\alpha}\cdot\sigma_{V}^{\alpha}
\end{equation}
which statistically mimic the input noise on the time scale of the entire session. In
case of the recordings described in the main text\cite{Bardella2020,Pani2022} we expect that quenched noise terms can be ignored.

\subsubsection{Ground state of the action and order parameter}
\label{groundstate}

The variational principle identifies a distribution $\eta$ called the ground state of the action (GS), which would be the one from which the mutielectrode interface draws the states we observe at the single trial level. Note that the GS of the action is a distribution in the space of kernels $\{-1,1\}^{VS}$, and hence a natural order parameter would be the kernel of the GS
\begin{equation}
\langle M\rangle_{\eta}:=\left\{ \langle\sigma_{i}^{\alpha}\rangle_{\eta} \in \mathbb{R}: \,i\in V,\,\alpha\in S\right\}. 
\end{equation}
The kernel of the GS is of particular interest since one can directly obtain the average momentum kernel $ \langle \partial M\rangle_{\eta}$ and all kernels derived from linear operations. It also allows to determine and subtract the steady state of the "hold" phase before the motor plan is observed (see main text). Notice that averages of first-order observables, such as the offset, or row and column averages, can be computed directly from the average kernel because they are related to it by a linear relationship. Additionally, if the connected correlation matrices are negligible then the correlation matrices can be deduced from the average kernel since $C_{0}$ and $Q_{0}$ depend on the averages. The amplitude of the kernel cell fluctuations satisfies the relation
\begin{equation}
\langle [\sigma_{i}^{\alpha} - \langle\sigma_{i}^{\alpha}\rangle_{\eta} ]^{2}\rangle_{\eta}=1-\langle\sigma_{i}^{\alpha}\rangle_{\eta}^{2}
\end{equation}
However, the relationship does not apply in general, 
\begin{multline}
\langle [\sigma_{i}^{\alpha}- \langle\sigma_{i}^{\alpha} \rangle_{\eta} ][\sigma_{j}^{\beta}- \langle \sigma_{j}^{\beta}\rangle_{\eta}]\rangle_{\eta}=\\
=\langle\sigma_{i}^{\alpha}\sigma_{j}^{\beta}\rangle_{\eta} -\langle\sigma_{i}^{\alpha}\rangle_{\eta}\langle\sigma_{j}^{\beta}\rangle_{\eta} -\langle\sigma_{i}^{\alpha}\rangle_{\eta} \langle\sigma_{j}^{\beta}\rangle_{\eta} + \langle\sigma_{i}^{\alpha}\rangle_{\eta} \langle\sigma_{j}^{\beta}\rangle_{\eta}
=\langle\sigma_{i}^{\alpha}\sigma_{j}^{\beta}\rangle_{\eta}-\langle\sigma_{i}^{\alpha}\rangle_{\eta} \langle\sigma_{j}^{\beta}\rangle_{\eta}\label{eq:ddds}
\end{multline}
Thus, the average kernel may not be a sufficient order parameter to fully describe the GS $\eta$, and we should also look at second-order variables. To verify this we can compute the ensemble covariance matrices 
\begin{equation}
\langle \delta C\rangle_{\eta}:=\langle C\rangle_{\eta} - \langle M\rangle_{\eta}\langle M\rangle_{\eta}^{\dagger}/T, \ \ \ 
\langle \delta Q\rangle_{\eta}:=\langle Q\rangle_{\eta} - \langle M\rangle_{\eta}^{\dagger}\langle M\rangle_{\eta}/N.
\label{eq:covariances}
\end{equation}
If these matrices are non-trivial it means that the correlations cannot be reconstructed from the average kernel. Thus, the most general order parameter in this approximation (two-body non-relativistic) is the \textbf{hypermatrix}, composed by the average kernel $\langle M\rangle_{\eta}$ and the average correlation matrices $\langle Q\rangle_{\eta}$ and $\langle C\rangle_{\eta}$.

\subsubsection{Repeated experiments and ensemble average}

Let us see how we might model an actual experimental observation of these matrices. A possible method is to replicate an experiment $n$ times and find a way to average the results in such a way to estimate the ensemble average. First, we need to index the trials with the label $k$, the span of the index is denoted by
\begin{equation}
W:=\left\{ 1\leq k\leq n\right\},
\end{equation}
the empirical ensemble collecting the actually observed kernels, called recording session in \cite{Pani2022}, can be represented as follows
\begin{equation}
\mathcal{W}:=\{M_{k}\in\left\{ -1,1\right\} ^{VS}:k\in W\},
\end{equation}
where $M_{k}$ is the $k-$th trial of the session
\begin{equation}
M_{k}:=\{\varphi_{ik}^{\alpha}\in\left\{ 0,1\right\} :\,i\in V,\,\alpha\in S\}.
\end{equation}
We added the new trial index $k$ in the lower part of the $\varphi$ symbol, though in principle it should have been placed on top as the idea of repeated experiments would suggest a kind of time variable. Since the trials are usually designed to be independent, the order in which they are performed should not matter for the $k$ index. This is an ideal situation that should be verified for each experimental set, but it is also a reasonable approximation of the experimenter's intentions. To confront the experiments and precisely define the empirical averages, we need one last ingredient: to choose the proper synchronization of the experimental kernels. Then we introduce the integer vector
\begin{equation}
\nu_{W}:=\left\{ \nu_{W}\in\mathbb{Z}:\,k\in W\right\} 
\end{equation}
that collects the relative time shifts of the trials, ie for the $k-$th trial the $\alpha$ index is shifted by $\nu_{k}$ units of clock time, that is equivalent to apply the substitution
\begin{equation}
\alpha\rightarrow\alpha+\nu_{k}
\end{equation}
We formally indicate the application of the timeshifts to the empirical ensemble with
\begin{equation}
\mathcal{W}\rightarrow\mathcal{W}\left(\nu_{W}\right).
\end{equation}
We argue that for some $\nu_{W}$, the average on the empirical set converges to the ensemble average in the ideal limit of infinite repetitions of the same experiments
\begin{equation}
\langle\mathcal{O}\left(M\right)\rangle_{\eta}=\lim_{n\rightarrow\infty}\ \frac{1}{n}\sum_{M\in\mathcal{W}\left(\nu_{W}\right)}\mathcal{O}\left(M\right).
\end{equation}
For example, the hypermatrix shown in Figure \ref{fig:1} is obtained by choosing $\nu_{W}$ in such a way to sinchronize the replicas of the experiment with respect to the movement onset (see Section \ref{Apparatus and task}), while in \cite{Pani2022}, the alignment is by Go signal. In principle, one should also consider more complex kinds of synchronizations, like alignments that are based on maximizing the correlations between the trials. Let us introduce the overlap matrix for the session
\begin{equation}
\mathcal{Q}\left(\nu_{W}\right):=\{\mathcal{Q}_{kk'}\left(\nu_{W}\right)\in\left[0,1\right]:\,k,k'\in W\},
\end{equation}
where the entries are the overlaps between the trials $k$ and $k'$ with timeshifts $\nu_{W}$, 
\begin{equation}
\mathcal{Q}_{kk'}\left(\nu_{W}\right):=\frac{1}{T}\sum_{\alpha\in S}\,\frac{1}{N}\sum_{i\in V}\,\varphi_{ik}^{\alpha+\nu_{k}}\,\varphi_{ik'}^{\alpha+\nu_{k'}}.
\end{equation}
For example, we may want to align the samples respect to some vector $\nu_{W}^{*}$ such that $\nu_{k}^{go}\leq\nu_{k}^{*}\leq\nu_{k}^{mov}$ and that it maximizes the norm of the overlap matrix
\begin{equation}
\left\Vert \mathcal{Q}\left(\nu_{W}^{*}\right)\right\Vert _{2}^{2}=\sup_{\nu_{W}}\ \left\Vert \mathcal{Q}\left(\nu_{W}\right)\right\Vert _{2}^{2},
\end{equation}
anyway, notice that the minimization of such kind of functionals could become soon impractical for large datasets. We will show an example of rank reduction method (by renormalization) in Section 3.4.7.

\subsubsection{Inference methods}

Since the work of Schneidman et al.\cite{Schneidman2006}, the possibility of reconstructing the coupligs has become a major goal in computational neuroscience, and powerful inference methods are now available \cite{Nguyen2017, Carleo2019, Zdeborova2016}. Let consider a relativistic theory in the mixed space $\ell\in\Lambda$ and let consider a free field model with input kernel $I$. If we assume the action is that of a free (non-interacting ) theory \cite{FranchiniREM2023}, then the parameters can be obtained easily from the average magnetizations by inverting the Callen equations \cite{Albert2014,Swendsen1984}
\begin{equation}
\lambda I_{\ell}=\tanh^{-1}\left(m_{\ell}\right),\ \ \ m_{\ell}:=\langle\sigma_{\ell}\rangle,
\end{equation}
Clearly there is an error that depends on the number of samples of the empirical ensemble average. Calling n the number of experiments, or replicas, on which the ensemble average is taken (e.g. the ”session trials” of \cite{Pani2022}), the error is as follows
\begin{equation}
\lambda\delta I_{\ell}=\frac{\cosh\left(\lambda I_{\ell}\right)}{\sqrt{n}}.
\end{equation}
We can go further and add two body interactions, in this case the problem becomes less trivial, and we have to deal with the so called inverse Ising problem, a classic inference problem \cite{Albert2014,Swendsen1984,Aurell2012,Nguyen2017}. Let switch on the two body interactions $F_{\ell\ell'}$ and introduce the “grand covariance” of the mixed space 
\begin{equation}
\mathcal{C}_{\ell\ell'}\left(\eta\right):=\langle\sigma_{\ell}\sigma_{\ell'}\rangle_{\eta}-\langle\sigma_{\ell}\rangle_{\eta}\langle\sigma_{\ell'}\rangle_{\eta}.
\end{equation}
The values of the coupling parameters are ultimately recovered by inverting the following system of (possibly non linear) equations:\cite{Nguyen2017}
\begin{equation}
\mathcal{C}_{\ell\ell'}=\frac{\partial m_{\ell}}{\partial I_{\ell'}}=\frac{\partial^{2}\Psi}{\partial I_{\ell}\partial I_{\ell'}},
\end{equation}
that must be solved and then inverted to find the $F$ matrix. There are various methods to do so \cite{Albert2014,Swendsen1984,Aurell2012}, and an excellent survey is that of Nguyen et al.\cite{Nguyen2017}. There are also several useful approximate formulas that only require to invert the grand covariance, the simplest is
\begin{equation}
F_{\ell\ell'}^{\mathrm{LR}}=-\left(\mathcal{C}^{-1}\right)_{\ell\ell'}
\end{equation}
that correspond to the so-called “naive” mean-field theory \cite{Nguyen2017}. More advanced formulas depending on the inverse covariance are, for example, the TAP formula \cite{Nguyen2017}
\begin{equation}
F_{\ell\ell'}^{\mathrm{TAP}}=\frac{2\left(\mathcal{C}^{-1}\right)_{\ell\ell'}}{1+\sqrt{1-8\,m_{\ell}\,m_{\ell'}\left(\mathcal{C}^{-1}\right)_{\ell\ell'}}},
\end{equation}
the “independent-pair” approximation formula \cite{Nguyen2017}
\begin{equation}
F_{\ell\ell'}^{\mathrm{IP}}=\frac{1}{4}\ln\frac{\left[\left(1+m_{\ell}\right)\left(1+m_{\ell'}\right)+\mathcal{C}_{\ell\ell'}\right]\left[\left(1-m_{\ell}\right)\left(1-m_{\ell'}\right)+\mathcal{C}_{\ell\ell'}\right]}{\left[\left(1+m_{\ell}\right)\left(1-m_{\ell'}\right)-\mathcal{C}_{\ell\ell'}\right]\left[\left(1-m_{\ell}\right)\left(1+m_{\ell'}\right)-\mathcal{C}_{\ell\ell'}\right]},
\end{equation}
and the Sessak–Monasson formula \cite{Nguyen2017,SessakMonasson}
\begin{equation}
F_{\ell\ell'}^{\mathrm{SM}}=F_{\ell\ell'}^{\mathrm{IP}}-\left(\mathcal{C}^{-1}\right)_{\ell\ell'}-\frac{\mathcal{C}_{\ell\ell'}}{\left(1-m_{\ell}^{2}\right)\left(1-m_{\ell'}^{2}\right)-\mathcal{C}_{\ell\ell'}^{2}},
\end{equation}
especially suited in the limit of small covariances. Notice that the presence of the pair interactions also modify the expression of the external fields. Introducing the Legendre transform of the free energy respect to the magnetizations,
\begin{equation}
\Gamma:=\max_{I_{\Lambda}\in\mathbb{R}^{\Lambda}}\,\left\{ \,I_{\Lambda}\cdot m_{\Lambda}+\Psi\right\} ,
\end{equation}
the equations for both parameters are obtained from \cite{Nguyen2017}
\begin{equation}
I_{\ell}=\frac{\partial\Gamma}{\partial m_{\ell}},\ \ \ \left(\mathcal{C}^{-1}\right)_{\ell\ell'}=\frac{\partial\Gamma}{\partial m_{\ell}\partial m_{\ell'}}.
\end{equation}
Using the double index the equations for the $F$ parameters are
\begin{equation}
\mathcal{C}_{ij}^{\,\alpha\beta}=\frac{\partial^{2}\Psi}{\partial I_{i}^{\alpha}\partial I_{j}^{\beta}}.
\end{equation}
since in the non-relativistic approximation we ignored correlations where both upper and lower pairs of indices are different, that can be translated in
\begin{equation}
\mathcal{C}_{ij}^{\,\alpha\beta}=\mathcal{C}_{ij}^{\,\alpha\alpha}\,\mathbb{I}\left(\alpha=\beta\right)+\mathcal{C}_{ii}^{\,\alpha\beta}\,\mathbb{I}\left(i=j\right),
\end{equation}
the relations with the elements of the covariance matrices are
\begin{equation}
\delta c_{ij}\left(\eta\right)=\frac{1}{T}\sum_{\alpha\in S}\mathcal{C}_{ij}^{\,\alpha\alpha},\ \ \ \delta q^{\alpha\beta}\left(\eta\right)=\frac{1}{N}\sum_{i\in V}\mathcal{C}_{ii}^{\,\alpha\beta}.
\end{equation}
This reduces from $T^{2}N^{2}$ to $NT(N+T)$ the number of parameters that sholuld be actually computed to reconstruct the action, greatly enhancing the computational tractability in the non-relativistic case.

\subsubsection{Renormalization}
\label{Renormalization}

We conclude the general theoretical part with a simple renormalization \cite{DiCastro1969,Wilson1983} scheme based on \cite{Franchini2023} that will be useful to link theory with experimental observations. From this subsection we switch again to the lattice gas representation. Consider a joint kernel partition as in Section 3 of Franchini 2023\cite{Franchini2023}, with two levels (equivalent to one step Replica Symmetry Breaking: RSB1). Let $N_{1}$, $N_{2}$, $T_{1}$ and $T_{2}$ be numbers such that $N=N_{1}N_{2}$ and $T=T_{1}T_{2}$, and let 
\begin{equation}
\begin{split}
&V_{0}=\left\{ 1\leq  i_{1}\leq N_{1}\right\} ,\  V_{i_{1}}= \ \left\{ 1 \leq i_{2}\leq N_{2}\right\} , \  S_{0}=\left\{ 1\leq\alpha_{1}\leq T_{1}\right\} ,\  S_{\alpha_{1}}=\left\{ 1\leq\alpha_{2}\leq T_{2}\right\} .    
\end{split}
\end{equation}
The kernel can be rewritten according to the new multiscale index
\begin{equation}
\Omega=\left\{ \Omega_{\,i_{1}}^{\alpha_{1}}\in\left\{ 0,1\right\} :\,i_{1}\in V_{0},\,\alpha_{1}\in S_{0}\right\} 
\end{equation}
where we introduced the sub-kernels 
\begin{equation}
\Omega_{i_{1}}^{\alpha_{1}}:=\left\{ \varphi_{\,i_{1}i_{2}}^{\alpha_{1}\alpha_{2}}\in\left\{ 0,1\right\} :\,i_{2}\in V_{i_{1}},\,\alpha_{2}\in S_{\alpha_{1}}\right\} 
\end{equation}
the field is renormalized according to a map such that
\begin{equation}
\hat{\varphi}_{i_{1}}^{\alpha_{1}}:={\textstyle \mathcal{R}(\,\Omega_{i_{1}}^{\alpha_{1}})}\in\left\{ 0,1\right\} 
\end{equation}
to regain some binary variable, i.e. $\hat{\varphi}_{i_{1}}^{\alpha_{1}}$ will be one if within the cell $V_{i_{1}}S_{\alpha_{1}}$ the condition set by the renormalization map is verified, and zero otherwise. By construction, the relationship between the two variables is such that
\begin{equation}
\Omega_{i_{1}}^{\alpha_{1}}=\hat{\varphi}_{i_{1}}^{\alpha_{1}}\Omega_{i_{1}}^{\alpha_{1}}
\end{equation}
We can define the renormalized kernel as follows:
\begin{equation}
\hat{\Omega}:=\left\{ \hat{\varphi}_{\,i_{1}}^{\alpha_{1}}\in\left\{ 0,1\right\} :\,i_{1}\in V_{0},\,\alpha_{1}\in S_{0}\right\} 
\end{equation}
Since the action structure is symmetrical between space and time, we can also perform the calculations on the potential term alone. We apply the multiscale index 
\begin{equation}
\sum_{i\in V}\sum_{j\in V}A_{ij}\,\sum_{\alpha \in S}\varphi_{i}^{\alpha}\varphi_{j}^{\alpha}=\sum_{i_{1}\in V_{0}}\sum_{j_{1}\in V_{0}}\sum_{i_{2}\in V_{i_{1}}}\sum_{j_{2}\in V_{j_{1}}}A_{i_{1}i_{2}j_{1}j_{2}}\sum_{\alpha_{1}\in S_{0}}\sum_{\alpha_{2}\in S_{\alpha1}}\varphi_{i_{1}i_{2}}^{\alpha_{1}\alpha_{2}}\varphi_{j_{1}j_{2}}^{\alpha_{1}\alpha_{2}}
\end{equation}
and then the renormalization map 
\begin{equation}
\sum_{i\in V}\sum_{j\in V}A_{ij}\,\sum_{\alpha \in S}\varphi_{i}^{\alpha}\varphi_{j}^{\alpha}=\sum_{i_{1}\in V_{0}}\sum_{j_{1}\in V_{0}}\sum_{\alpha_{1}\in S_{0}}\hat{A}_{i_{1}j_{1}}^{\alpha_{1}}(\Omega)\,\hat{\varphi}_{i_{1}}^{\alpha_{1}}\hat{\varphi}_{j_{1}}^{\alpha_{1}}
\end{equation}
For example, for a bin renormalization
\begin{equation}
\hat{\varphi}_{i_{1}}^{\alpha_{1}}=\mathbb{I}(\Omega_{i_{1}}^{\alpha_{1}}\neq0)
\end{equation}
the effective interaction will be given by
\begin{equation}
\hat{A}_{i_{1}j_{1}}^{\alpha_{1}}(\Omega):=\sum_{i_{2}\in V_{i_{1}}}\sum_{j_{2}\in V_{j_{1}}}A_{i_{1}i_{2}j_{1}j_{2}}\sum_{\alpha_{2}\in S_{\alpha_{1}}}\varphi_{i_{1}i_{2}}^{\alpha_{1}\alpha_{2}}\varphi_{j_{1}j_{2}}^{\alpha_{1}\alpha_{2}}
\end{equation}
while for a renormalization by decimation (Kadanoff renormalization)\cite{Kadanoff2011,Efrati2014Real-spaceMechanics,Wilson1983}, 
\begin{equation}
\hat{\varphi}_{i_{1}}^{\alpha_{1}}=\varphi_{i_{1}1}^{\alpha_{1}1}
\end{equation}
we will have that
\begin{equation}
\hat{A}_{i_{1}j_{1}}^{\alpha_{1}}(\Omega):=\sum_{i_{2}\in V_{i_{1}}\setminus\{1\}}\sum_{j_{2}\in V_{j_{1}}\setminus\{1\}}A_{i_{1}i_{2}j_{1}j_{2}}\sum_{\alpha_{2}\in S_{\alpha_{1}}\setminus\{1\}}\varphi_{i_{1}i_{2}}^{\alpha_{1}\alpha_{2}}\varphi_{j_{1}j_{2}}^{\alpha_{1}\alpha_{2}}
\end{equation}
We separate the stationary term (if any)
\begin{equation}
\hat{A}_{i_{1}j_{1}}^{\alpha_{1}}(\Omega):=\hat{A}{}_{i_{1}j_{1}}+\delta\hat{A}_{i_{1}j_{1}}^{\alpha_{1}}(\Omega)
\end{equation}
The stationary term corresponds to the renormalized coupling matrix; we can thus rewrite the action potential term by separating the renormalized part from the fluctuation
\begin{equation}
\sum_{i\in V}\sum_{j\in V}A_{ij}\,\sum_{\alpha}\varphi_{i}^{\alpha}\varphi_{j}^{\alpha} =\sum_{i_{1}\in V_{0}}\sum_{j_{1}\in V_{0}}\hat{A}{}_{i_{1}j_{1}}\sum_{\alpha_{1}\in S_{0}}\hat{\varphi}_{i_{1}}^{\alpha_{1}}\hat{\varphi}_{j_{1}}^{\alpha_{1}}+\sum_{i_{1}\in V_{0}}\sum_{j_{1}\in V_{0}}\sum_{\alpha_{1}\in S_{0}}\delta\hat{A}_{i_{1}j_{1}}^{\alpha_{1}}(\Omega)\,\hat{\varphi}_{i_{1}}^{\alpha_{1}}\hat{\varphi}_{j_{1}}^{\alpha_{1}}\label{eq:yyyy-1}
\end{equation}
Doing the same with the kinetic term
\begin{equation}
\hat{B}_{i_{1}}^{\alpha_{1}\beta_{1}}(\Omega):=\sum_{\alpha_{2}\in S_{\alpha_{1}}}\sum_{\beta_{2}\in S_{\alpha_{1}}}B^{\alpha_{1}\alpha_{2}\beta_{1}\beta_{2}}\sum_{i_{2}\in V_{i_{1}}}\varphi_{i_{1}i_{2}}^{\alpha_{1}\alpha_{2}}\varphi_{i_{1}i_{2}}^{\beta_{1}\beta_{2}}
\end{equation}
and separating the uniform term
\begin{equation}
\hat{B}_{i_{1}}^{\alpha_{1}\beta_{1}}(\Omega):=\hat{B}^{\alpha_{1}\beta_{1}}+\delta\hat{B}_{i_{1}}^{\alpha_{1}\beta_{1}}(\Omega)
\end{equation}
the treatment is completely symmetrical, leading to 
\begin{equation}
\sum_{\alpha\in S}\sum_{\beta\in S}B^{\alpha\beta}\sum_{i\in V}\varphi_{i}^{\alpha}\varphi_{i}^{\beta}
=\sum_{\alpha_{1}\in S_{0}}\sum_{\beta_{1}\in S_{0}}\hat{B}^{\alpha_{1}\beta_{1}}\sum_{i_{1}\in V_{0}}\hat{\varphi}_{i_{1}}^{\alpha_{1}}\hat{\varphi}_{i_{1}}^{\beta_{1}}+\sum_{\alpha_{1}\in S_{0}}\sum_{\beta_{1}\in S_{0}}\sum_{i_{1}\in V_{0}}\hat{B}_{i_{1}}^{\alpha_{1}\beta_{1}}(\Omega)\,\hat{\varphi}_{i_{1}}^{\alpha_{1}}\hat{\varphi}_{i_{1}}^{\beta_{1}}\label{eq:yyyy-1-1}
\end{equation}
The action in the renormalized variables will therefore have a perturbation
\begin{multline}
G\left(A,B\right)=\sum_{\Omega\in\left\{ 0,1\right\} ^{VS}}\exp\left[-\lambda\mathcal{A}\left(\Omega|\,A,B\right)\right]
=\\
=\sum_{\hat{\Omega}\in\left\{ 0,1\right\} ^{V_{0}S_{0}}}\exp\,[-\lambda\mathcal{A}\,(\hat{\Omega}|\,\hat{A},\hat{B})-\lambda\delta\mathcal{A}(\hat{\Omega}|\,A,B)] = G\,(\hat{A},\hat{B})\langle\exp\,[-\lambda\delta\mathcal{A}\,(\hat{\Omega}|\,A,B)]\rangle_{\hat{\eta}}
\end{multline}
where $\hat{\eta}$ is the GS of the renormalized action. In general, this expression depends on the details of the couplings within the renormalized cell. The perturbation of the action is formally defined as 
\begin{equation}
\delta\mathcal{A}\,(\hat{\Omega}|\,A,B):=-\frac{1}{\lambda}\ \log\sum_{\Omega\in\mathcal{K}\left(\hat{\Omega}\right)}\exp\,[-\lambda\Gamma\,(\Omega,\hat{\Omega}|A,B)]
\end{equation}
where the sum is on those $\Omega$ that if renormalized are equal
to $\hat{\Omega}$, i.e. 
\begin{equation}
\mathcal{K}\,(\hat{\Omega}):=\{ \Omega\in\left\{ 0,1\right\} ^{VS}\,:\mathcal{R}\left(\Omega\right)=\hat{\Omega}\} 
\end{equation}
and the function $\Gamma$ is defined as follows:
\begin{equation}
\Gamma\,(\Omega,\hat{\Omega}|A,B):=\sum_{i_{1}\in V_{0}}\sum_{j_{1}\in V_{0}}\sum_{\alpha_{1}\in S_{0}}\delta\hat{A}_{i_{1}j_{1}}^{\alpha_{1}}(\Omega)\,\hat{\varphi}_{i_{1}}^{\alpha_{1}}\hat{\varphi}_{j_{1}}^{\alpha_{1}}
+\sum_{\alpha_{1}\in S_{0}}\sum_{\beta_{1}\in S_{0}}\sum_{i_{1}\in V_{0}}\delta\hat{B}_{i_{1}}^{\alpha_{1}\beta_{1}}(\Omega)\,\hat{\varphi}_{i_{1}}^{\alpha_{1}}\hat{\varphi}_{i_{1}}^{\beta_{1}}
\end{equation}
Thus, renormalization operations can change also the structure of the action. For example, consider the potential part: we can approximate the renormalized coupling fluctuations with a stationary Random Energy Model (REM universality, see Arous \& Kuptsov\cite{Arous2009} or Section 6 of Franchini 2023\cite{Franchini2023, FranchiniREM2023} for a practical example in kernel language)
\begin{equation}
\delta\hat{A}_{i_{1}j_{1}}^{\alpha_{1}}(\Omega)\approx J{}_{i_{1}j_{1}}\,(\Omega)\sqrt{\Delta_{i_{1}j_{1}}}
\end{equation}
the partition function can be approximated as follows 
\begin{multline}
\sum_{\Omega\in\mathcal{K}\left(\hat{\Omega}\right)}\exp\left[-\lambda\sum_{i_{1}\in V_{0}}\sum_{j_{1}\in V_{0}}\sum_{\alpha_{1}\in S_{0}}\delta\hat{A}_{i_{1}j_{1}}^{\alpha_{1}}(\Omega)\,\hat{\varphi}_{i_{1}}^{\alpha_{1}}\hat{\varphi}_{j_{1}}^{\alpha_{1}}\right]\approx\\
\approx\sum_{\Omega\in\mathcal{K}\left(\hat{\Omega}\right)}\exp\left[-\lambda\sum_{i_{1}\in V_{0}}\sum_{j_{1}\in V_{0}}J{}_{i_{1}j_{1}}\,(\Omega)\sqrt{\Delta_{i_{1}j_{1}}}\,\sum_{\alpha_{1}\in S_{0}} \hat{\varphi}_{i_{1}}^{\alpha_{1}}\hat{\varphi}_{j_{1}}^{\alpha_{1}}\right]=\\
=\exp\left(\,\hat{\lambda}^{2}T_{1}^{2}\sum_{i_{1}\in V_{0}}\sum_{j_{1}\in V_{0}}\Delta_{i_{1}j_{1}}\,\hat{\phi}_{i_{1}j_{1}}^{\,2}\right)\label{eq:ghlo}
\end{multline}
where in the second row we applied the PPP-REM \cite{Franchini2023} average and $\hat{\lambda}$ is the renormalized temperature. In essence, this type of mean field approximation introduces a linear term in the renormalization map
\begin{equation}
A_{ij}\rightarrow\hat{A}_{i_{1} j_{1}}-\hat{\lambda} T_{1} \Delta_{i_{1} j_{1}}\,\hat{\phi}_{i_{1} j_{1}}+\ ...
\end{equation}
which results in a quadratic term added to the action
\begin{equation}
\sum_{i\in V}\sum_{j\in V}A_{ij}\,\phi_{ij}\rightarrow\sum_{i_{1}\in V_{0}}\sum_{j_{1}\in V_{0}}\hat{A}{}_{i_{1} j_{1}}\,\hat{\phi}_{i_{1} j_{1}}-\hat{\lambda} T_{1} \sum_{i_{1}\in V_{0}}\sum_{j_{1}\in V_{0}}\Delta_{i_{1} j_{1}}\,\hat{\phi}_{i_{1} j_{1}}^{\,2}+\ ...
\end{equation}
Then, in first approximation we could ignore the corrections terms in the PMd experimets with Utha96, due to the small magnitude of the correlations. Notice that this could also explain other deviations from the max entropy principle like those shown in Figure 2 of Meshulam et al.\cite{Meshulam2021}. More accurate renormalization schemes based on multi-scale analysis can be computed following the methods of Franchini 2023\cite{Franchini2023, Franchini2021,FranchiniREM2023} and many others methods as well\cite{Wilson1983,Kadanoff2011,Efrati2014Real-spaceMechanics,Tiberi2022, AngeliniRG2023}, although in general the exact shape of the perturbations depend on the details of the system and on the instrumental limits and systematics, and to push further it is therefore necessary to introduce more specified information about the couplings and the kinetic properties of the system, both of the neocortex and the sensor.

\subsection{Compactification}

We conclude the striclty theoretical part by introducing the compactified observables described in the first sections of Franchini 2023 \cite{Franchini2023} and their physical significance. Although the compactified variables are easily evaluated, even from the kernel in matrix form, it is important to understand that these are ideally designed to deal with limit kernels, such as those obtained from systems that are close to the thermodynamic limit, or that are observed for extremely long times, typically many times the mean time to reach equilibrium in the case of statistical mechanics problems. Therefore, they are  designed to analyze very large kernels, in which both rows and columns are in such numbers to have approximately a continuous support of averages and correlations, and their use in experimental contexts should be carefully evaluated on a case-by-case basis. An interesting perspective is that in a LFT that describes neurons or cortical columns there are no reasons to remove the additional excitations that some forms of regularization introduce in continuous FTs (e.g. "Fermion doubling," chirality breaking, etc.), which are considered "artifacts" which should be removed. In fact, all of these artifacts would potentially be legitimate in an LFT describing neural activity, and in principle might even have functional roles for computations. Interestingly, in contrast to LFTs regularizing quantum theories (e.g. Quark Confinement Dynamics, QCD\cite{Wilson1974}), which are generally considered approximate theories, in the case of neurons regularization is achieved by default by the fact that they are distinct objects, and that they have a refractory time that establishes the minimum relevant time scale: a network of neurons is naturally regularized in the ultraviolet. 

\subsubsection{Compactified representation}

Taking into account that is designed for the thermodynamic limit, the compact kernel representation allows an elegant redefinition of the previous quantities also at finite intervals. On the other hand there are several advantages in considering the compactified variables, like the possibility of directly comparing the matrices with the transposed system, and thus constructing observables that compare spatial properties with temporal ones. Moreover, it worth notice that this is the representation used by Lovasz in his seminal book on kernels and graphs\cite{Lovasz2012}. Since any quantity defined before has a compact counterpart, for this section only we will redefine the previous symbols to match the new representation. We start with $V$ and $S$ 
\begin{equation}
V:=\left[0,1\right],\ \ \ S:=\left[0,1\right].
\end{equation}
The generic instant of time is denoted by the real variable $t\in S$, the generic point of space by the variable $x\in V$. We partition $V$ and $S$ into sub-intervals of lengths $1/N$ and $1/T$ respectively 
\begin{equation}
V_{i}:=\left[\,i/N,\,\left(i-1\right)/N\right],\ \ \ S_{\alpha}:=\left[\,\alpha/T,\,\left(\alpha-1\right)/T\right]
\end{equation}
in this way there is a correspondence between the vertex $i$ and the segment $V_{i}$, and the same is true fo $\alpha$ and  $S_{\alpha}$. It is important to notice that if the reference model is the discrete one, the ordering of the points within the volumes $V_{i}$ and $S_{\alpha}$ describe scales below the clock time, and should be irrelevant for the purpose of describing the task: therefore, the $\theta$ ordering of the $\alpha$ index introduced before remains the only meaningful theta-map.

\subsubsection{Compactified kernel (Graphon)}

Let us introduce the compactified kernel, which in the case of activities is exactly a "graphon" of the type treated by Lovasz in his book\cite{Lovasz2012} 
\begin{equation}
{\Omega}:\left[0,1\right]^{2}\rightarrow\left\{ 0,1\right\}, \ \ \ \  {\Omega}:=\{\Omega\left(x,t\right)\in\left\{ -1,1\right\} :\,\left(x,t\right)\in\left[0,1\right]^{2}\}
\end{equation}
for $N$ and $T$ finite the compact kernel is a step-function
\begin{equation}
{\Omega}\left(x,t\right):=\sum_{\alpha\in S}\sum_{i\in V}{\Omega}_{i}^{\alpha}\left(x,t\right)
\end{equation}
individual spins are represented by subkernels
\begin{equation}
{\Omega}_{i}^{\alpha}\left(x,t\right):=\varphi_{i}^{\alpha}\,\mathbb{I}\left(x\in V_{i}\right)\, \ \ \ \ \mathbb{I}\left(t\in S_{\alpha}\right)\in\left\{ -1,1\right\} 
\end{equation}
Similarly, we define the compacted version of the magnetization kernel
\begin{equation}
{M}:\left[0,1\right]^{2}\rightarrow\left\{ -1,1\right\},\ \ \ \  {M}:=\{M\left(x,t\right)\in\left\{ -1,1\right\} :\,\left(x,t\right)\in\left[0,1\right]^{2}\}
\end{equation}
the step function is defined as before
\begin{equation}
{M}\left(x,t\right):=\sum_{\alpha\in S}\sum_{i\in V}{M}_{i}^{\alpha}\left(x,t\right)
\end{equation}
But the sub-kernels representing individual spins are now steps between -1 and 1 
\begin{equation}
{M}_{i}^{\alpha}\left(x,t\right):=\sigma_{i}^{\alpha}\,\mathbb{I}\left(x\in V_{i}\right)\,\mathbb{I}\left(t\in S_{\alpha}\right)\in\left\{ -1,1\right\} 
\end{equation}
Since the kernels are linked by the relationship 
\begin{equation}
{M}=2{\Omega}-1
\end{equation}
and that we have already discussed the main differences, in the following we consider only the kernel of magnetization, so as to refer directly to the notation of Franchini 2023\cite{Franchini2023}.

\subsubsection{Averages and correlations}

The transition from discrete to compact representations is made by replacing the sums over V and S with integrals over the segment 
\begin{equation}
\ \frac{1}{N}\sum_{i\in V}\rightarrow\int_{x\in{V}}dx,\ \ \ \frac{1}{T}\sum_{\alpha\in S}\rightarrow\int_{t\in{S}}dt
\end{equation}
and the kernel in matrix form with the compacted kernel
\begin{equation}
M_{i}^{\alpha}\rightarrow{M}\left(x,t\right).
\end{equation}
In the case of the neocortex this can be used to make a connection between the “macroscopic” continuous neural field theories (like those used in functional NMR data analysis) and the underlying microscopic theory (the BCI recordings that we address in this paper). We assume that any differential volume $V(x)$ of the continuous theory contains a large number of cells and is obseved for a time interval $S\left(t\right)$ wide enough to reach equilibrium. Then the compactified kernel is defined as the offset of the differential volume centered in $x$ at time $t$ 
\begin{equation}
\Omega\left(x,t\right):=\left\{ \Omega_{i}^{\alpha}:\,i\in V\left(x\right),\,\alpha\in S\left(t\right)\right\} 
\end{equation}
in the thermodynamic limit. Notice that the support of the kernel offset is discrete, and could be mapped on a Potts model with dense spectrum. We can redefine all the previous quantities in terms of the compactified kernel: the offset is 
\begin{equation}
M_{0}:=\int_{0}^{1}dx\int_{0}^{1}dt\,{M}\left(x,t\right),
\end{equation}
averages are now functions on the segment. The average of the column is
\begin{equation}
{\mu}:=\left\{ {\mu}\left(t\right)\in\left[-1,1\right]:\,t\in\left[0,1\right]\right\} ,\ \ \ {\mu}\left(t\right):=\int_{0}^{1}dx\,{M}\left(x,t\right)
\end{equation}
the average of the row also is
\begin{equation}
{m}:=\left\{ {m}\left(x\right)\in\left[-1,1\right]:\,x\in V\right\} ,\ \ \ {m}\left(x\right):=\int_{0}^{1}dt\,{M}\left(x,t\right).
\end{equation}
The distributions of the averages 
\begin{equation}
p_{m}\left(s\right):=\int_{0}^{1}dx\,\delta\left(s-m\left(x\right)\right),\ \ \ p_{\mu}\left(s\right):=\int_{0}^{1}dt\,\delta\left(s-\mu\left(t\right)\right)
\end{equation}
The compactified correlation matrix is 
\begin{equation}
{C}:=\left\{ {c}\left(x_{1},x_{2}\right)\in\left[-1,1\right]:\,\left(x_{1},x_{2}\right)\in\left[0,1\right]^{2}\right\}, \ \ \ \ {c}\left(x_{1},x_{2}\right):=\int_{0}^{1}dt\,{M}\left(x_{1},t\right){M}\left(x_{2},t\right)
\end{equation}
The compactified overlap matrix is
\begin{equation}
{Q}:=\left\{ {q}\left(t_{1},t_{2}\right)\in\left[-1,1\right]:\,\left(t_{1},t_{2}\right)\in\left[0,1\right]^{2}\right\}, \ \ \ \ {q}\left(t_{1},t_{2}\right):=\int_{0}^{1}dx\,{M}\left(x,t_{1}\right){M}\left(x,t_{2}\right)
\end{equation}
The relationships between correlations and kernels is again
\begin{equation}
{M}\,{M}^{\dagger}={C},\ \ \ {M}^{\dagger}{M}={Q},
\end{equation}
but notice that in compactified notation there is no need of normalizating the product of the kernel operators. The mean field approximation is
\begin{equation}
{C}_{0}:={m}\otimes{m},\ \ \ {Q}_{0}:={\mu}\otimes{\mu}, \ \ \ \ {C}_{0}\left(x_{1},x_{2}\right)={m}\left(x_{1}\right){m}\left(x_{2}\right),\ \ \ {Q}_{0}\left(t_{1},t_{2}\right)={\mu}\left(t_{1}\right){\mu}\left(t_{2}\right)
\end{equation}
In the end, the correlation matrices are
\begin{equation}
{C}^{*}={C}-{C}_{0},\ \ \ {Q}^{*}={Q}-{Q}_{0},
\end{equation}
and the levels distributions are
\begin{equation}
p_{C^{*}}\left(s\right):=\int_{0}^{1}dx_{1}\int_{0}^{1}dx_{2}\,\delta\left[s-{c}^{*}\left(x_{1},x_{2}\right)\right], \ \ \ \ \ p_{Q^{*}}\left(s\right):=\ \int_{0}^{1}dt_{1}\,\int_{0}^{1}dt_{2}\,\delta\left[s-{q}^{*}\left(\,t_{1},t_{2}\right)\right]
\end{equation}
Up to this point the compactified notation seems identical to the matrix description, but notice that the averages are now supported by the same compact set (the segment $\left[0,1\right]$) and also the correlation matrices (the square with side $\left[0,1\right]$). This allows direct comparison of row and column averages, and the overlap matrix with the correlation matrix.

\subsubsection{The commutator}

An interesting quantity is certainly the kernel commutator 
\begin{equation}
[\,{M},{M}^{\dagger}]={C}-{Q}
\end{equation}
which is dependent on the $\theta$ map. In general, this map is established by the spatial part of the function whose commutation we want to test. We can introduce a norm for the commutator in terms of the distance between correlations and overlap, which should be minimized on the possible $\theta$ maps, which in a sense identifies the "best" association between space and time according to the chosen norm 
\begin{equation}
\inf_{\theta}\ \left\Vert {C}-{Q}\right\Vert
\end{equation}
The norm to use is a point to consider carefully. In Franchini 2023\cite{Franchini2023} we use the cut norm, but to study the correlations of a sparse subset of vertices one can also look at the mean of the total variation and many other quantities. For our present aims the classical Euclidean norm of order $k$ will be fine 
\begin{equation}
\left\Vert {C}-{Q}\right\Vert _{k}^{k}=\int_{0}^{1}ds_{1}\int_{0}^{1}ds_{2}\left|\,{C}\left(\theta\left(s_{1}\right),\theta\left(s_{2}\right)\right)-{Q}\left(s_{1},s_{2}\right)\right|^{k}
\end{equation}
as it can be related to the Wasserstein distance. In any case, before establishing the best-fit norm it will necessary to clarify whether the map $\theta$ acts on the set of the vertices or on its compacted counterpart: let $\theta:V\rightarrow V$ and let ${\theta^{*}}:[0,1]\rightarrow [0,1]$. Since ${V}$ has continuous cardinality, the number of ${\theta'}$ maps is infinitely larger than the number of $\theta$ maps, and in general a minimization on the continuous map will lead to smaller distances
\begin{equation}
\inf_{{\theta^{*}}}\ \left\Vert {C}-{Q}\right\Vert \leq\inf_{\theta}\ \left\Vert {C}-{Q}\right\Vert 
\end{equation}
Although by the assumptions made above, the map to be considered should be the discrete $\theta$ map. This may have its own relevance in case the matrices (which are kernels themselves) are step functions with different numbers of steps, i.e., when $T$ is different from $N$. In this case, to practically compare the two matrices, one must in fact define an appropriate binning.

\subsubsection{Ergodicity and commutation}

From the kernel commutator operator before we can introduce another kind of ergodicity, which is rapidly discussed also at the beginning of Franchini 2023 \cite{Franchini2023}. We say that the kernel commutes within tolerance $\epsilon$ of order $k$ if the correlation and overlap matrices satisfy 
\begin{equation}
\inf_{\theta}\ \left\Vert {C}-{Q}\right\Vert _{k}\leq\epsilon
\end{equation}
The reason lies in the relationship with the Wasserstein distance of order two: in fact, in the free-field description (in which the connected correlations are neglected) it is possible to prove a correspondence between the distance of the matrices ${C}_{0}$, ${Q}_{0}$ and that between the distributions of the averages $p_{m}$ and $p_{\mu}$. This is particularly easy to prove in the k=2. Indeed, notice that
\begin{multline}
\int_{0}^{1}ds_{1}\int_{0}^{1}ds_{2}\left|\,m\left(\theta\left(s_{1}\right)\right)m\left(\theta\left(s_{2}\right)\right)-\mu\left(s_{1}\right)\mu\left(s_{2}\right)\right|^{2}=\\
=\int_{0}^{1}ds_{1}\,m\left(s_{1}\right)^{2}\int_{0}^{1}ds_{2}\,m\left(s_{2}\right)^{2}+\int_{0}^{1}ds_{1}\,\mu\left(s_{1}\right)^{2}\int_{0}^{1}ds_{2}\,\mu\left(s_{2}\right)^{2}+\\
-2\int_{0}^{1}ds_{1}\,\mu\left(s_{1}\right)m\left(\theta\left(s_{2}\right)\right)\int_{0}^{1}ds_{2}\,\mu\left(s_{2}\right)m\left(\theta\left(s_{2}\right)\right)
\end{multline}
So the order two norm of the commutator can be written explicitly
\begin{equation}
\left\Vert {C}_{0}-{Q}_{0}\right\Vert _{2}^{2}=\left|\int_{0}^{1}ds\,m\left(s\right)^{2}\right|^{2}+\left|\int_{0}^{1}ds\,\mu\left(s\right)^{2}\right|^{2}-2\left|\int_{0}^{1}ds\,\mu\left(s\right)m\left(\theta\left(s\right)\right)\right|^{2}
\end{equation}
and note that the dependence on the map $\theta$ is only in the last term. Similarly, we can calculate the distance functional 
\begin{equation}
W_{2}\int_{0}^{1}ds\left|\,m\left(\theta\left(s\right)\right)-\mu\left(s\right)\right|^{2}=\int_{0}^{1}ds\,m\left(s\right)^{2}+\int_{0}^{1}ds\,\mu\left(s\right)^{2}-2\int_{0}^{1}ds\,\mu\left(s\right)m\left(\theta\left(s\right)\right)
\end{equation}
the last term contains the dependence of the map, and is the square root of the last commutator term, it follows that the $\theta$ map that minimizes the distance $W_{2}$ is the same one that also minimizes the commutator norm. Moreover, in the case of the Wasserstein distance in dimension one the theta map is determined by matching the order statistics (one orders the two sequences by decreasing values, which establishes the map, and then applies the inverse map of the temporal order statistics to both of them to bring the $\alpha$ back into the right sequence). Intuitively, seems that the correspondence between the maps holds for each $k$ if we neglect the connected correlations, it follows that in a first-order approximation the $\theta$ map that minimizes the commutator is the order statistics, and the commutator norm is correctly estimated in this way. If, on the other hand, the connected correlation matrices are not negligible, then there may be maps that lead to distances even lower than those computed with the order statistics.

\subsubsection{Compactified action, Theta maps and ergodicity}

Let us briefly sketch the action functional also in the compactified formalism of Franchini 2023 \cite{Franchini2023}. The Taylor expansion for the action is 
\begin{multline}
\mathcal{A}(M|F)=\int_{0}^{1}dx_{1}\int_{0}^{1}dt_{1}\ F\left(x_{1},t_{1}\right)\,M\left(x_{1},t_{1}\right)+\\
+\int_{0}^{1}dx_{1}\int_{0}^{1}dx_{2}\int_{0}^{1}dt_{1}\,\int_{0}^{1}dt_{2}\ F\left(x_{1},x_{2},t_{1},t_{2}\right)\,M\left(x_{1},t_{1}\right)\,M\left(x_{2},t_{2}\right)+\ ...
\end{multline}
The kernel operator corresponding to the two-body non-relativistic approximation (with memory) considered in this paper is
\begin{equation}
F_{2}\left(x_{1},x_{2},t_{1},t_{2}\right)=A\left(x_{1},x_{2}\right)\,\delta\left(t_{1}-t_{2}\right)+B\left(t_{1},t_{2}\right)\,\delta\left(x_{1}-x_{2}\right)
\end{equation}
We can separate the input contribution of the action (free field)
\begin{equation}
\mathcal{I}(M|I):=\int_{0}^{1}dx_{1}\int_{0}^{1}dt_{1}\ I\left(x_{1},t_{1}\right)M\left(x_{1},t_{1}\right)
\end{equation}
from the two body terms that depends on the correlation matrices
\begin{multline}
\mathcal{A}_{0}(M|A,B):=\int_{0}^{1}dx_{1}\int_{0}^{1}dx_{2}\ A\left(x_{1},x_{2}\right)\int_{0}^{1}dt_{1}\,M\left(x_{1},t_{1}\right)\,M\left(x_{2},t_{1}\right)+\\
+\int_{0}^{1}dt_{1}\,\int_{0}^{1}dt_{2}\ B\left(t_{1},t_{2}\right)\int_{0}^{1}dx_{1}\,M\left(x_{1},t_{1}\right)\,M\left(x_{1},t_{2}\right)
\end{multline}
The formal limit of our action is the sum of these terms
\begin{equation}
\mathcal{A}(M|A,B,I):=\mathcal{I}(M|I)+\mathcal{A}_{0}(M|A,B).
\end{equation}
Since ther compactification put everythig on dense segments of the same size, the normalization constants in the formulas for correlations and overlaps disappear (are equal to one). The formulas for the correlation matrices are as follows:
\begin{equation}
c\left(x_{1},x_{2}\right)=\int_{0}^{1}dt_{1}\,M\left(x_{1},t_{1}\right)M\left(x_{2},t_{1}\right),\ q\left(t_{1},t_{2}\right)=\int_{0}^{1}dx_{1}\,M\left(x_{1},t_{1}\right)M\left(x_{1},t_{2}\right)
\end{equation}
since the normalization is one, the action simply becomes
\begin{equation}
\mathcal{A}_{0}(M|A,B):=\int_{0}^{1}dx_{1}\int_{0}^{1}dx_{2}\ A\left(x_{1},x_{2}\right)\,c\left(x_{1},x_{2}\right)+\int_{0}^{1}dt_{1}\int_{0}^{1}dt_{2}\ B\left(t_{1},t_{2}\right)\,q\left(t_{1},t_{2}\right).
\end{equation}
Consider a map $\theta$ that scrambles the points of $\left[0,1\right]$ \cite{Franchini2023},
\begin{equation}
\theta:\left[0,1\right]\rightarrow\left[0,1\right].
\end{equation}
Let also introduce a notation for the kernel dependence from such theta map
\begin{equation}
M:=\{\,M\left[\theta\left(s_{1}\right),s_{2}\right]\in\left[0,1\right]:\left(s_{1},s_{2}\right)\in\left[0,1\right]^{2}\,\}
\end{equation}
and the corresponding notation for the correlation matrix: 
\begin{equation}
C:=\{\,c\left[\theta\left(s_{1}\right),\theta\left(s_{2}\right)\right]\in\left[0,1\right]:\left(s_{1},s_{2}\right)\in\left[0,1\right]^{2}\,\}
\end{equation}
Then, the kernel commutator according to the map $\theta$ is defined as follows:
\begin{equation}
[\,M,M^{\dagger}\,]:=M^{\dagger}M-MM^{\dagger}=C-Q
\end{equation}
If some $\theta$ exists such that the commutator is zero then 
\begin{equation}
q\left(s_{1},s_{2}\right)=c[\theta\left(s_{1}\right),\theta\left(s_{2}\right)]
\end{equation}
for any pair $\left(t_{1},t_{2}\right)\in\left[0,1\right]^{2}$ for some $\theta$, then we can define
\begin{equation}
B'\left(s_{1},s_{2}\right)=A[\theta\left(s_{1}\right),\theta\left(s_{2}\right)]+B\left(s_{1},s_{2}\right)
\end{equation}
and the action would be stationary in space, with renormalized time couplings, 
\begin{equation}
\mathcal{A}_{0}(M|A,B)=\int_{0}^{1}ds_{1}\int_{0}^{1}ds_{2}\ B'\left(s_{1},s_{2}\right)\,q\left(s_{1},s_{2}\right).
\end{equation}
We obtained an action that only depends on the overlap. A symmetric reasoning with the correlation on behalf of the overlap leads to a similar expression: if we apply the theta map to the time variable,
\begin{equation}
c\left(s_{1},s_{2}\right)=q[\theta\left(s_{1}\right),\theta\left(s_{2}\right)],
\end{equation}
then also in this case we can define the renormalized space couplings
\begin{equation}
A'\left(s_{1},s_{2}\right)=A\left(s_{1},s_{2}\right)]+B[\theta\left(s_{1}\right),\theta\left(s_{2}\right)],
\end{equation}
the action is stationary in the time variable
\begin{equation}
\mathcal{A}_{0}(M|A,B)=\int_{0}^{1}ds_{1}\int_{0}^{1}ds_{2}\ A'\left(s_{1},s_{2}\right)\,c\left(s_{1},s_{2}\right).
\end{equation}
Therefore, we interpret the commutation as a sign of ergodicity, although of a quite different type than that discussed in Section \ref{ergodicity breaking estimators}. This concludes the theoretical section.

\clearpage

\clearpage

\section{Experimental methods}
\label{ExperiMethods}

\subsection{Cortical minitubes}
\label{Cortical minitube model}

So far, the most accepted theory for the anatomical and functional organization of the retina is the columnar model (see Figure 5) \cite{Mountcastle1997,Jones2000,Buxhoeveden2002}, and similar assemblies of neurons are observable trough the whole neocortex, at least at anatomical level. Anyway, since in the retina there is also a well established corresponding functional organization, that has still not been shown for the whole neocortex, in the following we will use the name "minitubes" to indicate only the anatomical structures that are seen from histological inspection (Figure 7). Then, let $\mathbb{L}_{3}$ be a cubic lattice and let $xyz\in\mathbb{L}_{3}$ such that $z$ represents, for example, the average height from the surface of the cortex at which a given cortical layer is located.

\subsubsection{Decimated kernel}
\label{decimated kernel}

Let $xy$ be the position of the center of gravity of the cortical minitube section in the horizontal plane. To model the minitube layers we will define a partition of the space $\mathbb{R}^{3}$ into volumes
of equal size according to the lattice cells, for simplicity, we will approximate the cortical minitubes with square-based minitubes. Notice that the present charting of neocortex is not accounting for the neural connections, that may have any topology and are encoded in the interaction matrix $A$. The reason for using an euclidean reference frame is to allow comparisons with existing histological and fMRI, and other data\cite{Breitenberg}. Also, it may highligth effects due to possible extracellular fields and currents \cite{Buzsaki2012}, whose correlations may follow euclidean topology. The layers of the minitubes are thus represented by the lattice cells
\begin{equation}
U_{xyz}:=U_{x}U_{y}U_{z}\subset\mathbb{R}^{3}
\end{equation}
Now, calling $v\left(i\right)\in\mathbb{R}^{3}$ the position of the
nucleus of the $i-$th neuron, we can group by the volume in which
they are located
\begin{equation}
V_{xyz}:=\left\{ i\in V:\,v\left(i\right)\in U_{xyz}\right\} 
\end{equation}
each of these groups of neurons will have its own associated kernel
\begin{equation}
\Omega_{xyz}:=\left\{ \varphi_{i}^{\alpha}\in\left\{ 0,1\right\} :\,i\in V_{xyz},\,\alpha\in S\right\} .
\end{equation}
At this point one could further group the neurons, first by index $z$, so as to form the cortical minitubes. The vertices belonging to the minitube are
\begin{equation}
V_{xy}:=\bigcup_{z\in\mathbb{L}}\,V_{xyz}
\end{equation}
that is the set of neurons that constitutes the minitube at position $xy$. The kernel is
\begin{equation}
\Omega_{xy}:=\{\Omega_{xyz}\in\left\{ 0,1\right\} ^{V_{xyz}}:\,z\in\mathbb{L},\,\alpha\in S\}
\end{equation}
and describes the activity of the single cortical minitube in $xy$. Some interfaces, such as Neuropixel or deep multielectrode shanks, allow direct observations of this activity. The minitubes are in the end grouped again to form the cortex structures and areas, 
\begin{equation}
V:=\bigcup_{xy\in\mathbb{L}_{2}}\,V_{xy}
\end{equation}
and the original kernel can thus be expressed in terms of the minitubes:
\begin{equation}
\Omega=\{\Omega_{xy}^{\alpha}\in\left\{ 0,1\right\} ^{V_{xy}}:\,xy\in\mathbb{L}_{2},\,\alpha\in S\},
\end{equation}
so that it represents a two-dimensional lattice of cortical minitubes\cite{Mountcastle1997,Jones2000,Buxhoeveden2002,Segev2006,Segev1998,Lubke2007},
a system in 2+1+1 dimensions. For the above we can consider the experimental kernel for a specific tubular layer
\begin{equation}
\Omega_{z}:=\{\Omega_{xyz}^{\alpha}\in\left\{ 0,1\right\} ^{V_{xyz}}:\,xy\in\mathbb{L}_{2},\,\alpha\in S\},
\end{equation}
where, again, $xyz\in\mathbb{L}_{3}$ are the spatial coordinates in a cubic lattice such that $z$ represents the average height from surface of the cortex at which a given layer is located, and $xy$ is the
position of the minitube section in the horizontal plane. The points are organized in a planar sub-lattice $x'y'\in\mathbb{L}'_{2}$ (of the observed cortical layer $z$) whose step is much greater than the diameter of the individual minitube, so that the activities recorded at the various points belong with high probability to different and well-spaced minitubes. 
At this point, to model the spacing between the probing points, we apply a renormalization by decimation on $\Omega$, and obtain the decimated activity kernel
\begin{equation}
\hat{\Omega}:=\{\hat{\varphi}_{x'y'}^{\alpha}\in\left\{ 0,1\right\} :\,x'y'\in\mathbb{L}'_{2},\,\alpha\in S\},\ \ \ \hat{\varphi}_{x'y'}^{\alpha}:=\mathbb{I}(\Omega_{x'y'z}^{\alpha}\neq0).
\label{eq:Methods_experikernel}
\end{equation} 
this is the electrode kernel of eq \ref{eq:experikernel} shown in main text. This kernel is intended to model approximately the sensor recording, net of systematic errors and approximations. According to our arguments it should be comparable with a renormalized theory. Notice that this renormalization happens only in space and hence the information coming from the digitalization of neuronal signals is largely preserved (as far as the signals inside a channel or multi units do not overlap too much in time). $\hat{\Omega}$ leads to the experimental hypermatrix of Figure \ref{fig:1}. Finally, notice that the present charting of neocortex is not accounting for the anatomical neural connections, that may have any topology and are encoded in the interaction matrix $A$. Clearly, determining the exact effective theory that can describe the dynamics of the columns and their excitations will require careful analysis of the body of knowledge about the structure of the neocortex and the interface itself, but these manipulations demonstrate that a treatment in terms of field theory is possible, at least in this formalism. Moreover, given the particular architecture of the cortex, it is possible that the topology of such a theory is essentially either mean-field or two-dimensional, and with layers of cortex behaving as interacting fields, just as in elementary particle theory. This could greatly facilitate the analytical construction of effective theories.

\subsection{Neural recordings with Utah 96}
\label{Neural recordings Methods}

\subsubsection{Subjects}
\label{Subjects}
Two male rhesus macaque monkeys (Macaca mulatta, Monkeys P and C), weighing 9 and 9.5 kg, respectively, were employed for the task shown as case study. Animal care, housing, surgical procedures and experiments conformed to European (Directive 86/609/ECC and 2010/63/UE) and Italian (D.L. 116/92 and D.L. 26/2014) laws and were approved by the Italian Ministry of Health. Monkeys were pair-housed with cage enrichment. They were fed daily with standard primate chow that was supplemented with nuts and fresh fruits if necessary. During recording days, the monkeys received their daily water supply during the experiments.

\subsubsection{Apparatus and task}
\label{Apparatus and task}
The monkeys were seated in front of a black isoluminant background ($<0.1 cd/m2$) of a 17-inch touchscreen monitor (LCD, 800 x 600 resolution), inside a darkened, acoustic-insulated room. A non-commercial software package, CORTEX (http://www.nimh.gov.it), was used to control the presentation of the stimuli and the behavioural responses. Figure 1 and 3 panel c show the scheme of the task: a Go-signal reaching task. Each trial started with the appearance of a central target (CT) (red circle, diameter 1.9 cm). The monkeys had to reach and hold the CT. After a variable holding time (400–900 ms, 100 ms increments) a peripheral target (PT) (red circle, diameter 1.9 cm) appeared randomly in one of two possible locations (right/left, D1/D2) and the CT disappeared (Go signal). After the Go signal the subjects had to reach and hold the PT for a variable time (400-800 ms, 100 ms increments) to receive juice. The time between the presentation of the Go signal and the onset of the hand movement (M\_on) is the Reaction time (RT). White circles around the central target were used as feedback for the animals to indicate the touch.

\subsubsection{Extraction and processing of neuronal data}
\label{Extraction and processing of neuronal data}
A multielectrode array (Blackrock Microsystems, Salt Lake City) with 96 electrodes (Utah 96, spacing 0.4 mm) was surgically implanted in the left dorsal premotor cortex (PMd; the references used after opening the dura were the arcuate sulcus and pre-central dimple) to acquire unfiltered electric field potentials (UFP; i.e., the raw signal) sampled at 24.4 kHz (Tucker Davis Technologies, Alachua, FL). As described in previous work from our group\cite{Pani2022}, we extracted single neurons activities from the raw signal by employing the spike sorting toolbox KiloSort3\cite{Pachitariu2016} with the following parameters: Thresholds: [9 9] (thresholds for template-matching on spike detection); Lambda: 10 (bias factor of the individual spike amplitude towards the cluster mean); Area Under the Curve split: 0.9 (threshold for cluster splitting); Number of blocks: 5 (amount of blocks channels are divided into for estimating probe drift). The output was manually curated in Phy (v2.0; 17) to merge clusters that were mistakenly separated by the automated sorter. From this procedure we obtained a binary spike raster with a time resolution of 1ms (1 for a spike, 0 for no spikes) for each single-trial of the experiment. Each single-trial raster was then put into the form of the kernel $\hat{\Omega}$ of eq \ref{eq:Methods_experikernel}.

\subsubsection{Neural dynamics underlying movement generation in PMd}
\label{Results CMT}

We chose this task as a use-case for its simplicity as it involves only two experimental conditions. In this way, the results obtained in our LFT context are directly comparable with those obtained previously using common approaches that rely on covariance analysis \cite{Churchland2012,Mattia2013,Pani2022,Chandrasekaran2017,Kaufman2016,Clawson}. We extracted the kernel $\langle\hat{\Omega}\rangle$ in relation to the movement onset (M\_on), considering an  epoch of 1s before and after the event. By doing so, the distributions of the behavioral events of the task (the Go signal and M\_on) are included (see Figure \ref{fig:1}). It has been demonstrated that, during the time preceding the movement, PMd neurons express strong modulations associated with movement control \cite{Weinrich1982,Mattia2013,Churchland2010,Bardella2020,Pani2022,Bardella2024}. The hypermatrices computed for the two experimental conditions are shown in Figure \ref{fig:1} (see also Extended Data, ED). The JS matrices exhibit striking features, and by comparing them across movement directions, one can retrieve most of the hallmarks of PMd neural dynamics. The first is the strong increase of synchronous activity peaking within the 200 ms interval preceding the M\_on (black markers in Figure \ref{fig:1}) that correspond to the functional state of the system linked to the incoming movement generation. Indeed, the motor planning of actions in PMd is recognized to be encoded at the population level in the form of synchronization patterns that exhibit a strong modulation around 200 ms before the onset of movement\cite{Pani2022,Bardella2020,Kaufman2016,Kaufman2014,Churchland2010,Churchland2012,Shenoy2013,Churchland2006,Mattia2013,Ames2014,Elsayed2016,Mirabella2011,Weinrich1982,Bardella2024,BattagliaMayer2014,Caminiti1991,Caminiti2017,Nambu2002,Middleton2001,Marconi2001,Johnson1996}.
The second is the specificity of PMd neurons for the direction of movement, which in the reported task could happen towards left or right (D1/D2). In Figure \ref{fig:1} and 8 (ED), this is evidenced by the more intense motifs of synchrony for one direction (D2) with respect to the other (D1). They emerge at the end of the motor plan maturation ($\sim $ within 200 ms before M\_on), continuing for at least 200 ms afterwards. In the ED section we report examples from a second subject and separately the components of the hypermatrix with additional details (e.g., the difference |D1-D2| for both $\langle\hat{\Pi}\rangle$ and $\langle\hat{\Phi}\rangle$.) Significantly, the dynamic contributions detectable from the JS matrix can be easily mapped in the spatial domain thanks to the hypermatrix arrangement, which emphasize the correspondences between the JS matrix, the kernels $\langle\hat{\Omega}\rangle$ and the spatial and temporal averages. For example, from the kernels in Figure \ref{fig:1} and the zoom of Figure 8, the firing patterns that elicit a specific configuration of dynamical synchrony can be identified. This reveals that the temporal correlations during the motor plan maturation are caused by a specific firing sequence in the kernel $\langle\hat{\Omega}\rangle$ (for both D1 and D2). Hence, we can infer that the maturation of the motor plan corresponds to different populations of neurons discharging with variable timings and intensities for D1 compared to D2. The JS matrix also demonstrates that the direction-specific correlations coincide with more intense firing for D2 compared to D1. In addition to direction-specific differences, relevant similarities are also appreciable. The cross emerging at the center of the JS matrix represents synchronization among neural ensembles that extends throughout the duration of the trial for both D1 and D2. Again, the neural assemblies responsible can be easily identified from the kernels $\langle\hat{\Omega}\rangle$. Future work will be needed to clarify more details. The spatial correlations are instead recoverable from the matrix $\langle\hat{\Phi}\rangle$. In our example, it can be noted how the combinations underlying the motor plan are preserved for both directions (same correlation values in $\langle\hat{\Phi}\rangle$ for both directions), while the direction-specific ones change. 

\subsubsection{Comparison with other methods}

Thus, with the hypermatrix representation, neural dynamics can be efficiently decomposed into its spatial and temporal contributions, and their roles in the studied task are easily mapped. From these remarks, we understand the striking traits of the hypermatrix: its completeness despite its simplicity. It conveys fundamental information about the system in a compact representation, without the need for complex numerical artifice. This is a substantial difference with other approaches frequently used to analyze neural activity (e.g. PCA or machine learning methods\cite{Pani2022,Kaufman2014,Elsayed2016,Gallego2017,Yu2009,Pandarinath2018} among the most popular). Although these methods have provided valuable insights, none of them offers a picture encompassing the temporal and the spatial attributes of the system at the same time.
In the case of PCA, for example, the temporal and spatial properties can be linked together only after a non-trivial, and most of the time arbitrary, sequence of numerical steps. Among others, these include a dimensionality reduction, i.e. choosing a number of PCs, and the subsequent projections onto the reduced space; and this requires computing the eigenvectors of the covariance matrix. In contrast, our theory only requires simple scalar products of the experimental rasters, eliminating the need for dimensionality reduction. In addition, the interpretations that conventional methods offer about the intrinsic nature of neural processes are strongly dependent on the chosen analysis pipeline and are far from being derived from the universal principles of a physical theory. This significantly impacts, for example, the definition that these methods can provide for the energy of the system, which remains vague and unformalized (such in the case of the manifold hypothesis\cite{Langdon2023} and the widespread PCA-based energy landscapes\cite{Gallego2017,Genkin2023}). We have instead shown that the kernel, its transpose, and the corresponding scalar products give an accurate and physical-based description of the energy functional of the system. Most importantly, our approach entails a formal communication between physics and neuroscience using as a language the governing equations of elementary particles. This allows the measurement of neural interactions through physically-grounded observables and its interpretation in terms of well-known laws. In our LFT framework, temporal and spatial correlations have a precise meaning, representing, respectively, the kinetic and potential energy terms of the recorded neurons. As detailed in Section \ref{LFT}, our energy functional is obtained through the parameters of the theory $A$, $B$ and $I$. 

\subsubsection{Test of the (renormalized) Neural LFT}

Generally speaking, the first requirement of a theory is that it should be possible to estimate the variables that describe it from experimental data (more formally, inverting the model). In our case, the set $A$, $B$ and $I$, may be recovered inverting the hypermatrix. To do so, it is necessary to resort to a class of well-defined methods that go by the name of inverse Ising problems\cite{Nguyen2017,Tersenghi2014}. The same class of methods has been used by Tkacik et al. \cite{Tkacik2009} to estimate the couplings of the Ising Hamiltonian with which they modeled the salamander retina recordings. This could also apply to the use-case here discussed, but at the price of a remarkable computational burden, mostly due to the very high rank of the JS matrix. To this respect, a viable way to lighten it could be to properly bin (renormalize) the process according to a larger clock time $\tau$. This would yield a JS matrix of smaller rank, without losing too much information. Following these considerations, we applied a bin renormalization to the kernel on a time step of 10 $ms$ (at the level of individual trial), reducing of a factor 100 the number of kernel cells to deal with. From Figures 11 and 12 is evident that the kernel and the patterns in the covariance matrices are almost unaffected by the chosen renormalization, at least for this type of behavioral task. We were able to compute the grand covariance of the renormalized kernel: the distributions of the matrix entries is shown in Figure 13. We see that the red distribution follows the expected normal-product peak centered on zero due to the product of independent Gaussian fluctuations, that is also in the blue and green distributions. But notice that the most correlated pairs deviating from the normal product distribution are only in blue and green. This shows that if we ignore correlation below a certain threshold (that in this case is around 5 \%) then we can approximate the activity with the simplified "non-relativistic" action proposed in this paper.  Notice that the deviations contributing to the overlap matrix are still much smaller than those contributing to the correlation matrix, and should produce only small deviations from the max entropy model of Schneidman et al. \cite{Schneidman2006}.

\subsection{Perspectives}

\subsubsection{Microscopic models}
\label{Micromodels}

It has been proposed that a movement may be carried out by the suppression of some steady signal that ends the holding or "non-movement state" and triggers the movement\cite{Pani2022}. This idea is in line with the shared view whereby a command initiated in other regions is executed locally in the PMd, which is part of a larger network subserving motor control based on frontal, parietal, subcortical, cerebellar and spinal structures. According to our formalism, we can state that the part of the brain deciding the movement sends the command to the PMd in the form of a spatially structured external field that is stationary throughout the execution of the computation. In analogy with magnetic systems, such an external field configures the phase toward which the population of neurons will try to balance. It can be hypothesized that the neural computation underlying the so-called motor plan is performed in the convergence to the system’s equilibrium: at the time $\alpha$ in which the external input changes, the system converges to the phase (valley) selected by the new input. This can be modeled with the magnetization profile of a one-dimensional Ising chain subject to some external field. If the field is suddenly switched on at time $\alpha_{0}$ the Lagrangian contains a one-dimensional Ising kinetic term in $\alpha_{0}$: this is to force the stationary dynamic with an average interspike period $\tau$ that is deduced from time covariance matrix $\delta Q\left(\eta\right)$ (see Figures in section \ref{ExperiMethods}). This simple interface model in one dimension was introduced and solved by Robert and Widom in \cite{Robert1984} adapting methods from Percus, Tejero, and others \cite{Percus1977,Tejero1987,Derrida1986}. One can confront the shape of the transient field with that predicted by \cite{Robert1984}. This mechanism also sets the typical relaxation timescale of the process. In this scenario it would be possible to construct analytically solvable models with locally stationary external input, like the aforementioned model, which could faithfully represent local circuitry. For example, one could formally model the circuit sketched in Pani et al.\cite{Pani2022} and check it against experimental data.

\subsubsection{Movement and the glassy phase}
\label{Glassymove}

Like the salamander retina, also the PMd (ora other cortices) might be structurally capable of exhibiting glassy phases, however, it is not necessarily the case that these are physiological within the "computation" of movement, nor that they play a central role in sending the system off balance (at least until consciousness is into play). For example, unlike the retina, which is a structure strictly devoted to "inputs" to be passed to the central nervous system (that in the case of \cite{Tkacik2009} is also detached from it), we recorded from a system that should mainly process and produce an "output" to the muscles or other areas. 
If the neural system responsible for movement is in a glassy phase, (not going to equilibrium quickly), it might be unable to consistently convey motor commands. As a result, the executed movements may deviate from the intended actions of the animal, leading to inaccuracies such as missing the targets or unintentional actions. Moreover, the time covariance matrix (see Figures 12, 17 and 20) support the idea that the movement is not glassy: the overlap covariance matrix does not contain the movement (except in the refractory profile) and the firing patterns of the neuron are consistent with a noise model of the kind considered in \cite{FranchiniREM2023}. Notice that the refractory period induces a structure in the overlap covariance matrix that is approximately stationary, and that the approximate symmetry of the overlap between trials shows that replica symmetry is only slightly broken. Following the ideas of \cite{Toulouse1986}, that see the learning as a selection of possible states of the system, we would expect more "glassy" behavior behavior during the initial stages of training, when the monkey still has not entirely learned the task requirements. This could be studied by calculating, for example, the overlap between kernels of sessions separated by large time intervals, but the known degradation problems of Utah interfaces could mask fine-grained differences. Also, it could be possible that glassy activity may insurge in similar conditions as those considered for ex-vivo salamander retina. For example, it would be of extreme interest to study the exceptionally rare recording of a dying brain published in \cite{Pani2018DY}, which is from the same monkey studied in \cite{Pani2022}.

\subsubsection{Computing physical LFTs with brain organoids}
\label{Organoids}

In addition to the orthodox purpose of reading and interpreting activity of natural neural networks \textit{in vivo}, even more interesting applications have been made possible from recent advances in growing, shaping and interfacing biological neural tissue. The most striking example is perhaps the digital interfacing of brain organoids \cite{Zheng2022,Sharf2022}, a method that has already reached a fairly good technical level as demonstrated in T. Sharf et al. 2022 \cite{Sharf2022}. In short, brain organoid modeling is an advanced technique for studying brain development, physiology, function and disease
occurrence (see Zheng et al. 2022\cite{Zheng2022} review for an interesting overview). The experimental possibilities in this regard would certainly be of far reach, less expensive on both ethical and material sides, and would also provide a safer guide for studying animal and human brains \textit{\textit{in vivo}}. There are now concrete possibilities of building hybrid circuits by connecting artificial neural networks and brain organoids \cite{Zheng2022}  through currently available interfaces, that could then be trained in the binary LFT language. Also, natural neural networks have shown to work on more efficient energetic basis and to learn from fewer examples. For example, shaping natural neural networks into useful neural circuitry \cite{Zheng2022} may allow to realize in practice the ideas described in \cite{Halverson2022} and use natural neurons to run physical LFT simulations.

\begin{figure}[h!]
\begin{center}

\includegraphics[scale=0.76]{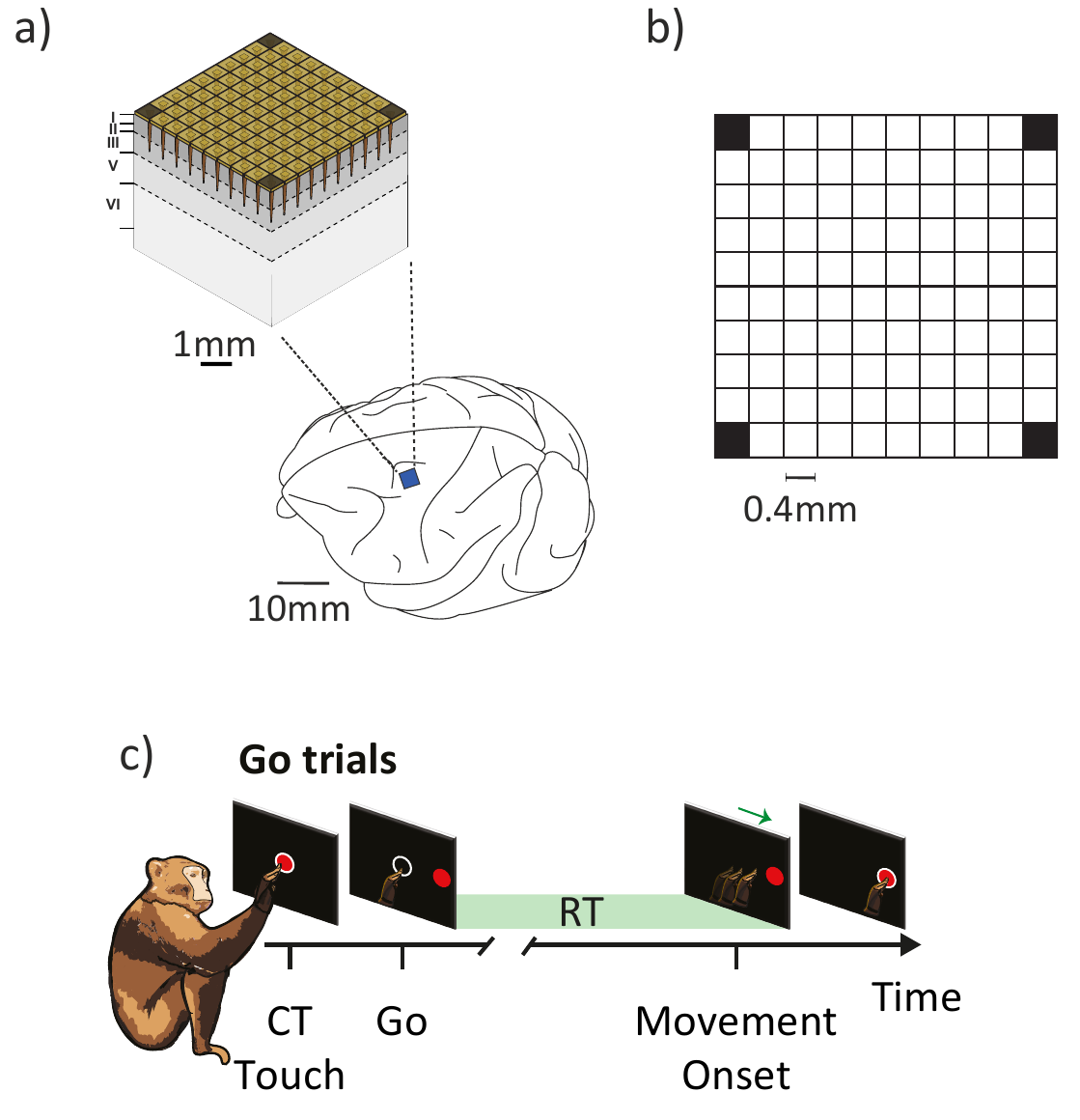}
\caption*{\textbf{Figure 3. Experimental recording of neural activity and behavioral task}. a) Cortical minitube sampling of the Utah 96 array: each electrode pitch is $\sim 40\mu m$ so that the listening volume of each electrode can be reasonably assumed of the order of the distance between the electrodes ($\sim 400\mu m$)\cite{Hill2014}. In the case of PMd the Utah 96 samples activity from around the inner Baillager band\cite{Rapan2021,Opris2011} at around 1.5 mm penetration. 
\textbf{b) Decimated lattice $x'y'$} of the electrode kernel $\hat{\Omega}$ for Utah 96. Each lattice cell can be either silent or active, as described in the section \ref{Cortical minitube model}.
\textbf{c) Behavioral task} that required to move the arm toward a peripheral target (\textbf{Go trials}) that could appear in  one of two directions of movement (D1 or D2). Monkeys had to reach and hold the peripheral target to get the reward. RT: reaction time; CT: central target; Go: Go signal appearance. 
}
\label{fig:3}
\end{center}
\end{figure}

\begin{figure}[h!]
\centering
\includegraphics[width=1\linewidth]{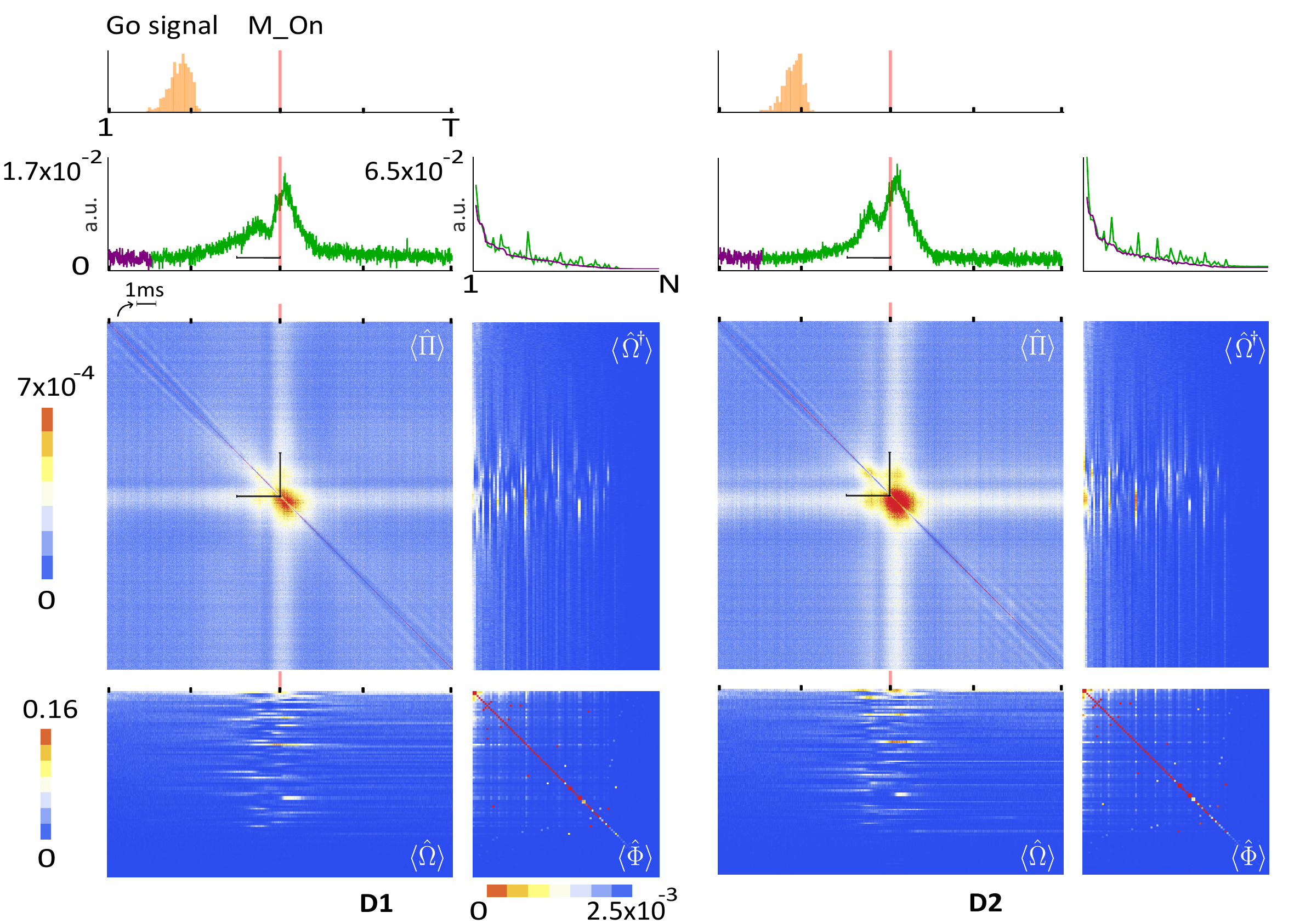}
\newline
\caption*{
\textbf{Figure 4. Experimental hypermatrices of Monkey P}. The figure shows the first order observables of the theory and the hypermatrix from the electrode kernel $\hat{\Omega}$ of eq. (\ref{eq:experikernel}) for PMd data (here for D1 and D2). Neural activity is aligned $[-1, 1]$ s to the M\_On to include the distributions of the stimuli (the Go signal, orange distribution and M\_On, magenta. $T=2s$.). The uppermost panels represent the $I$ of eq. (\ref{eq:action}) in the form of time markers for the stimuli presented during the task. Green traces above the $\Pi$ matrix are the time evolution of the spatially-averaged activity. Green traces above the transposed  $\hat{\Omega}$ are instead the time-average activity for each N.  Purple traces are the “baseline” observables computed in the first 250 ms, which , as expected, are indistinguishable for both conditions. The kernels and $\Phi$ are sorted according to the activity in the first 250 ms of D1, before the appearance of any Go signal. Black ticks are every 500 ms. Black segments are 200 ms wide.  
}
\label{fig:4}
\end{figure}

\begin{figure}[h!]
\begin{center}

\includegraphics[scale=1.5]{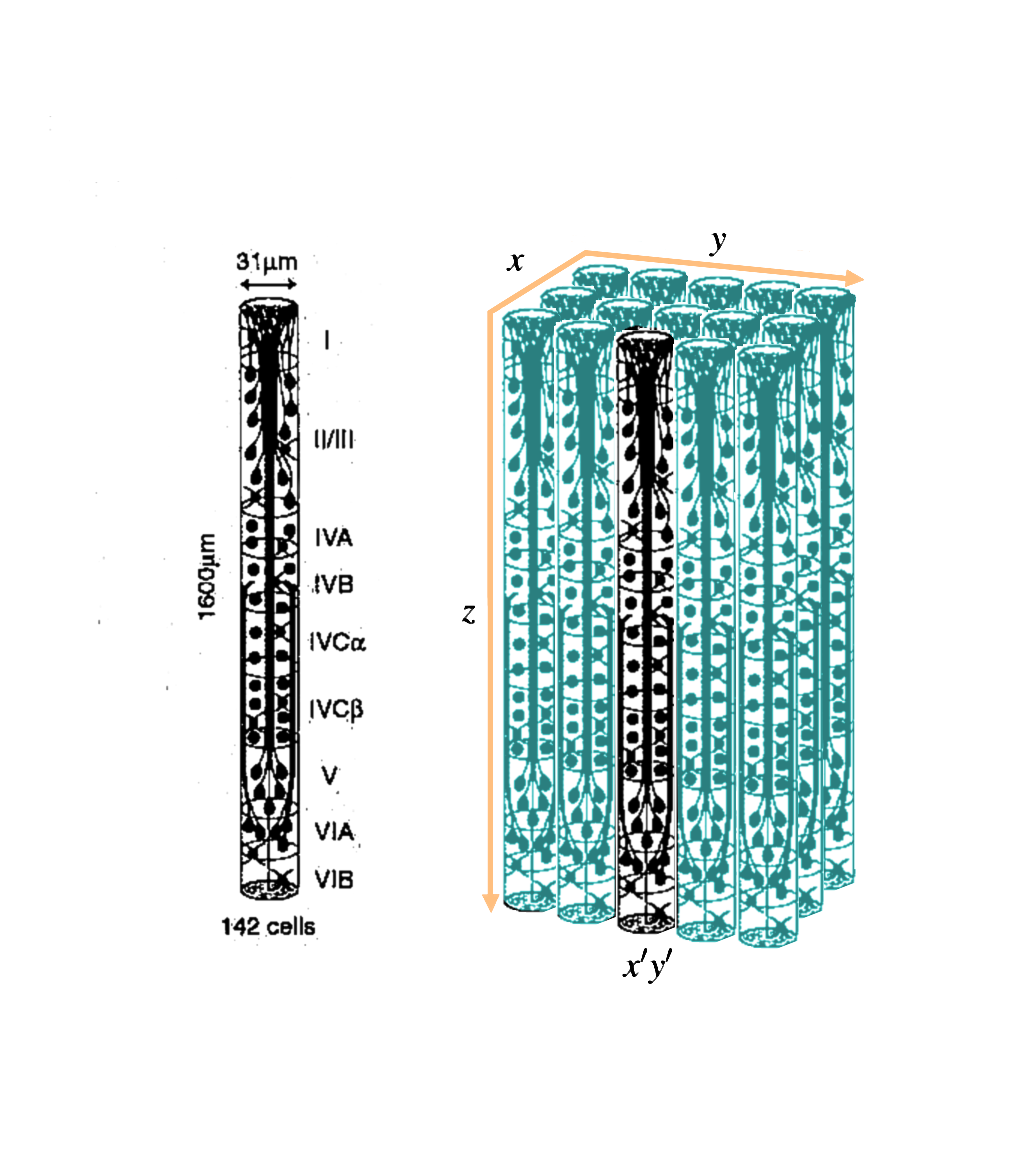}
\newline
\newline
\caption*{\textbf{Figure 5.} Example of (A) Columnar organization of retina and the decimation procedure from $x,y$ to $x',y'$. Diagram elaborated from Figure 2 of Jones 2000\cite{Jones2000}. We re-scaled the vertical dimension $z$ for improved visualization.
}
\label{fig:5}
\end{center}
\end{figure}

\begin{figure}[h!]
\centering
\includegraphics[width=0.9\linewidth]{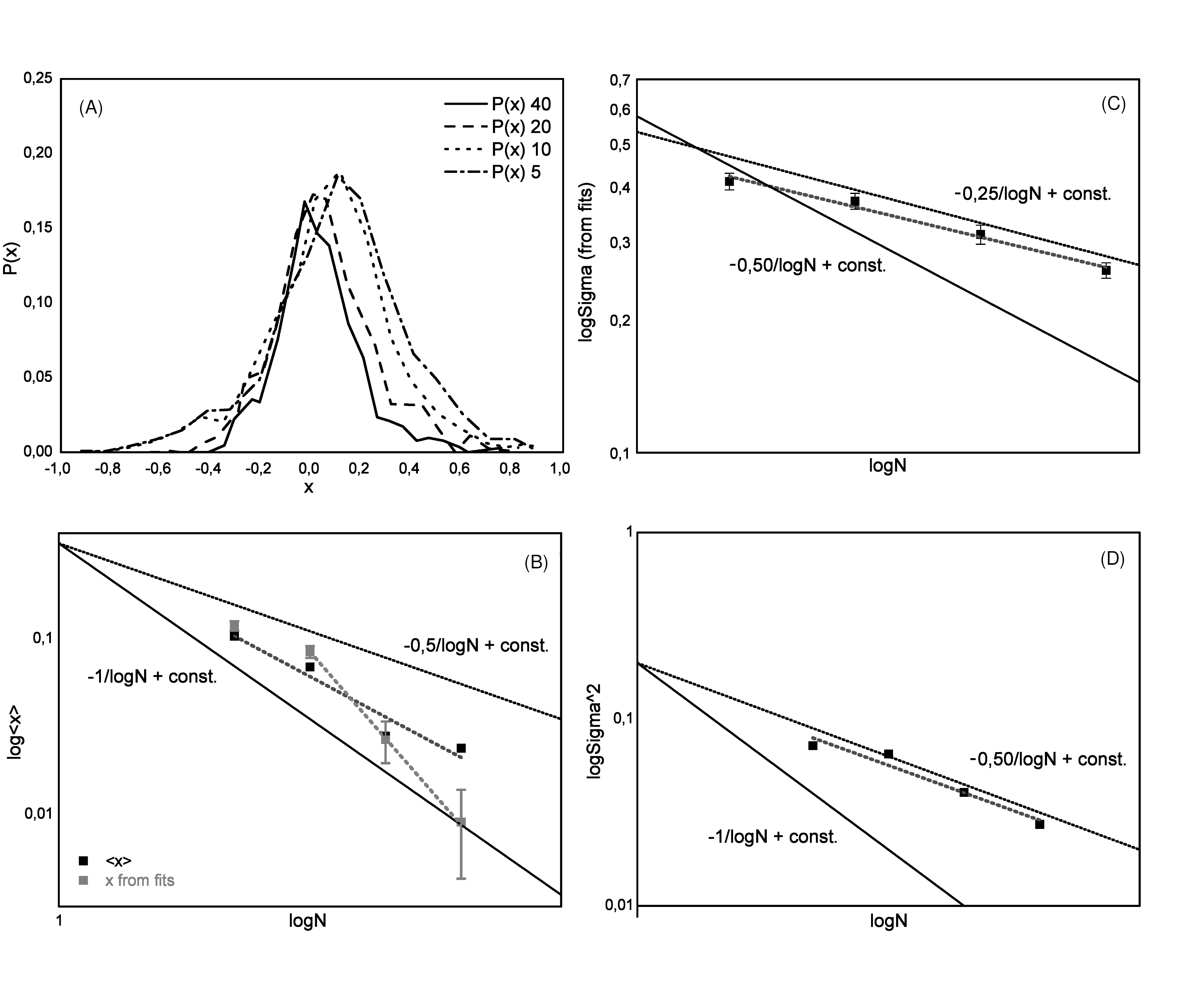}
\caption*{
\textbf{Figure 6.}
We extracted the distributions of the reconstructed couplings $J_{ij}$ from Figure 1f of Tkacik et al. 2009\cite{Tkacik2009} with G3data (A) and computed the scaling of the parameters, first from Gaussian fits (B-C) and then from the first two moments of the distributions (B-D). Both methods confirm that the couplings are approximately Gaussian with scaling exponent $\alpha=1/2$. We remark that finding the parameters of the theory using scaling techniques like this is typical of LFT analysis of elementary particle theory, where they are typically used to estimate particle masses and other observables..
}
\label{fig:tack-tack}
\end{figure}

\begin{figure}[h!]
\begin{center}

\includegraphics[scale=0.6]{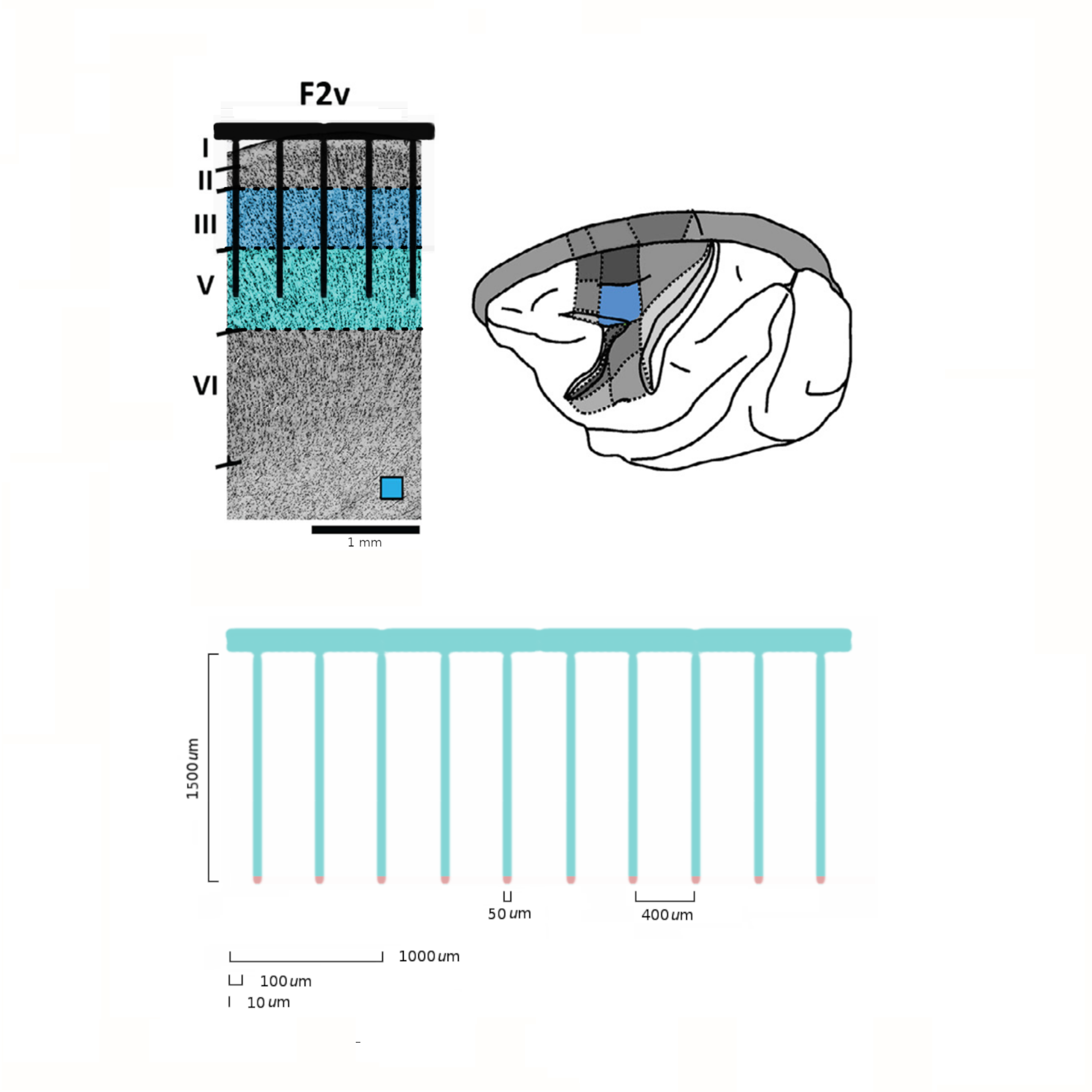}
\caption*{
\textbf{Figure 7.} Utah 96 compared with the an example of PMd histological taken from \cite{Rapan2021}. The Utah 96 BCI\cite{Leber2017LongElectrodes,Chandrasekaran2021,Bullard2019} is a silicon-based microelectrode array in the form of a rectangular or square grid in a10x10 pattern (the total channels are actually 96, as the vertices of the square have no record, see Figure 3). Each needle is 1.5 mm long, with a diameter of 80 $\mu$ m at the base tapering to the tip around 40-50 $\mu$ m. The electrodes are electrically insulated from neighboring electrodes by a glass moat surrounding the base. The electrode tips are coated with platinum to facilitate charge transfer into the nerve tissue, and the electrode stems are insulated with silicon nitride. In the recording used in \cite{Pani2022} the grid is square, and measures 4.2 mm, with 96 silicon microelectrodes and a spacing of 0.4 mm. In the case of non-human primates PMd, it should record neural activity from the inner Baillager band\cite{Rapan2021,Opris2011}.}
\label{fig:7}
\end{center}
\end{figure}

\begin{figure}[h!]
\centering
\includegraphics[width=0.825\linewidth]{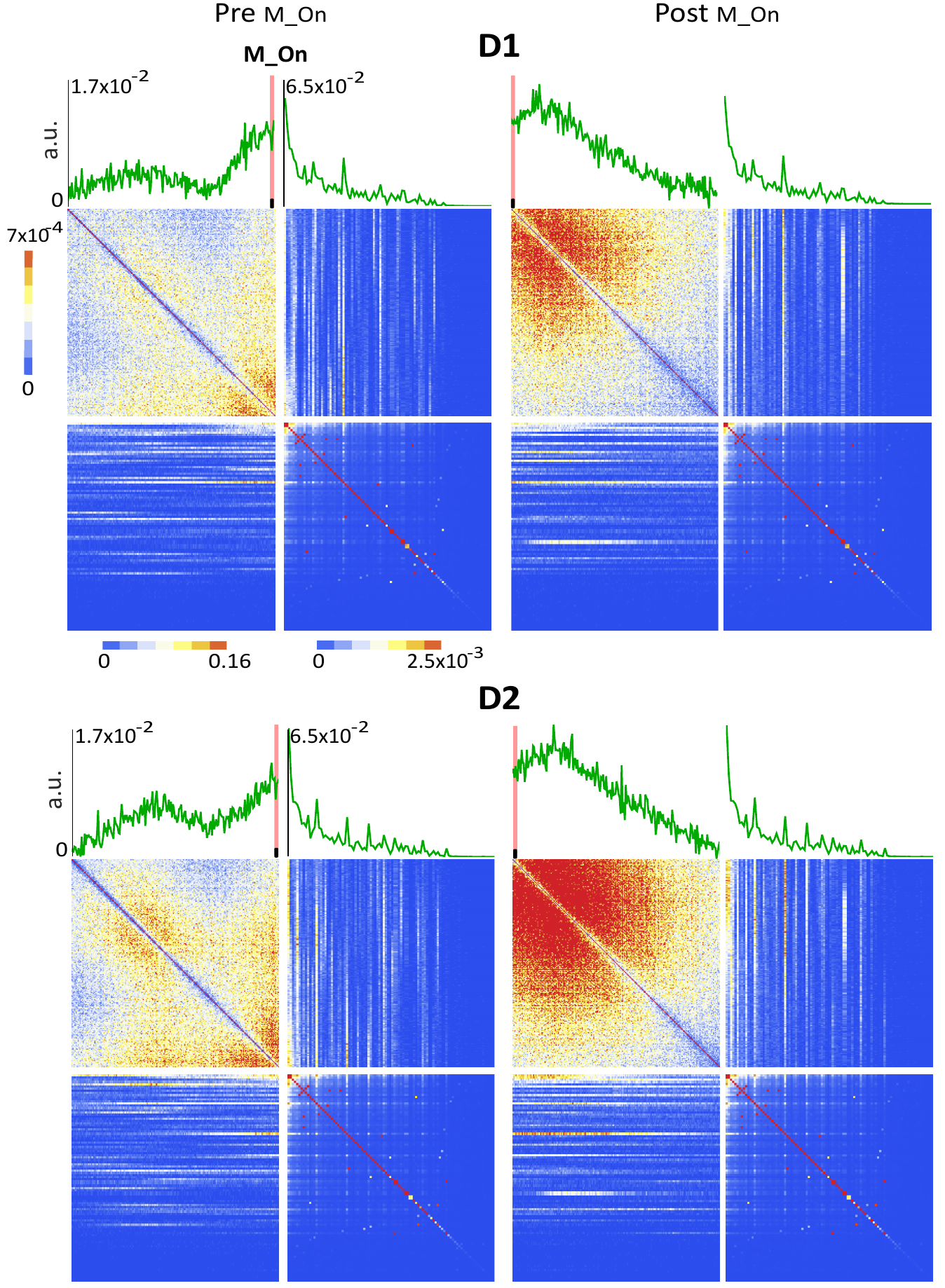}
\newline
\
\caption*{
\textbf{Figure 8. Hypermatrix detail}. The figures report details of the hypermatrices of Figure 1 (Monkey P) for two epochs of the task: -200 ms before (Pre Mov\_on panel) and after (Post Mov\_on panel) the Mov\_on for D1 and D2. Axes scales and color labels are the same as Figure 1. The details of the dynamical synchronization patterns changing between D1 and D2 and the corresponding kernels configurations are evident.
}
\label{fig:8}
\end{figure}

\begin{figure}[h!]
\centering
\includegraphics[width=1.0\linewidth]{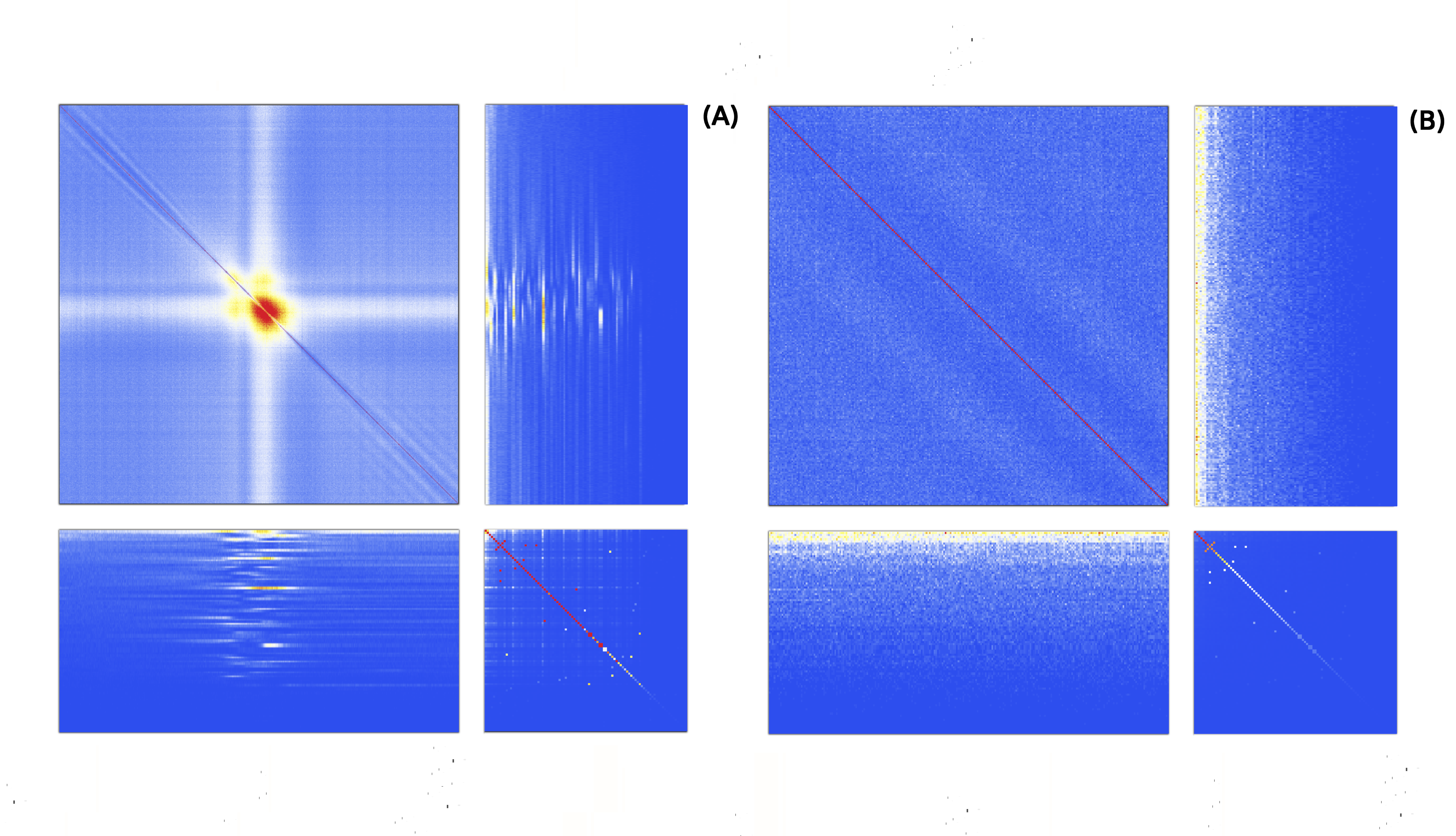}
\newline
\newline
\caption*{
\textbf{Figure 9. Comparison with background}. We can compare the experimental hypermatrix of D1+D2 (A) with the same observable computed in the first 250$ms$ only (B) that is the region highlighted in purple in Figure 1. The most correlated channel pairs in the spatial correlation matrix are still visible. 
}
\label{fig:9}
\end{figure}

\clearpage

\begin{figure}[h!]
\begin{center}
\includegraphics[scale=0.8]{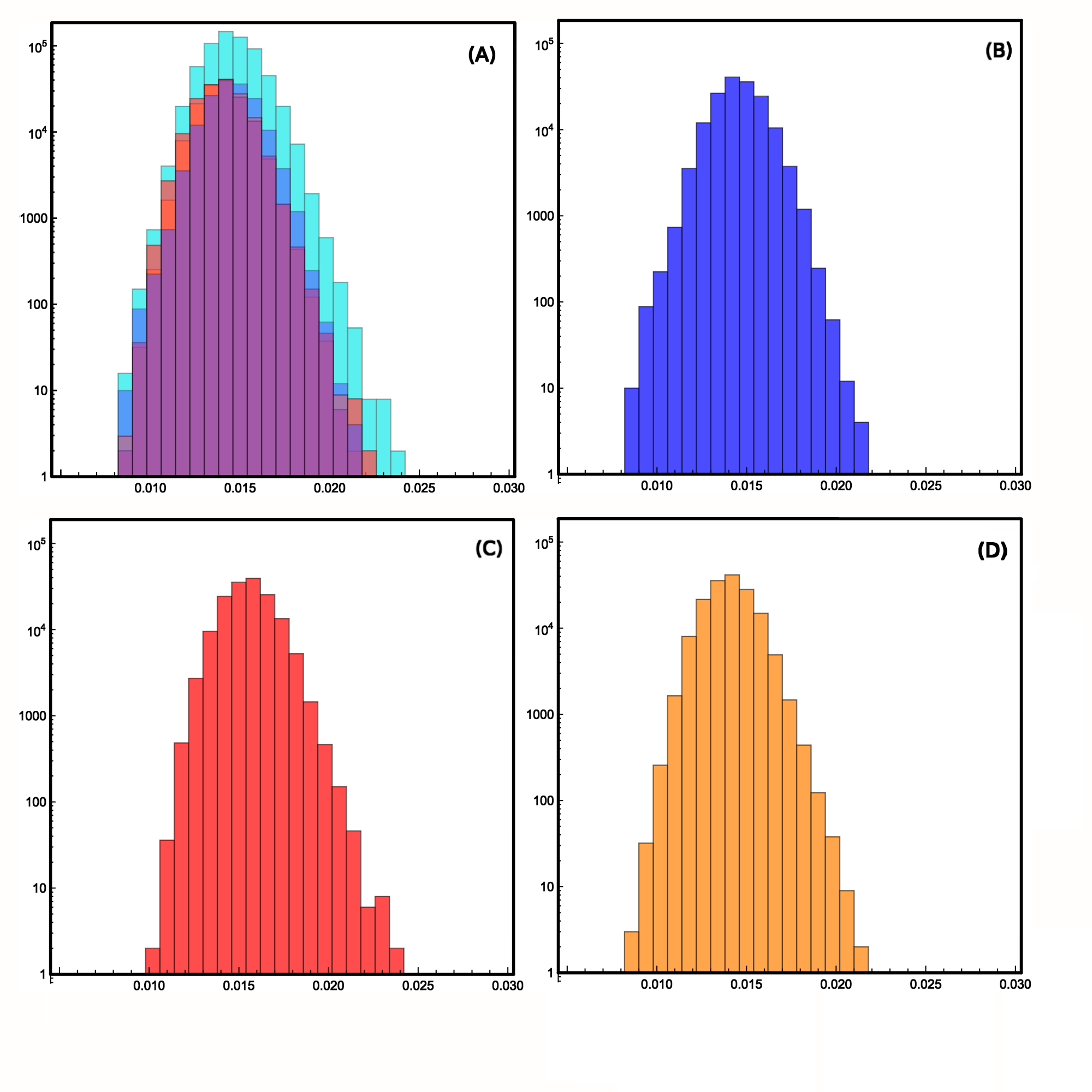}
\caption*{
\textbf{Figure 10. Overlap distributions}: comparison of the distributions of the overlap between various set of experimental trials (eq. 149). (A) Comparison between D1+D2, D1 and D2 and the interoverlap (D1 trials vs D2 trials). (B) D1. (C) D2. (D) interoverlap.
}
\label{fig:10}
\end{center}
\end{figure}

\clearpage

\begin{figure}[h!]
\begin{center}
\includegraphics[scale=0.63]{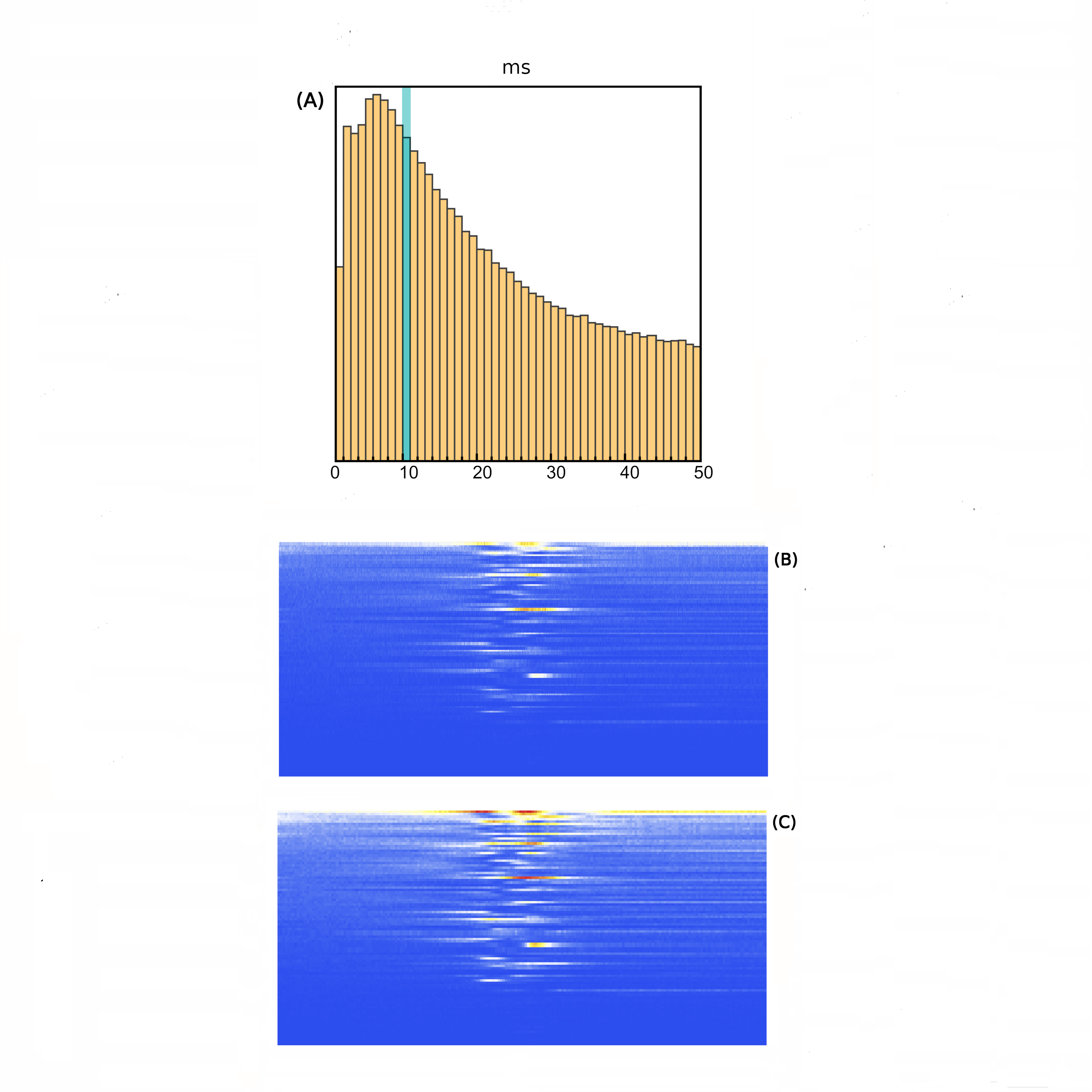}
\caption*{
\textbf{Figure 11. Renormalization}: (A) Interspike interval (ISI) distribution for Monkey P considering both direction of Movement together (D1+D2): the renormalization time is chosen at 10ms. (B-C) Comparison between the D1+D2 kernel (B) and its renormalized version (C). We can see that the two kernels are similar. Notice the amplification of the signal in the renormalized kernel, due to increase in spike density. 
}
\label{fig:11}
\end{center}
\end{figure}

\begin{figure}[h!]
\begin{center}
\includegraphics[scale=0.6]{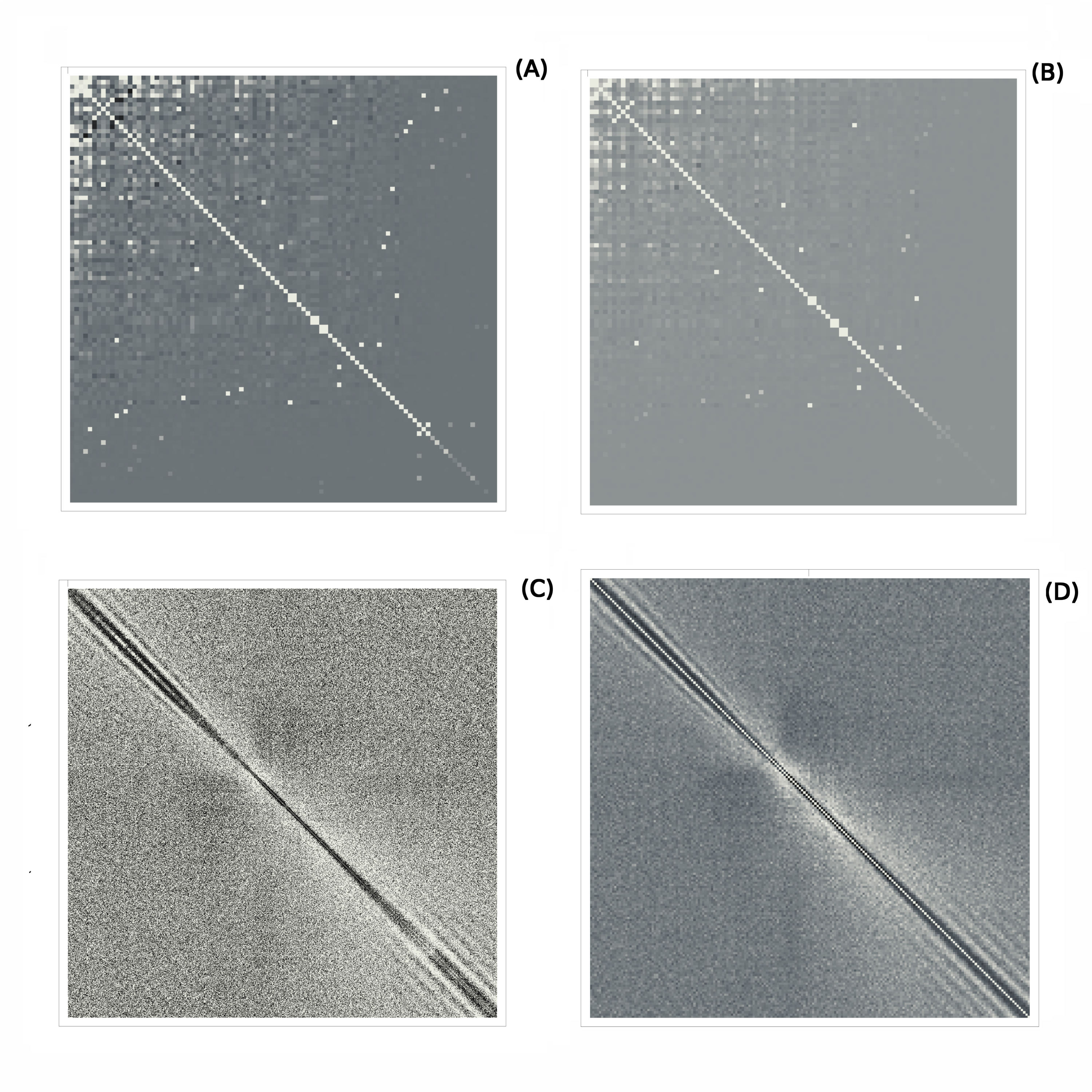}
\caption*{
\textbf{Figure 12. Renormalization}: Comparison between the D1+D2 covariance matrices (A-C) and their renormalized version (B-D). We can see how the patterns are preserved in both cases.
}
\label{fig:12}
\end{center}
\end{figure}

\begin{figure}[h!]
\begin{center}
\includegraphics[scale=0.85]{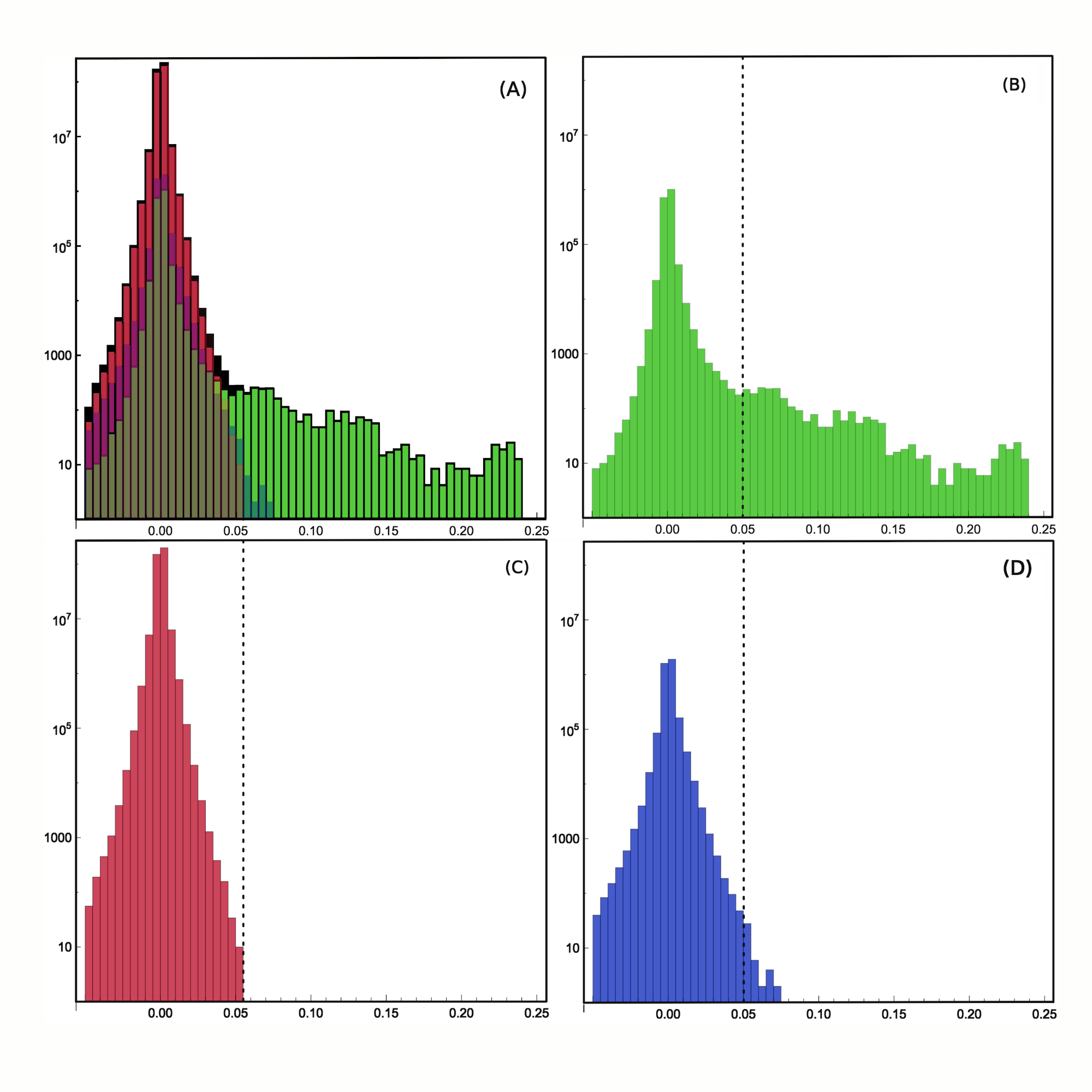}
\caption*{
\textbf{Figure 13. Grand-covariance test} We analyze the entries of the renormalized grand covariance, that is the connected two-body correlation between the points of the renormalized kernel. Panel (A) is the full distribution of the grand covariance, panel (B) in green and (D) in blue are distributions containing only elements that contributes to the correlation and overlap matrices defined in Eq.s 25 and 26, the red distribution (C) contains all the other points. We see that the red distribution follows the expected normal-product peak centered on zero due to the product of independent Gaussian fluctuations, that is also in the blue and green distributions. But notice that the most correlated pairs deviating from the normal product distribution are only in blue and green. 
}
\label{fig:13}
\end{center}
\end{figure}

\begin{figure}[h!]
\begin{center}
\includegraphics[scale=0.4]{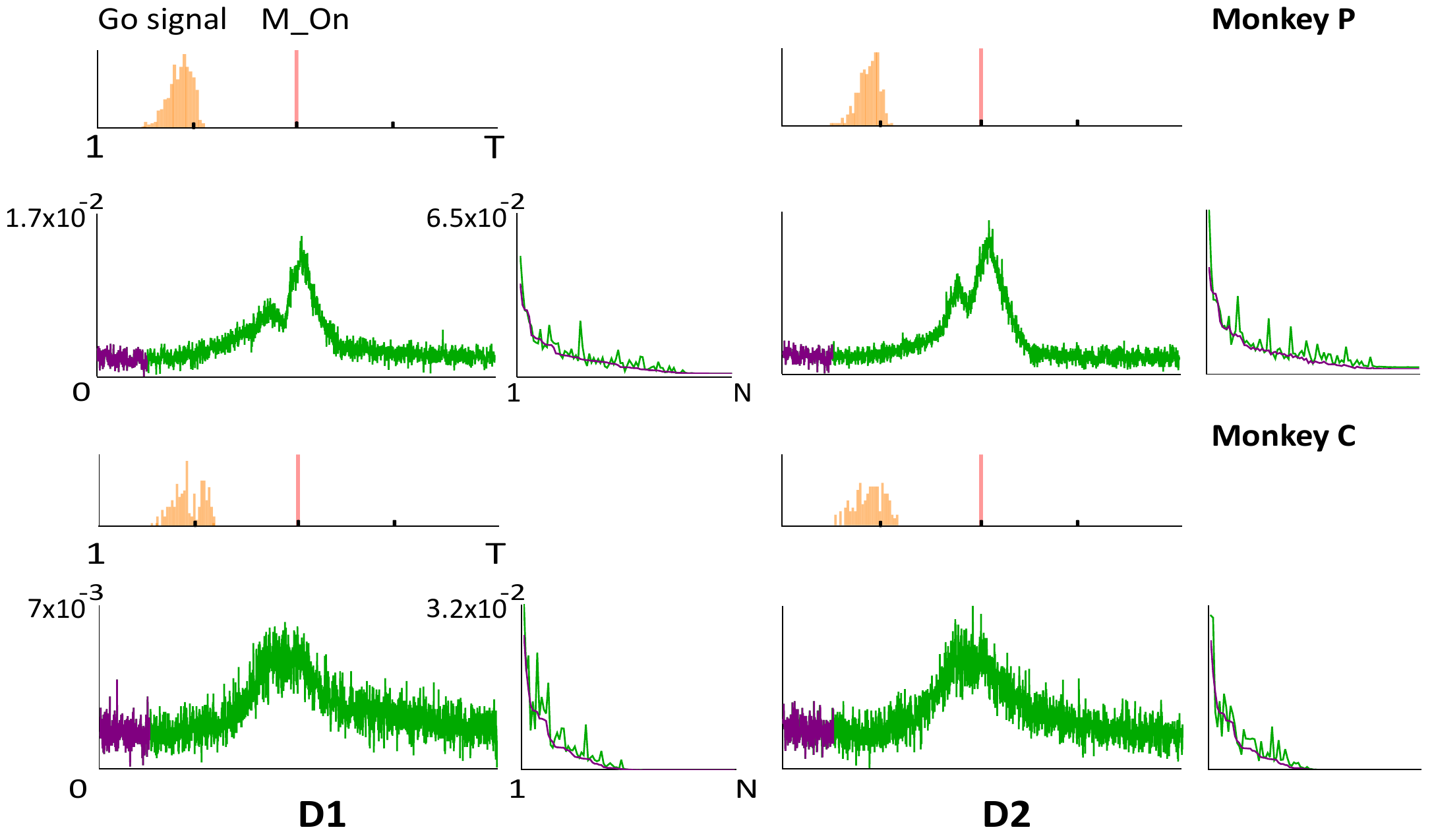}
\caption*{
\textbf{Figure 14. Experimental first order observables}: upper panels $\langle \hat{I} \rangle $; lower panels  $\langle\hat{\omega}\rangle$ and $\langle\hat{f}\rangle$. The time window is centered [-1, +1]s to the onset of Movement (M\_on). Alignment include the distributions of the stimuli: Go signal (orange distribution), M\_on (magenta). Total number of trials: $n_{tr}^{P}=800$, $n_{tr}^{C}=404$. Number of neurons recorded: $n_{neur}^{P}=166$, $n_{neur}^{C}=71$. T = 2s. Number of recording electrodes of the Utah array: N = 96 for both monkeys. Ticks are every 500 ms.
}
\label{fig:14}
\end{center}
\end{figure}

\clearpage

\begin{figure}[h!]
\begin{center}
\includegraphics[scale=0.4]{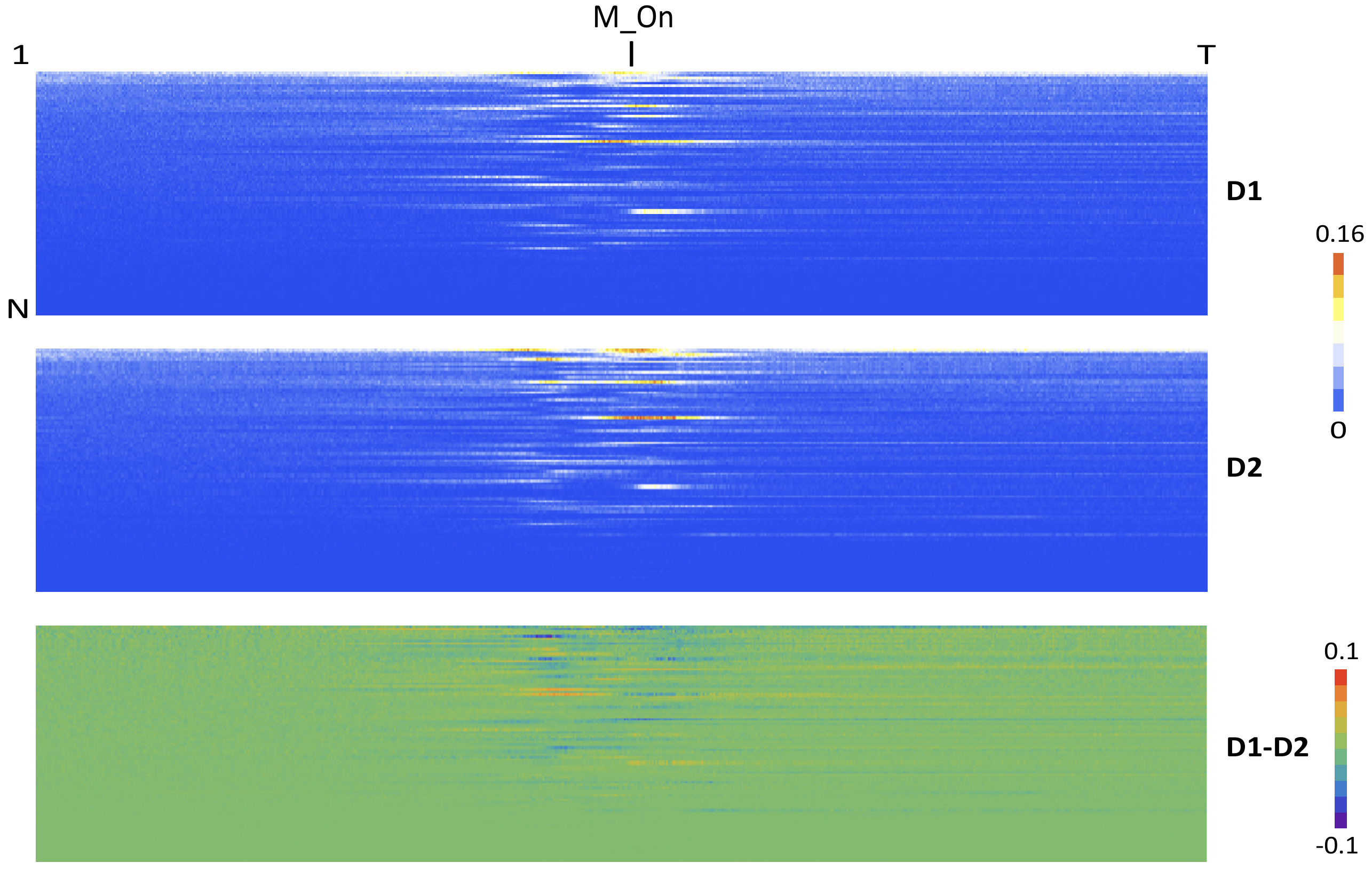}
\caption*{
\textbf{Figure 15. Electrode kernels for Monkey P}. The time window is centered [-1, +1]s to the M\_on. $n_{tr}^{P}=800$; $n_{neur}^{P}=166$. T = 2s; N = 96.
}
\label{fig:15}
\end{center}
\end{figure}

\begin{figure}[h!]
\begin{center}
\includegraphics[scale=0.4]{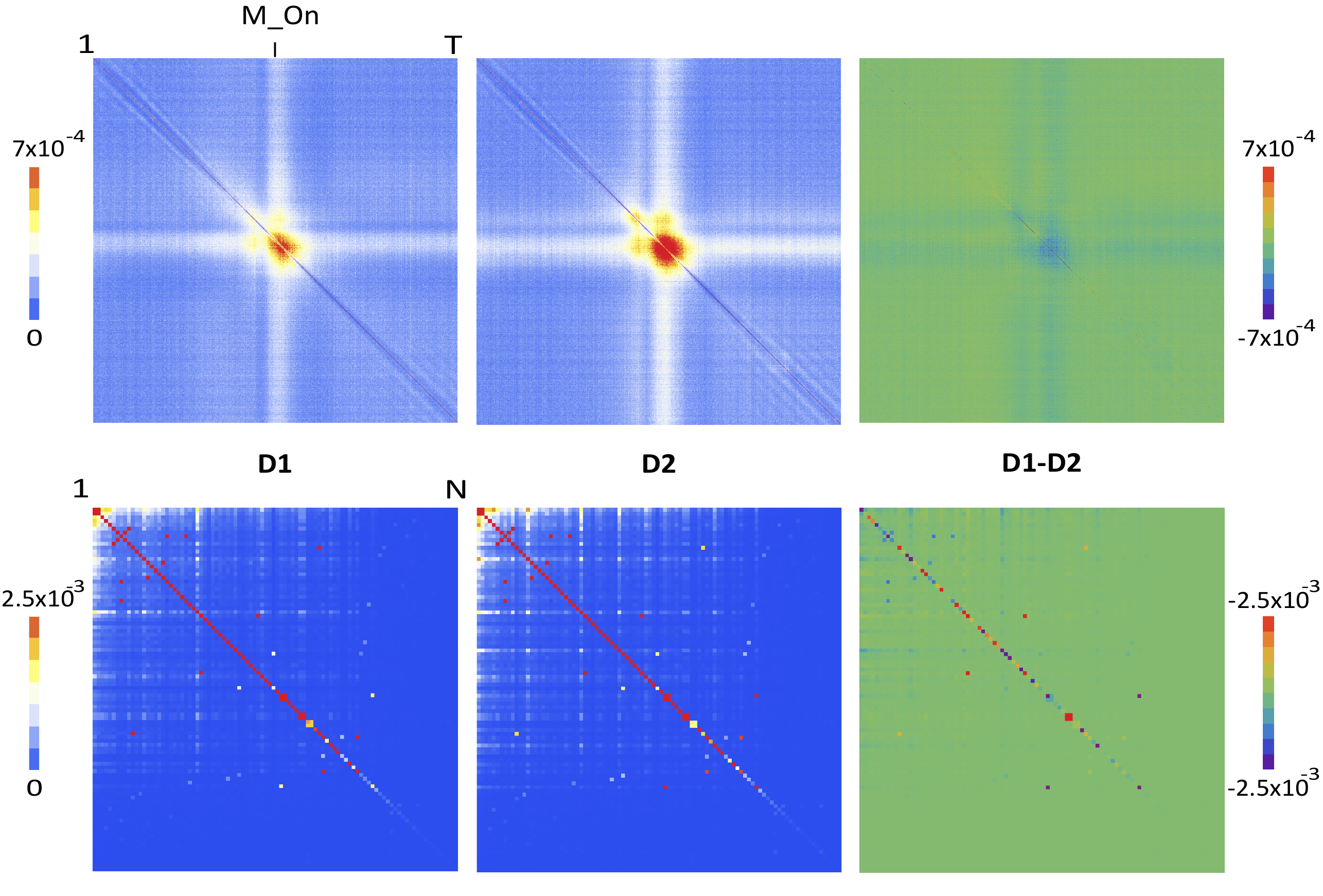}
\caption*{
\textbf{Figure 16. Experimental matrices for Monkey P}: upper panel $\langle\hat{\Pi}\rangle$; lower panel $\langle\hat{\Phi}\rangle$. $n_{tr}^{P}=800$; $n_{neur}^{P}=166$. T = 2s; N = 96.
}
\label{fig:16}
\end{center}
\end{figure}

\begin{figure}[h!]
\begin{center}
\includegraphics[scale=0.4]{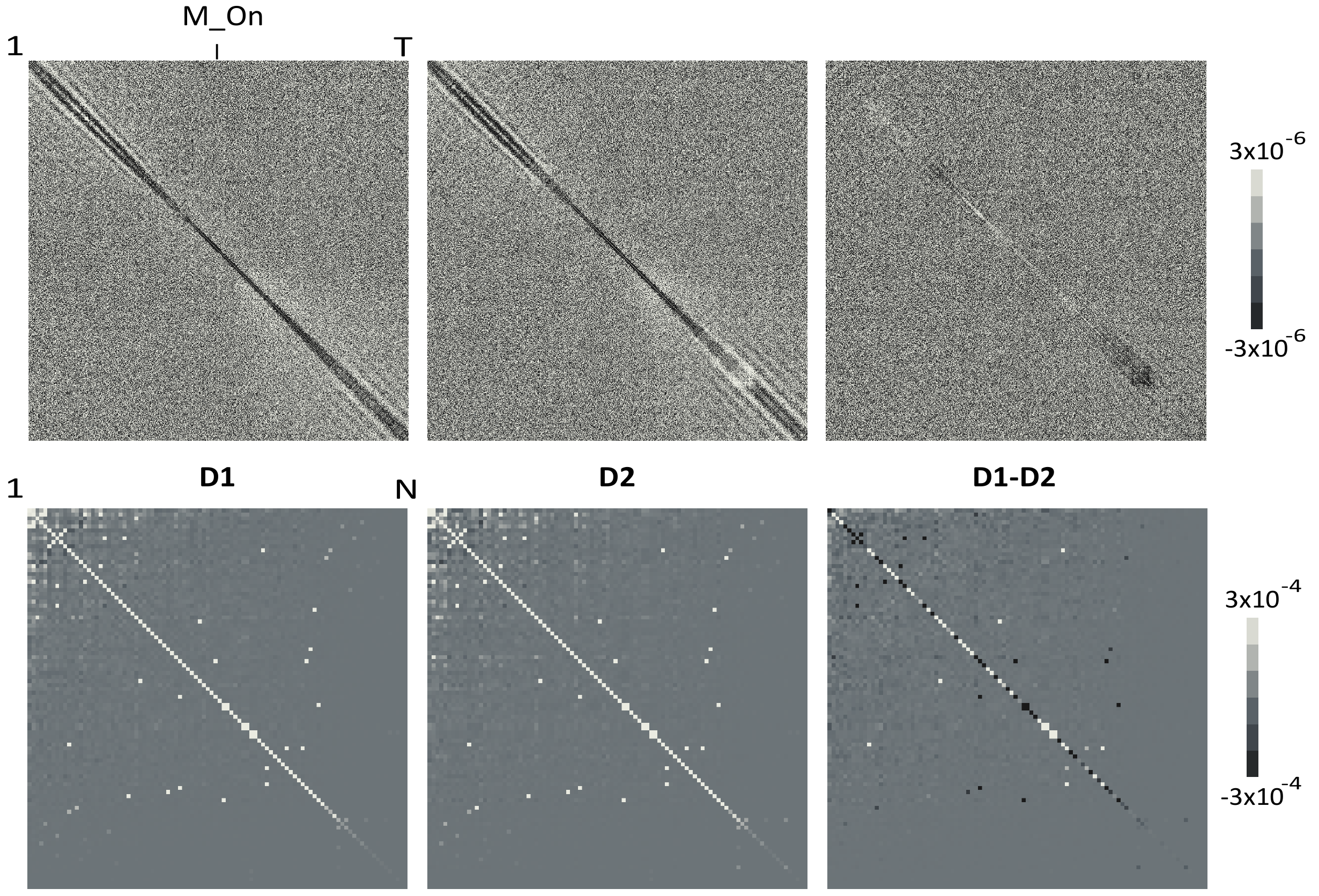}
\caption*{\textbf{Figure 17. Experimental ensemble covariance matrices for Monkey P}: upper panel $\langle \delta \hat{Q}\rangle$; lower panel$\langle \delta \hat{C}\rangle$; from eq.\ref{eq:covariances}. $n_{tr}^{C}=800$ $n_{neur}^{P}=166$. T = 2s; N = 96.
}
\label{fig:17}
\end{center}
\end{figure}

\begin{figure}[h!]
\begin{center}
\includegraphics[scale=0.4]{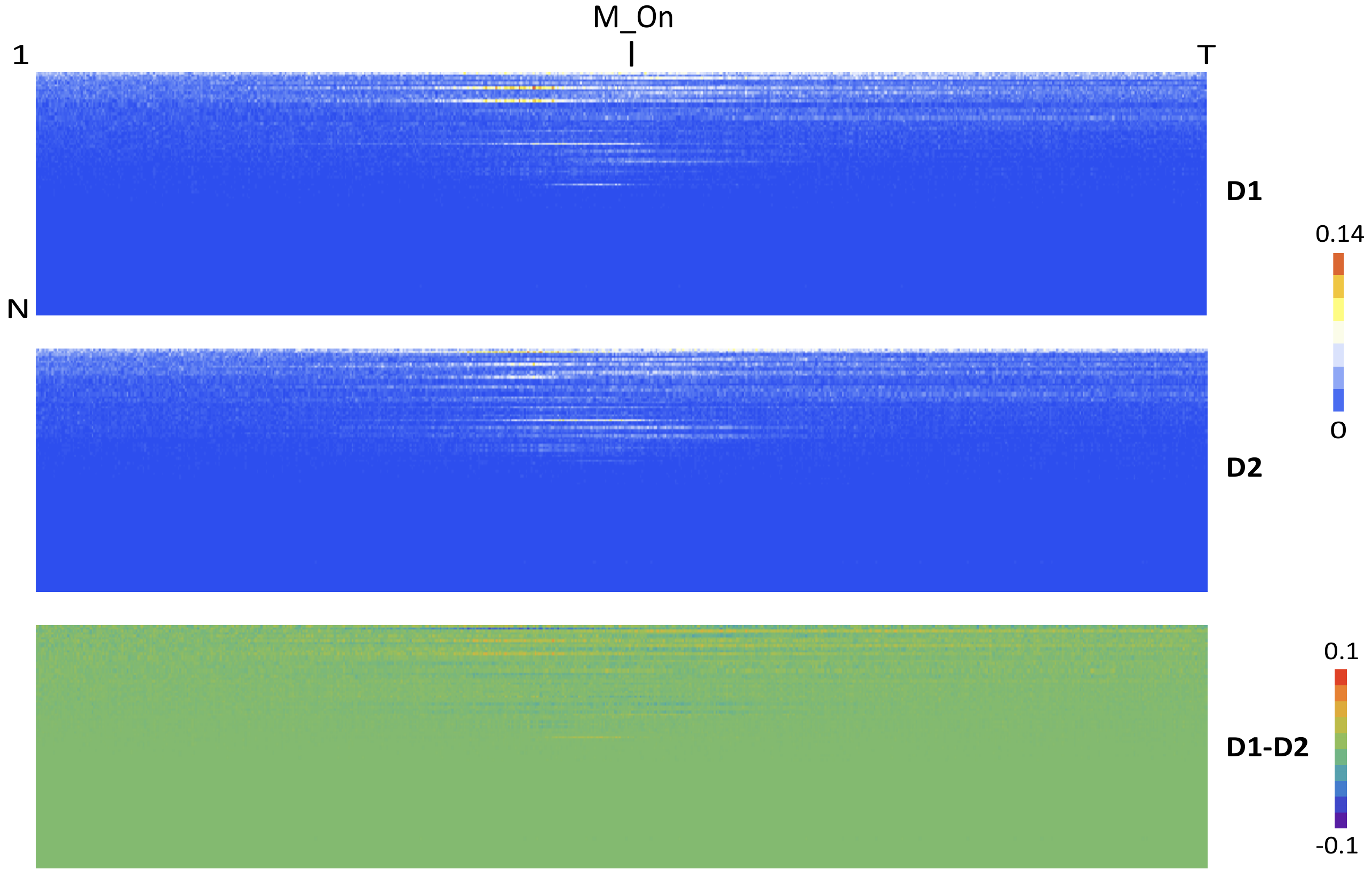}
\caption*{\textbf{Figure 18. Electrode kernels for Monkey C:} The time window is centered [-1, +1]s to the M\_on. $n_{tr}^{C}=404$, $n_{neur}^{P}=71$. T = 2s; N = 96.
}
\label{fig:18}
\end{center}
\end{figure}

\begin{figure}[h!]
\begin{center}
\includegraphics[scale=0.4]{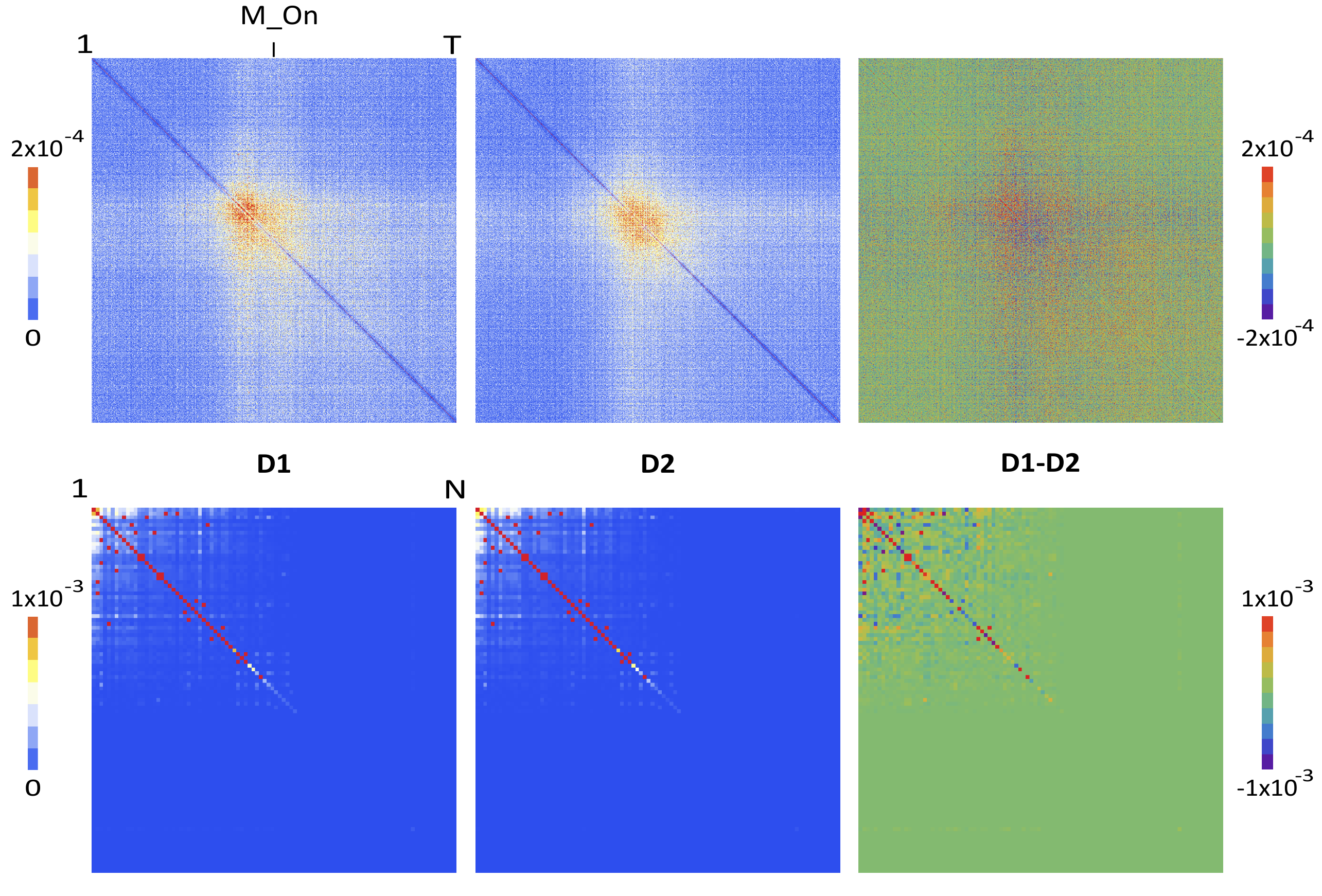}
\caption*{
\textbf{Figure 19. Experimental matrices for Monkey C}: upper panel $\langle\hat{\Pi}\rangle$; lower panel $\langle\hat{\Phi}\rangle$. $n_{tr}^{C}=404$, $n_{neur}^{P}=71$. T = 2s; N = 96. 
}
\label{fig:19}
\end{center}
\end{figure}

\begin{figure}[h!]
\begin{center}
\includegraphics[scale=0.4]{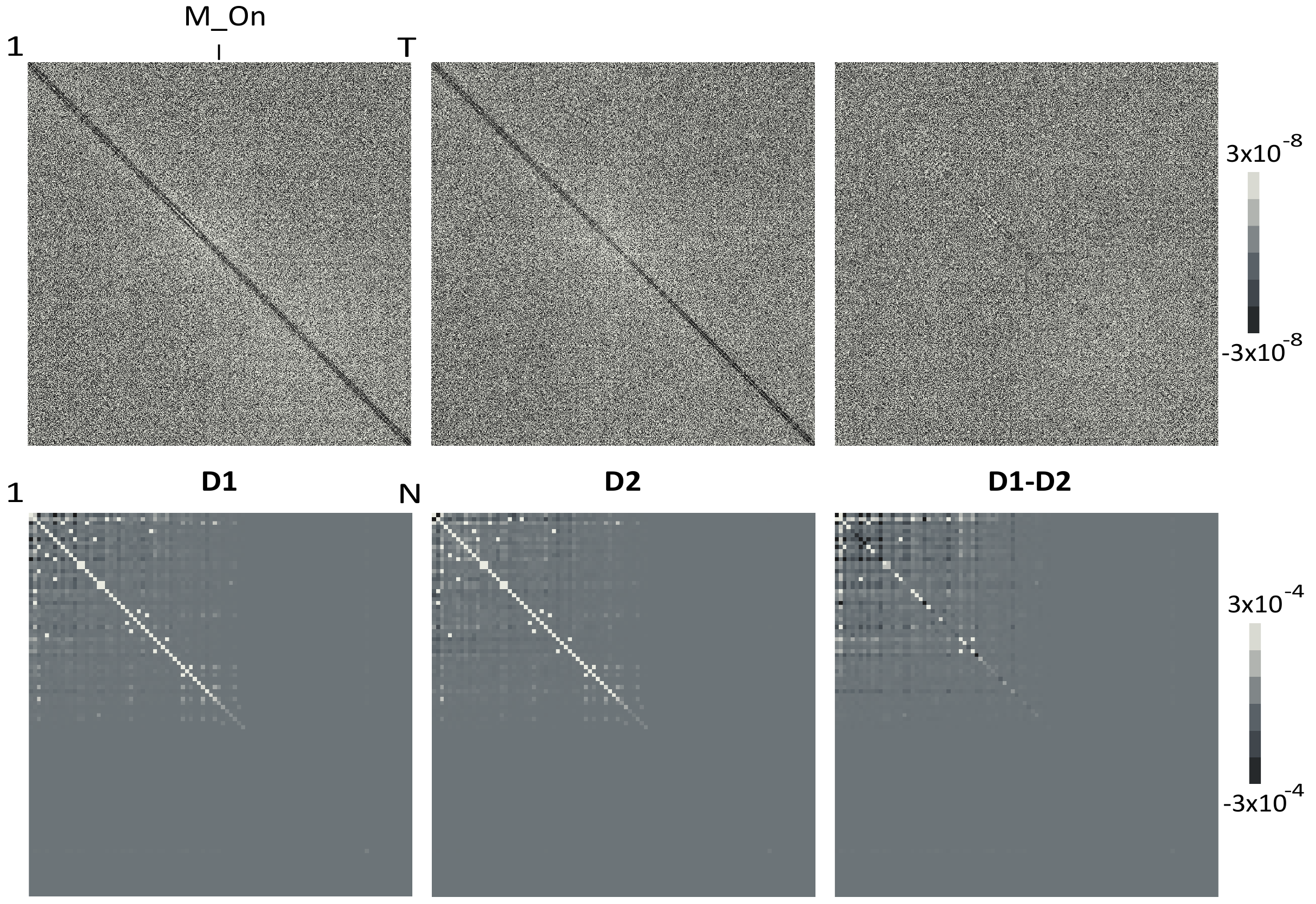}
\caption*{
\textbf{Figure 20. Experimental ensemble covariance matrices for Monkey C}: upper panel $\langle \delta \hat{Q}\rangle$; lower panel $\langle \delta \hat{C}\rangle$; from eq.(80) SI. $n_{tr}^{C}=404$ $n_{neur}^{P}=71$. T = 2s; N = 96.
}
\label{fig:20}
\end{center}
\end{figure}

\clearpage

\phantomsection

\clearpage

\clearpage

\section*{Acknowledgements}

We thank Cheng Shi (University of Basel), Alessandro Treves (SISSA Trieste), Giorgio Parisi (Accademia dei Lincei) and Karl Friston (University College London) for interesting discussions. Partially supported by grants from Sapienza University (PH11715C823A9528 to Stefano Ferraina and RM12117A8AD27DB1 to Paolo Pani), from EBRAINS-Italy PNRR 2023 grant (to Stefano Ferraina).

\section*{Data availability statement}
Data are available from the corresponding author(s) upon reasonable request.

\section*{Competing interests}
The authors report no declarations of interest.

\clearpage

\phantomsection

\bibliography{Ref_Neuroglass}

\begin{thebibliography}{100}
\urlstyle{rm}
\expandafter\ifx\csname url\endcsname\relax
  \def\url#1{\texttt{#1}}\fi
\expandafter\ifx\csname urlprefix\endcsname\relax\def\urlprefix{URL }\fi
\expandafter\ifx\csname doiprefix\endcsname\relax\def\doiprefix{DOI: }\fi
\providecommand{\bibinfo}[2]{#2}
\providecommand{\eprint}[2][]{\url{#2}}

\bibitem{Pandarinath2018}
\bibinfo{author}{Pandarinath, C.} \emph{et~al.}
\newblock \bibinfo{journal}{\bibinfo{title}{{Inferring single-trial neural population dynamics using sequential auto-encoders}}}.
\newblock {\emph{\JournalTitle{Nature Methods 2018 15:10}}} \textbf{\bibinfo{volume}{15}}, \bibinfo{pages}{805--815}, \url{10.1038/s41592-018-0109-9} (\bibinfo{year}{2018}).

\bibitem{Barack2021}
\bibinfo{author}{Barack, D.~L.} \& \bibinfo{author}{Krakauer, J.~W.}
\newblock \bibinfo{journal}{\bibinfo{title}{{Two views on the cognitive brain}}}.
\newblock {\emph{\JournalTitle{Nature Reviews Neuroscience 2021 22:6}}} \textbf{\bibinfo{volume}{22}}, \bibinfo{pages}{359--371}, \url{10.1038/s41583-021-00448-6} (\bibinfo{year}{2021}).

\bibitem{Stavisky2021}
\bibinfo{author}{Stavisky, S.~D.} \emph{et~al.}
\newblock \bibinfo{journal}{\bibinfo{title}{{Robust and accurate decoding of hand kinematics from entire spiking activity using deep learning}}}.
\newblock {\emph{\JournalTitle{Journal of Neural Engineering}}} \textbf{\bibinfo{volume}{18}}, \bibinfo{pages}{026011}, \url{10.1088/1741-2552/ABDE8A} (\bibinfo{year}{2021}).

\bibitem{Pang2023}
\bibinfo{author}{Pang, J.~C.} \emph{et~al.}
\newblock \bibinfo{journal}{\bibinfo{title}{{Geometric constraints on human brain function}}}.
\newblock {\emph{\JournalTitle{Nature 2023 618:7965}}} \textbf{\bibinfo{volume}{618}}, \bibinfo{pages}{566--574}, \url{10.1038/s41586-023-06098-1} (\bibinfo{year}{2023}).

\bibitem{Faskowitz2023}
\bibinfo{author}{Faskowitz, J.} \emph{et~al.}
\newblock \bibinfo{journal}{\bibinfo{title}{{Commentary on Pang et al. (2023) Nature}}}.
\newblock {\emph{\JournalTitle{bioRxiv}}} \bibinfo{pages}{2023.07.20.549785}, \url{10.1101/2023.07.20.549785} (\bibinfo{year}{2023}).

\bibitem{Gardner2022}
\bibinfo{author}{Gardner, R.~J.} \emph{et~al.}
\newblock \bibinfo{journal}{\bibinfo{title}{{Toroidal topology of population activity in grid cells}}}.
\newblock {\emph{\JournalTitle{Nature 2022 602:7895}}} \textbf{\bibinfo{volume}{602}}, \bibinfo{pages}{123--128}, \url{10.1038/s41586-021-04268-7} (\bibinfo{year}{2022}).

\bibitem{Shi2023}
\bibinfo{author}{Shi, Y.~L.}, \bibinfo{author}{Zeraati, R.}, \bibinfo{author}{Levina, A.} \& \bibinfo{author}{Engel, T.~A.}
\newblock \bibinfo{journal}{\bibinfo{title}{{Spatial and temporal correlations in neural networks with structured connectivity}}}.
\newblock {\emph{\JournalTitle{Physical Review Research}}} \textbf{\bibinfo{volume}{5}}, \bibinfo{pages}{013005}, \url{10.1103/PHYSREVRESEARCH.5.013005/FIGURES/21/MEDIUM} (\bibinfo{year}{2023}).

\bibitem{Genkin2021}
\bibinfo{author}{Genkin, M.}, \bibinfo{author}{Hughes, O.} \& \bibinfo{author}{Engel, T.~A.}
\newblock \bibinfo{journal}{\bibinfo{title}{{Learning non-stationary Langevin dynamics from stochastic observations of latent trajectories}}}.
\newblock {\emph{\JournalTitle{Nature Communications 2021 12:1}}} \textbf{\bibinfo{volume}{12}}, \bibinfo{pages}{1--9}, \url{10.1038/s41467-021-26202-1} (\bibinfo{year}{2021}).

\bibitem{Pinotsis2023}
\bibinfo{author}{Pinotsis, D.~A.} \& \bibinfo{author}{Miller, E.~K.}
\newblock \bibinfo{journal}{\bibinfo{title}{{In vivo ephaptic coupling allows memory network formation}}}.
\newblock {\emph{\JournalTitle{Cerebral Cortex}}} \url{10.1093/CERCOR/BHAD251} (\bibinfo{year}{2023}).

\bibitem{Bullard2019}
\bibinfo{author}{Bullard, A.}
\newblock \bibinfo{title}{{Feasibility of Using the Utah Array for Long-Term Fully Implantable Neuroprosthesis Systems}}.
\newblock \bibinfo{type}{Tech. Rep.} (\bibinfo{year}{2019}).

\bibitem{Z.Wei2020}
\bibinfo{author}{Wei, Z.} \emph{et~al.}
\newblock \bibinfo{journal}{\bibinfo{title}{{A comparison of neuronal population dynamics measured with calcium imaging and electrophysiology}}}.
\newblock {\emph{\JournalTitle{PLoS Computational Biology}}} \textbf{\bibinfo{volume}{16}}, \url{10.1371/journal.pcbi.1008198} (\bibinfo{year}{2020}).

\bibitem{Chandrasekaran2021}
\bibinfo{author}{Chandrasekaran, S.} \emph{et~al.}
\newblock \bibinfo{journal}{\bibinfo{title}{{Historical perspectives, challenges, and future directions of implantable brain-computer interfaces for sensorimotor applications}}}.
\newblock {\emph{\JournalTitle{Bioelectronic medicine}}} \textbf{\bibinfo{volume}{7}}, \url{10.1186/S42234-021-00076-6} (\bibinfo{year}{2021}).

\bibitem{Pani2022}
\bibinfo{author}{Pani, P.} \emph{et~al.}
\newblock \bibinfo{journal}{\bibinfo{title}{{Neuronal population dynamics during motor plan cancellation in nonhuman primates}}}.
\newblock {\emph{\JournalTitle{Proceedings of the National Academy of Sciences of the United States of America}}} \textbf{\bibinfo{volume}{119}}, \bibinfo{pages}{e2122395119}, \url{10.1073/pnas.2122395119} (\bibinfo{year}{2022}).

\bibitem{Stringer2019}
\bibinfo{author}{Stringer, C.} \emph{et~al.}
\newblock \bibinfo{journal}{\bibinfo{title}{{Spontaneous behaviors drive multidimensional, brainwide activity}}}.
\newblock {\emph{\JournalTitle{Science (New York, N.Y.)}}} \textbf{\bibinfo{volume}{364}}, \url{10.1126/SCIENCE.AAV7893} (\bibinfo{year}{2019}).

\bibitem{Pachitariu2017}
\bibinfo{author}{Pachitariu, M.} \emph{et~al.}
\newblock \bibinfo{journal}{\bibinfo{title}{{Suite2p: beyond 10,000 neurons with standard two-photon microscopy}}}.
\newblock {\emph{\JournalTitle{bioRxiv}}} \bibinfo{pages}{061507}, \url{10.1101/061507} (\bibinfo{year}{2017}).

\bibitem{Wilson1974}
\bibinfo{author}{Wilson, K.~G.}
\newblock \bibinfo{journal}{\bibinfo{title}{{Confinement of quarks}}}.
\newblock {\emph{\JournalTitle{Physical Review D}}} \textbf{\bibinfo{volume}{10}}, \bibinfo{pages}{2445}, \url{10.1103/PhysRevD.10.2445} (\bibinfo{year}{1974}).

\bibitem{Balian1974}
\bibinfo{author}{Balian, R.}, \bibinfo{author}{Drouffe, J.~M.} \& \bibinfo{author}{Itzykson, C.}
\newblock \bibinfo{journal}{\bibinfo{title}{{Gauge fields on a lattice. I. General outlook}}}.
\newblock {\emph{\JournalTitle{Physical Review D}}} \textbf{\bibinfo{volume}{10}}, \bibinfo{pages}{3376}, \url{10.1103/PhysRevD.10.3376} (\bibinfo{year}{1974}).

\bibitem{Lee1983}
\bibinfo{author}{Lee, T.~D.}
\newblock \bibinfo{journal}{\bibinfo{title}{{Can time be a discrete dynamical variable?}}}
\newblock {\emph{\JournalTitle{Physics Letters B}}} \textbf{\bibinfo{volume}{122}}, \bibinfo{pages}{217--220}, \url{10.1016/0370-2693(83)90687-1} (\bibinfo{year}{1983}).

\bibitem{Lee1987}
\bibinfo{author}{Lee, T.~D.}
\newblock \bibinfo{journal}{\bibinfo{title}{{Difference equations and conservation laws}}}.
\newblock {\emph{\JournalTitle{Journal of Statistical Physics}}} \textbf{\bibinfo{volume}{46}}, \bibinfo{pages}{843--860}, \url{10.1007/BF01011145/METRICS} (\bibinfo{year}{1987}).

\bibitem{Parisi1989}
\bibinfo{author}{Parisi, G.}
\newblock \emph{\bibinfo{title}{{Statistical Field Theory}}} (\bibinfo{publisher}{Addison-Wesley}, \bibinfo{address}{Redwood City,CA, USA,}, \bibinfo{year}{1989}).

\bibitem{Wiese2009}
\bibinfo{author}{Wiese, U.-J.}
\newblock \bibinfo{title}{{An Introduction to Lattice Field Theory}}.
\newblock \bibinfo{type}{Tech. Rep.} \bibinfo{number}{11} (\bibinfo{year}{2009}).

\bibitem{Gupta2011}
\bibinfo{author}{Gupta, S.}
\newblock \bibinfo{title}{{Introduction to Lattice Field Theory}}.
\newblock \bibinfo{type}{Tech. Rep.}, \bibinfo{institution}{Asian Schoolon Lattice Field Theory TIFR}, \bibinfo{address}{Mumbai,India} (\bibinfo{year}{2011}).

\bibitem{Zohar2015}
\bibinfo{author}{Zohar, E.} \& \bibinfo{author}{Burrello, M.}
\newblock \bibinfo{journal}{\bibinfo{title}{{Formulation of lattice gauge theories for quantum simulations}}}.
\newblock {\emph{\JournalTitle{Physical Review D - Particles, Fields, Gravitation and Cosmology}}} \textbf{\bibinfo{volume}{91}}, \bibinfo{pages}{054506}, \url{10.1103/PHYSREVD.91.054506/FIGURES/1/MEDIUM} (\bibinfo{year}{2015}).

\bibitem{ParottoQCD2018}
\bibinfo{author}{Parotto, P.}
\newblock \bibinfo{journal}{\bibinfo{title}{{Parametrized Equation of State for QCD from 3D Ising Model}}}.
\newblock {\emph{\JournalTitle{Proceedings of Science}}} \textbf{\bibinfo{volume}{311}}, \bibinfo{pages}{036}, \url{10.22323/1.311.0036} (\bibinfo{year}{2018}).

\bibitem{FaccioliSFT}
\bibinfo{author}{Faccioli, P.}
\newblock \bibinfo{title}{{Lecture Course: Statistical Field Theory - YouTube}} (\bibinfo{year}{2020}).

\bibitem{Magnifico2021}
\bibinfo{author}{Magnifico, G.}, \bibinfo{author}{Felser, T.}, \bibinfo{author}{Silvi, P.} \& \bibinfo{author}{Montangero, S.}
\newblock \bibinfo{journal}{\bibinfo{title}{{Lattice quantum electrodynamics in (3+1)-dimensions at finite density with tensor networks}}}.
\newblock {\emph{\JournalTitle{Nature Communications}}} \textbf{\bibinfo{volume}{12}}, \url{10.1038/s41467-021-23646-3} (\bibinfo{year}{2021}).

\bibitem{Buice2007}
\bibinfo{author}{Buice, M.~A.} \& \bibinfo{author}{Cowan, J.~D.}
\newblock \bibinfo{journal}{\bibinfo{title}{{Field-theoretic approach to fluctuation effects in neural networks}}}.
\newblock {\emph{\JournalTitle{Physical Review E - Statistical, Nonlinear, and Soft Matter Physics}}} \textbf{\bibinfo{volume}{75}}, \bibinfo{pages}{051919}, \url{10.1103/PHYSREVE.75.051919/FIGURES/9/MEDIUM} (\bibinfo{year}{2007}).

\bibitem{Buice2013}
\bibinfo{author}{Buice, M.~A.} \& \bibinfo{author}{Chow, C.~C.}
\newblock \bibinfo{journal}{\bibinfo{title}{{Beyond mean field theory: statistical field theory for neural networks}}}.
\newblock {\emph{\JournalTitle{Journal of Statistical Mechanics: Theory and Experiment}}} \textbf{\bibinfo{volume}{2013}}, \bibinfo{pages}{P03003}, \url{10.1088/1742-5468/2013/03/P03003} (\bibinfo{year}{2013}).

\bibitem{Fagerholm2021}
\bibinfo{author}{Fagerholm, E.~D.}, \bibinfo{author}{Foulkes, W.~M.}, \bibinfo{author}{Friston, K.~J.}, \bibinfo{author}{Moran, R.~J.} \& \bibinfo{author}{Leech, R.}
\newblock \bibinfo{journal}{\bibinfo{title}{{Rendering neuronal state equations compatible with the principle of stationary action}}}.
\newblock {\emph{\JournalTitle{Journal of Mathematical Neuroscience}}} \textbf{\bibinfo{volume}{11}}, \bibinfo{pages}{1--15}, \url{10.1186/S13408-021-00108-0/FIGURES/2} (\bibinfo{year}{2021}).

\bibitem{Gosselin2022}
\bibinfo{author}{Gosselin, P.}, \bibinfo{author}{Lotz, A.} \& \bibinfo{author}{Wambst, M.}
\newblock \bibinfo{journal}{\bibinfo{title}{{Statistical Field Theory and Networks of Spiking Neurons}}}.
\newblock {\emph{\JournalTitle{arXiv}}}  (\bibinfo{year}{2020}).

\bibitem{Halverson2022}
\bibinfo{author}{Halverson, J.}
\newblock \bibinfo{journal}{\bibinfo{title}{{Building Quantum Field Theories Out of Neurons}}}.
\newblock {\emph{\JournalTitle{arXiv}}}  (\bibinfo{year}{2021}).

\bibitem{Tiberi2022}
\bibinfo{author}{Tiberi, L.} \emph{et~al.}
\newblock \bibinfo{journal}{\bibinfo{title}{{Gell-Mann-Low Criticality in Neural Networks}}}.
\newblock {\emph{\JournalTitle{Physical Review Letters}}} \textbf{\bibinfo{volume}{128}}, \bibinfo{pages}{168301}, \url{10.1103/PHYSREVLETT.128.168301/FIGURES/1/MEDIUM} (\bibinfo{year}{2022}).

\bibitem{Gornitz1992}
\bibinfo{author}{Gornitz, T.}, \bibinfo{author}{Graudenz, D.} \& \bibinfo{author}{v.~Weizsacker, C.~F.}
\newblock \bibinfo{journal}{\bibinfo{title}{{Quantum field theory of binary alternatives}}}.
\newblock {\emph{\JournalTitle{International Journal of Theoretical Physics}}} \textbf{\bibinfo{volume}{31}}, \bibinfo{pages}{1929--1959}, \url{10.1007/BF00671965/METRICS} (\bibinfo{year}{1992}).

\bibitem{Deutsch2004}
\bibinfo{author}{Deutsch, D.}
\newblock \bibinfo{journal}{\bibinfo{title}{{Qubit Field Theory}}}.
\newblock {\emph{\JournalTitle{arXiv}}}  (\bibinfo{year}{2004}).

\bibitem{Singh2020}
\bibinfo{author}{Singh, H.}
\newblock \emph{\bibinfo{title}{{Exploring Quantum Field Theories with Qubit Lattice Models}}}.
\newblock Ph.D. thesis (\bibinfo{year}{2020}).

\bibitem{Franchini2023}
\bibinfo{author}{Franchini, S.}
\newblock \bibinfo{journal}{\bibinfo{title}{{Replica Symmetry Breaking without replicas}}}.
\newblock {\emph{\JournalTitle{Annals of Physics}}} \textbf{\bibinfo{volume}{450}}, \url{10.1016/j.aop.2023.169220} (\bibinfo{year}{2023}).

\bibitem{Franchini2021}
\bibinfo{author}{Franchini, S.}
\newblock \bibinfo{journal}{\bibinfo{title}{{A simplified Parisi ansatz}}}.
\newblock {\emph{\JournalTitle{Communications in Theoretical Physics}}} \textbf{\bibinfo{volume}{73}}, \bibinfo{pages}{055601}, \url{10.1088/1572-9494/ABDE32} (\bibinfo{year}{2021}).

\bibitem{FranchiniREM2023}
\bibinfo{author}{Franchini, S.}
\newblock \bibinfo{journal}{\bibinfo{title}{{A simplified Parisi Ansatz II: REM Universality}}}.
\newblock {\emph{\JournalTitle{arXiv}}}  (\bibinfo{year}{2023}).

\bibitem{Concetti2018}
\bibinfo{author}{Concetti, F.}
\newblock \bibinfo{journal}{\bibinfo{title}{{The Full Replica Symmetry Breaking in the Ising Spin Glass on Random Regular Graph}}}.
\newblock {\emph{\JournalTitle{Journal of Statistical Physics}}} \textbf{\bibinfo{volume}{173}}, \bibinfo{pages}{1459--1483}, \url{10.1007/S10955-018-2142-6/METRICS} (\bibinfo{year}{2018}).

\bibitem{Mezard1987}
\bibinfo{author}{Mezard, M.}, \bibinfo{author}{Parisi, G.} \& \bibinfo{author}{Virasoro, M.~A.}
\newblock \emph{\bibinfo{title}{{Spin glass theory and beyond}}} (\bibinfo{year}{1987}).

\bibitem{Mezard2009}
\bibinfo{author}{Mezard, M.} \& \bibinfo{author}{Montanari, A.}
\newblock \emph{\bibinfo{title}{{Information, physics, and computation}}} (\bibinfo{year}{2009}).

\bibitem{Schneidman2006}
\bibinfo{author}{Schneidman, E.}, \bibinfo{author}{Berry, M.~J.}, \bibinfo{author}{Segev, R.} \& \bibinfo{author}{Bialek, W.}
\newblock \bibinfo{journal}{\bibinfo{title}{{Weak pairwise correlations imply strongly correlated network states in a neural population}}}.
\newblock {\emph{\JournalTitle{Nature}}} \textbf{\bibinfo{volume}{440}}, \bibinfo{pages}{1007--1012}, \url{10.1038/nature04701} (\bibinfo{year}{2006}).

\bibitem{Tkacik2009}
\bibinfo{author}{Tkacik, G.}, \bibinfo{author}{Schneidman, E.}, \bibinfo{author}{Berry~II, M.~J.} \& \bibinfo{author}{Bialek, W.}
\newblock \bibinfo{journal}{\bibinfo{title}{{Spin glass models for a network of real neurons}}}.
\newblock {\emph{\JournalTitle{arXiv}}}  (\bibinfo{year}{2009}).

\bibitem{Tkacik2013}
\bibinfo{author}{Tka{\v{c}}ik, G.} \emph{et~al.}
\newblock \bibinfo{journal}{\bibinfo{title}{{The simplest maximum entropy model for collective behavior in a neural network}}}.
\newblock {\emph{\JournalTitle{Journal of Statistical Mechanics: Theory and Experiment}}} \textbf{\bibinfo{volume}{2013}}, \url{10.1088/1742-5468/2013/03/P03011} (\bibinfo{year}{2013}).

\bibitem{Meshulam2021}
\bibinfo{author}{Meshulam, L.}, \bibinfo{author}{Gauthier, J.~L.}, \bibinfo{author}{Brody, C.~D.}, \bibinfo{author}{Tank, D.~W.} \& \bibinfo{author}{Bialek, W.}
\newblock \bibinfo{journal}{\bibinfo{title}{{Successes and failures of simple statistical physics models for a network of real neurons}}}.
\newblock {\emph{\JournalTitle{arXiv}}}  (\bibinfo{year}{2021}).

\bibitem{Tang2008}
\bibinfo{author}{Tang, A.} \emph{et~al.}
\newblock \bibinfo{journal}{\bibinfo{title}{{A maximum entropy model applied to spatial and temporal correlations from cortical networks in vitro}}}.
\newblock {\emph{\JournalTitle{Journal of Neuroscience}}} \textbf{\bibinfo{volume}{28}}, \bibinfo{pages}{505--518}, \url{10.1523/JNEUROSCI.3359-07.2008} (\bibinfo{year}{2008}).

\bibitem{Treves1991}
\bibinfo{author}{Treves, A.}
\newblock \bibinfo{journal}{\bibinfo{title}{{Are spin-glass effects relevant to understanding realistic auto-associative networks?}}}
\newblock {\emph{\JournalTitle{Journal of Physics A: Mathematical and General}}} \textbf{\bibinfo{volume}{24}}, \bibinfo{pages}{2645}, \url{10.1088/0305-4470/24/11/029} (\bibinfo{year}{1991}).

\bibitem{Hermann2012}
\bibinfo{author}{Hermann, G.} \& \bibinfo{author}{Touboul, J.}
\newblock \bibinfo{journal}{\bibinfo{title}{{Heterogeneous connections induce oscillations in large-scale networks}}}.
\newblock {\emph{\JournalTitle{Physical Review Letters}}} \textbf{\bibinfo{volume}{109}}, \bibinfo{pages}{018702}, \url{10.1103/PHYSREVLETT.109.018702/FIGURES/3/MEDIUM} (\bibinfo{year}{2012}).

\bibitem{Hopfield1982}
\bibinfo{author}{Hopfield, J.~J.}
\newblock \bibinfo{journal}{\bibinfo{title}{{Neural networks and physical systems with emergent collective computational abilities.}}}
\newblock {\emph{\JournalTitle{Proceedings of the National Academy of Sciences of the United States of America}}} \textbf{\bibinfo{volume}{79}}, \bibinfo{pages}{2554}, \url{10.1073/PNAS.79.8.2554} (\bibinfo{year}{1982}).

\bibitem{Amit1985}
\bibinfo{author}{Amit, D.~J.}, \bibinfo{author}{Gutfreund, H.} \& \bibinfo{author}{Sompolinsky, H.}
\newblock \bibinfo{journal}{\bibinfo{title}{{Storing Infinite Numbers of Patterns in a Spin-Glass Model of Neural Networks}}}.
\newblock {\emph{\JournalTitle{Physical Review Letters}}} \textbf{\bibinfo{volume}{55}}, \bibinfo{pages}{1530}, \url{10.1103/PhysRevLett.55.1530} (\bibinfo{year}{1985}).

\bibitem{Toulouse1986}
\bibinfo{author}{Toulouse, G.}, \bibinfo{author}{Dehaene, S.} \& \bibinfo{author}{Changeux, J.-P.}
\newblock \bibinfo{title}{{Spin glass model of learning by selection (Darwinism/categorizaton/Hebb synapse/ultrametricity/frustradon)}}.
\newblock \bibinfo{type}{Tech. Rep.} (\bibinfo{year}{1986}).

\bibitem{Treves1988}
\bibinfo{author}{Treves, A.} \& \bibinfo{author}{Amit, D.~J.}
\newblock \bibinfo{journal}{\bibinfo{title}{{Metastable states in asymmetrically diluted Hopfield networks}}}.
\newblock {\emph{\JournalTitle{Journal of Physics A: Mathematical and General}}} \textbf{\bibinfo{volume}{21}}, \bibinfo{pages}{3155}, \url{10.1088/0305-4470/21/14/016} (\bibinfo{year}{1988}).

\bibitem{Kwang2023}
\bibinfo{author}{Ryom, K.~I.} \& \bibinfo{author}{Treves, A.}
\newblock \bibinfo{journal}{\bibinfo{title}{{Speed Inversion in a Potts Glass Model of Cortical Dynamics}}}.
\newblock {\emph{\JournalTitle{PRX Life}}} \textbf{\bibinfo{volume}{1}}, \bibinfo{pages}{013005}, \url{10.1103/PRXLife.1.013005} (\bibinfo{year}{2023}).

\bibitem{Friston2010}
\bibinfo{author}{Friston, K.}
\newblock \bibinfo{journal}{\bibinfo{title}{{The free-energy principle: a unified brain theory?}}}
\newblock {\emph{\JournalTitle{Nature Reviews Neuroscience 2010 11:2}}} \textbf{\bibinfo{volume}{11}}, \bibinfo{pages}{127--138}, \url{10.1038/nrn2787} (\bibinfo{year}{2010}).

\bibitem{Friston2019}
\bibinfo{author}{Friston, K.}
\newblock \bibinfo{journal}{\bibinfo{title}{{A free energy principle for a particular physics}}}.
\newblock {\emph{\JournalTitle{arXiv}}}  (\bibinfo{year}{2019}).

\bibitem{Qiu2014}
\bibinfo{author}{Qiu, S.} \& \bibinfo{author}{Chow, C.}
\newblock \bibinfo{journal}{\bibinfo{title}{{Field theory for biophysical neural networks}}}.
\newblock {\emph{\JournalTitle{Proceedings of Science}}} \textbf{\bibinfo{volume}{Part F130500}}, \bibinfo{pages}{23--28} (\bibinfo{year}{2014}).

\bibitem{FeynmanQM}
\bibinfo{author}{Brown, L.~M.}
\newblock \emph{\bibinfo{title}{{Feynman’s thesis: A new approach to quantum theory}}} (\bibinfo{publisher}{World Scientific Publishing Co.}, \bibinfo{year}{2005}).

\bibitem{Huang2003}
\bibinfo{author}{Huang, K.}
\newblock \emph{\bibinfo{title}{{Statistical mechanics}}} (\bibinfo{publisher}{Wiley}, \bibinfo{year}{2003}).

\bibitem{Charbonneau2023}
\bibinfo{author}{Charbonneau, P.} \emph{et~al.}
\newblock \bibinfo{journal}{\bibinfo{title}{{Spin Glass Theory and Far Beyond: Replica Symmetry Breaking After 40 Years}}}.
\newblock {\emph{\JournalTitle{Spin Glass Theory and Far Beyond: Replica Symmetry Breaking after 40 Years}}} \bibinfo{pages}{1--740}, \url{10.1142/13341/SUPPL{\_}FILE/13341{\_}PREFACE.PDF} (\bibinfo{year}{2023}).

\bibitem{Goodfellow2016}
\bibinfo{author}{Goodfellow, I.}, \bibinfo{author}{Bengio, Y.~o.} \& \bibinfo{author}{Courville, A.~.}
\newblock \emph{\bibinfo{title}{{Deep Learning}}} (\bibinfo{publisher}{MIT Press}, \bibinfo{year}{2016}).

\bibitem{Leber2017LongElectrodes}
\bibinfo{author}{Leber, M.} \emph{et~al.}
\newblock \bibinfo{journal}{\bibinfo{title}{{Long term performance of porous platinum coated neural electrodes}}}.
\newblock {\emph{\JournalTitle{Biomedical Microdevices}}} \textbf{\bibinfo{volume}{19}}, \url{10.1007/s10544-017-0201-4} (\bibinfo{year}{2017}).

\bibitem{Mountcastle1997}
\bibinfo{author}{Mountcastle, V.~B.}
\newblock \bibinfo{journal}{\bibinfo{title}{{The columnar organization of the neocortex.}}}
\newblock {\emph{\JournalTitle{Brain}}} \textbf{\bibinfo{volume}{120}}, \bibinfo{pages}{701--722}, \url{10.1093/BRAIN/120.4.701} (\bibinfo{year}{1997}).

\bibitem{Jones2000}
\bibinfo{author}{Jones, E.~G.}
\newblock \bibinfo{journal}{\bibinfo{title}{{Microcolumns in the cerebral cortex}}}.
\newblock {\emph{\JournalTitle{Proceedings of the National Academy of Sciences of the United States of America}}} \textbf{\bibinfo{volume}{97}}, \bibinfo{pages}{5019--5021}, \url{10.1073/PNAS.97.10.5019/ASSET/3DA8C745-11C8-48BA-B49F-00678B6F22A0/ASSETS/GRAPHIC/PQ1001216003.JPEG} (\bibinfo{year}{2000}).

\bibitem{Buxhoeveden2002}
\bibinfo{author}{Buxhoeveden, D.~P.} \& \bibinfo{author}{Casanova, M.~F.}
\newblock \bibinfo{journal}{\bibinfo{title}{{The minicolumn hypothesis in neuroscience}}}.
\newblock {\emph{\JournalTitle{Brain}}} \textbf{\bibinfo{volume}{125}}, \bibinfo{pages}{935--951}, \url{10.1093/BRAIN/AWF110} (\bibinfo{year}{2002}).

\bibitem{Hatsopoulos2010}
\bibinfo{author}{Hatsopoulos, N.~G.}
\newblock \bibinfo{journal}{\bibinfo{title}{{Columnar organization in the motor cortex}}}.
\newblock {\emph{\JournalTitle{Cortex}}} \textbf{\bibinfo{volume}{46}}, \bibinfo{pages}{270--271}, \url{10.1016/J.CORTEX.2008.07.005} (\bibinfo{year}{2010}).

\bibitem{Georgopoulos2010}
\bibinfo{author}{Georgopoulos, A.~P.}, \bibinfo{author}{Merchant, H.}, \bibinfo{author}{Naselaris, T.} \& \bibinfo{author}{Amirikian, B.}
\newblock \bibinfo{journal}{\bibinfo{title}{{Mapping of the preferred direction in the motor cortex}}}.
\newblock {\emph{\JournalTitle{Proceedings of the National Academy of Sciences of the United States of America}}} \textbf{\bibinfo{volume}{104}}, \bibinfo{pages}{11068--11072}, \url{10.1073/PNAS.0611597104/ASSET/5FB2C5E4-5B15-498F-BCE5-65FC871F850A/ASSETS/GRAPHIC/ZPQ02607-6806-M03.JPEG} (\bibinfo{year}{2007}).

\bibitem{Opris2011}
\bibinfo{author}{Opris, I.}, \bibinfo{author}{Hampson, R.~E.}, \bibinfo{author}{Stanford, T.~R.}, \bibinfo{author}{Gerhardt, G.~A.} \& \bibinfo{author}{Deadwyler, S.~A.}
\newblock \bibinfo{journal}{\bibinfo{title}{{Neural Activity in Frontal Cortical Cell Layers: Evidence for Columnar Sensorimotor Processing}}}.
\newblock {\emph{\JournalTitle{Journal of cognitive neuroscience}}} \textbf{\bibinfo{volume}{23}}, \bibinfo{pages}{1507}, \url{10.1162/JOCN.2010.21534} (\bibinfo{year}{2011}).

\bibitem{Hill2014}
\bibinfo{author}{Hill, S.}
\newblock \emph{\bibinfo{title}{{Cortical Columns, Models of}}} (\bibinfo{publisher}{Springer, New York, NY}, \bibinfo{year}{2014}).

\bibitem{Potjans2014}
\bibinfo{author}{Potjans, T.~C.} \& \bibinfo{author}{Diesmann, M.}
\newblock \bibinfo{journal}{\bibinfo{title}{{The cell-type specific cortical microcircuit: Relating structure and activity in a full-scale spiking network model}}}.
\newblock {\emph{\JournalTitle{Cerebral Cortex}}} \textbf{\bibinfo{volume}{24}}, \url{10.1093/cercor/bhs358} (\bibinfo{year}{2014}).

\bibitem{Markowitz2015}
\bibinfo{author}{Markowitz, D.~A.}, \bibinfo{author}{Curtis, C.~E.} \& \bibinfo{author}{Pesaran, B.}
\newblock \bibinfo{journal}{\bibinfo{title}{{Multiple component networks support working memory in prefrontal cortex}}}.
\newblock {\emph{\JournalTitle{Proceedings of the National Academy of Sciences of the United States of America}}} \textbf{\bibinfo{volume}{112}}, \bibinfo{pages}{11084--11089}, \url{10.1073/PNAS.1504172112/SUPPL{\_}FILE/PNAS.1504172112.SAPP.PDF} (\bibinfo{year}{2015}).

\bibitem{Cain2016}
\bibinfo{author}{Cain, N.}, \bibinfo{author}{Iyer, R.}, \bibinfo{author}{Koch, C.} \& \bibinfo{author}{Mihalas, S.}
\newblock \bibinfo{journal}{\bibinfo{title}{{The Computational Properties of a Simplified Cortical Column Model}}}.
\newblock {\emph{\JournalTitle{PLoS Computational Biology}}} \textbf{\bibinfo{volume}{12}}, \url{10.1371/journal.pcbi.1005045} (\bibinfo{year}{2016}).

\bibitem{Hawkins2017}
\bibinfo{author}{Hawkins, J.}, \bibinfo{author}{Ahmad, S.} \& \bibinfo{author}{Cui, Y.}
\newblock \bibinfo{journal}{\bibinfo{title}{{A theory of how columns in the neocortex enable learning the structure of the world}}}.
\newblock {\emph{\JournalTitle{Frontiers in Neural Circuits}}} \textbf{\bibinfo{volume}{11}}, \bibinfo{pages}{295079}, \url{10.3389/FNCIR.2017.00081/BIBTEX} (\bibinfo{year}{2017}).

\bibitem{Chandrasekaran2017}
\bibinfo{author}{Chandrasekaran, C.}, \bibinfo{author}{Peixoto, D.}, \bibinfo{author}{Newsome, W.~T.} \& \bibinfo{author}{Shenoy, K.~V.}
\newblock \bibinfo{journal}{\bibinfo{title}{{Laminar differences in decision-related neural activity in dorsal premotor cortex}}}.
\newblock {\emph{\JournalTitle{Nature Communications 2017 8:1}}} \textbf{\bibinfo{volume}{8}}, \bibinfo{pages}{1--16}, \url{10.1038/s41467-017-00715-0} (\bibinfo{year}{2017}).

\bibitem{Paulk2022}
\bibinfo{author}{Paulk, A.~C.} \emph{et~al.}
\newblock \bibinfo{journal}{\bibinfo{title}{{Large-scale neural recordings with single neuron resolution using Neuropixels probes in human cortex}}}.
\newblock {\emph{\JournalTitle{Nature Neuroscience 2022 25:2}}} \textbf{\bibinfo{volume}{25}}, \bibinfo{pages}{252--263}, \url{10.1038/s41593-021-00997-0} (\bibinfo{year}{2022}).

\bibitem{Wilson1983}
\bibinfo{author}{Wilson, K.~G.}
\newblock \bibinfo{journal}{\bibinfo{title}{{The renormalization group and critical phenomena}}}.
\newblock {\emph{\JournalTitle{Reviews of Modern Physics}}} \textbf{\bibinfo{volume}{55}}, \bibinfo{pages}{583}, \url{10.1103/RevModPhys.55.583} (\bibinfo{year}{1983}).

\bibitem{Kadanoff2011}
\bibinfo{author}{Kadanoff, L.~P.}
\newblock \bibinfo{journal}{\bibinfo{title}{{Relating theories via renormalization}}}.
\newblock {\emph{\JournalTitle{Studies in History and Philosophy of Science Part B: Studies in History and Philosophy of Modern Physics}}} \textbf{\bibinfo{volume}{44}}, \bibinfo{pages}{22--39}, \url{10.1016/J.SHPSB.2012.05.002} (\bibinfo{year}{2013}).

\bibitem{Efrati2014Real-spaceMechanics}
\bibinfo{author}{Efrati, E.}, \bibinfo{author}{Wang, Z.}, \bibinfo{author}{Kolan, A.} \& \bibinfo{author}{Kadanoff, L.~P.}
\newblock \bibinfo{journal}{\bibinfo{title}{{Real-space renormalization in statistical mechanics}}}.
\newblock {\emph{\JournalTitle{Reviews of Modern Physics}}} \textbf{\bibinfo{volume}{86}}, \bibinfo{pages}{647--667}, \url{10.1103/RevModPhys.86.647} (\bibinfo{year}{2014}).

\bibitem{Niemeijer1973}
\bibinfo{author}{Niemeijer, T.} \& \bibinfo{author}{Van~Leeuwen, J.~M.}
\newblock \bibinfo{journal}{\bibinfo{title}{{Wilson Theory for Spin Systems on a Triangular Lattice}}}.
\newblock {\emph{\JournalTitle{Physical Review Letters}}} \textbf{\bibinfo{volume}{31}}, \bibinfo{pages}{1411}, \url{10.1103/PhysRevLett.31.1411} (\bibinfo{year}{1973}).

\bibitem{Niemeyer1974}
\bibinfo{author}{Niemeyer, T.} \& \bibinfo{author}{Van~Leeuwen, J.~M.}
\newblock \bibinfo{journal}{\bibinfo{title}{{Wilson theory for 2-dimensional Ising spin systems}}}.
\newblock {\emph{\JournalTitle{Physica}}} \textbf{\bibinfo{volume}{71}}, \bibinfo{pages}{17--40}, \url{10.1016/0031-8914(74)90044-5} (\bibinfo{year}{1974}).

\bibitem{ParisiRG2001}
\bibinfo{author}{Paris, G.}, \bibinfo{author}{Petronzio, R.} \& \bibinfo{author}{Rosati, F.}
\newblock \bibinfo{journal}{\bibinfo{title}{{Renormalization group approach to spin glass systems}}}.
\newblock {\emph{\JournalTitle{European Physical Journal B}}} \textbf{\bibinfo{volume}{21}}, \bibinfo{pages}{605--609}, \url{10.1007/S100510170171/METRICS} (\bibinfo{year}{2001}).

\bibitem{AngeliniRG2023}
\bibinfo{author}{Angelini, M.~C.}
\newblock \bibinfo{journal}{\bibinfo{title}{{Real-Space Renormalization group for spin glasses}}}.
\newblock {\emph{\JournalTitle{arXiv}}}  (\bibinfo{year}{2023}).

\bibitem{Wilczek2012}
\bibinfo{author}{Wilczek, F.}
\newblock \bibinfo{journal}{\bibinfo{title}{{Quantum Time Crystals}}}.
\newblock {\emph{\JournalTitle{Physical Review Letters}}} \textbf{\bibinfo{volume}{109}}, \bibinfo{pages}{160401}, \url{10.1103/PhysRevLett.109.160401} (\bibinfo{year}{2012}).

\bibitem{Zhang2017}
\bibinfo{author}{Zhang, J.} \emph{et~al.}
\newblock \bibinfo{journal}{\bibinfo{title}{{Observation of a discrete time crystal}}}.
\newblock {\emph{\JournalTitle{Nature 2017 543:7644}}} \textbf{\bibinfo{volume}{543}}, \bibinfo{pages}{217--220}, \url{10.1038/nature21413} (\bibinfo{year}{2017}).

\bibitem{Martinelli2023}
\bibinfo{title}{{2023{\_}03{\_}30{\_}Martinelli.mp4 - Google Drive}}.

\bibitem{Grinstein1985}
\bibinfo{author}{Grinstein, G.}, \bibinfo{author}{Jayaprakash, C.} \& \bibinfo{author}{He, Y.}
\newblock \bibinfo{title}{{Statistical Mechanics of Probabilistic Cellular Automata}}.
\newblock \bibinfo{type}{Tech. Rep.} \bibinfo{number}{23} (\bibinfo{year}{1985}).

\bibitem{Elze2014}
\bibinfo{author}{Elze, H.~T.}
\newblock \bibinfo{journal}{\bibinfo{title}{{Action principle for cellular automata and the linearity of quantum mechanics}}}.
\newblock {\emph{\JournalTitle{Physical Review A}}} \textbf{\bibinfo{volume}{89}}, \bibinfo{pages}{012111}, \url{10.1103/PhysRevA.89.012111} (\bibinfo{year}{2014}).

\bibitem{Hooft2014}
\bibinfo{author}{Hooft, G.~t.}
\newblock \emph{\bibinfo{title}{{The Cellular Automaton Interpretation of Quantum Mechanics}}} (\bibinfo{year}{2014}).

\bibitem{Fredkin1982}
\bibinfo{author}{Fredkin, E.} \& \bibinfo{author}{Toffoli, T.}
\newblock \emph{\bibinfo{title}{{Conservative logic}}}, vol.~\bibinfo{volume}{21} (\bibinfo{publisher}{Kluwer Academic Publishers-Plenum Publishers}, \bibinfo{year}{1982}).

\bibitem{Capobianco2011}
\bibinfo{author}{Capobianco, S.} \& \bibinfo{author}{Toffoli, T.}
\newblock \bibinfo{journal}{\bibinfo{title}{{Can anything from Noether's Theorem be salvaged for discrete dynamical systems?}}}
\newblock {\emph{\JournalTitle{Lecture Notes in Computer Science (including subseries Lecture Notes in Artificial Intelligence and Lecture Notes in Bioinformatics)}}} \textbf{\bibinfo{volume}{6714 LNCS}}, \bibinfo{pages}{77--88}, \url{10.1007/978-3-642-21341-0{\_}13/COVER} (\bibinfo{year}{2011}).

\bibitem{Cranmer2023}
\bibinfo{author}{Cranmer, K.}, \bibinfo{author}{Kanwar, G.}, \bibinfo{author}{Racani{\`{e}}re, S.}, \bibinfo{author}{Rezende, D.~J.} \& \bibinfo{author}{Shanahan, P.~E.}
\newblock \bibinfo{journal}{\bibinfo{title}{{Advances in machine-learning-based sampling motivated by lattice quantum chromodynamics}}}.
\newblock {\emph{\JournalTitle{Nature Reviews Physics 2023}}} \bibinfo{pages}{1--10}, \url{10.1038/s42254-023-00616-w} (\bibinfo{year}{2023}).

\bibitem{Kogut1975}
\bibinfo{author}{Kogut, J.} \& \bibinfo{author}{Susskind{\~{}}, L.}
\newblock \bibinfo{title}{{Hamiltonian formulation of Wilson's lattice gauge theories}}.
\newblock \bibinfo{type}{Tech. Rep.} (\bibinfo{year}{1975}).

\bibitem{Rapan2021}
\bibinfo{author}{Rapan, L.} \emph{et~al.}
\newblock \bibinfo{journal}{\bibinfo{title}{{Multimodal 3D atlas of the macaque monkey motor and premotor cortex}}}.
\newblock {\emph{\JournalTitle{NeuroImage}}} \textbf{\bibinfo{volume}{226}}, \url{10.1016/j.neuroimage.2020.117574} (\bibinfo{year}{2021}).

\bibitem{Bardella2020}
\bibinfo{author}{Bardella, G.}, \bibinfo{author}{Pani, P.}, \bibinfo{author}{Brunamonti, E.}, \bibinfo{author}{Giarrocco, F.} \& \bibinfo{author}{Ferraina, S.}
\newblock \bibinfo{journal}{\bibinfo{title}{{The small scale functional topology of movement control: Hierarchical organization of local activity anticipates movement generation in the premotor cortex of primates}}}.
\newblock {\emph{\JournalTitle{NeuroImage}}} \textbf{\bibinfo{volume}{207}}, \url{10.1016/j.neuroimage.2019.116354} (\bibinfo{year}{2020}).

\bibitem{NelsonMechanics}
\bibinfo{author}{Nelson, E.}
\newblock \bibinfo{journal}{\bibinfo{title}{{Review of stochastic mechanics}}}.
\newblock {\emph{\JournalTitle{Journal of Physics: Conference Series}}} \textbf{\bibinfo{volume}{361}}, \bibinfo{pages}{012011}, \url{10.1088/1742-6596/361/1/012011} (\bibinfo{year}{2012}).

\bibitem{Parisi1981}
\bibinfo{author}{Parisi, G.} \& \bibinfo{author}{WU, Y.}
\newblock \bibinfo{journal}{\bibinfo{title}{{Pertubation theory without gauge fixing}}}.
\newblock {\emph{\JournalTitle{Scientia Sinica}}} \textbf{\bibinfo{volume}{24}}, \bibinfo{pages}{483--}, \url{10.1360/YA1981-24-4-483} (\bibinfo{year}{1981}).

\bibitem{LSZ1955}
\bibinfo{author}{Lehmann, H.}, \bibinfo{author}{Symanzik, K.} \& \bibinfo{author}{Zimmermann, W.}
\newblock \bibinfo{journal}{\bibinfo{title}{{Zur Formulierung quantisierter Feldtheorien}}}.
\newblock {\emph{\JournalTitle{Il Nuovo Cimento}}} \textbf{\bibinfo{volume}{1}}, \bibinfo{pages}{205--225}, \url{10.1007/BF02731765/METRICS} (\bibinfo{year}{1955}).

\bibitem{Steinbrecher2008}
\bibinfo{author}{Steinbrecher, G.} \& \bibinfo{author}{Shaw, W.~T.}
\newblock \bibinfo{journal}{\bibinfo{title}{{Quantile mechanics}}}.
\newblock {\emph{\JournalTitle{European Journal of Applied Mathematics}}} \textbf{\bibinfo{volume}{19}}, \bibinfo{pages}{87--112}, \url{10.1017/S0956792508007341} (\bibinfo{year}{2008}).

\bibitem{Yamamoto2004}
\bibinfo{author}{Yamamoto, Y.}
\newblock \emph{\bibinfo{title}{{Fundamentals of Noise Processes}}} (\bibinfo{publisher}{Cambridge University Press}, \bibinfo{year}{2004}).

\bibitem{Rovelli1995}
\bibinfo{author}{Rovelli, C.} \& \bibinfo{author}{Smolin, L.}
\newblock \bibinfo{journal}{\bibinfo{title}{{Discreteness of area and volume in quantum gravity}}}.
\newblock {\emph{\JournalTitle{Nuclear Physics B}}} \textbf{\bibinfo{volume}{442}}, \bibinfo{pages}{593--619}, \url{10.1016/0550-3213(95)00150-Q} (\bibinfo{year}{1995}).

\bibitem{Guerra1975}
\bibinfo{author}{Guerra, F.}, \bibinfo{author}{Rosen, L.} \& \bibinfo{author}{Simon, B.}
\newblock \bibinfo{journal}{\bibinfo{title}{{The P({$\phi$}) 2 Euclidean Quantum Field Theory as Classical Statistical Mechanics}}}.
\newblock {\emph{\JournalTitle{The Annals of Mathematics}}} \textbf{\bibinfo{volume}{101}}, \bibinfo{pages}{111}, \url{10.2307/1970988} (\bibinfo{year}{1975}).

\bibitem{Gibbs1902}
\bibinfo{author}{Gibbs, J.~W.}
\newblock \bibinfo{journal}{\bibinfo{title}{{Elementary principles in statistical mechanics: Developed with especial reference to the rational foundation of thermodynamics}}}.
\newblock {\emph{\JournalTitle{Elementary Principles in Statistical Mechanics: Developed with Especial Reference to the Rational Foundation of Thermodynamics}}} \bibinfo{pages}{1--207}, \url{10.1017/CBO9780511686948} (\bibinfo{year}{2010}).

\bibitem{Finkelstein2003}
\bibinfo{author}{Finkelstein, D.~R.}
\newblock \bibinfo{journal}{\bibinfo{title}{{Ur Theory and Space-Time Structure}}}.
\newblock {\emph{\JournalTitle{Time, Quantum and Information}}} \bibinfo{pages}{397--407}, \url{10.1007/978-3-662-10557-3{\_}27} (\bibinfo{year}{2003}).

\bibitem{IsingLimit}
\bibinfo{author}{Caginalp, G.}
\newblock \bibinfo{journal}{\bibinfo{title}{{Thermodynamic properties of the phi/sup 4/ lattice field theory near the Ising limit}}}.
\newblock {\emph{\JournalTitle{Ann. Phys. (N.Y.); (United States)}}} \textbf{\bibinfo{volume}{126:2}}, \bibinfo{pages}{500--511}, \url{10.1016/0003-4916(80)90185-2} (\bibinfo{year}{1980}).

\bibitem{Lagrangemechanics}
\bibinfo{author}{Goldstein, H.}, \bibinfo{author}{Poole, C.~P.} \& \bibinfo{author}{Safko, J.~L.}
\newblock \emph{\bibinfo{title}{{Classical mechanics}}} (\bibinfo{publisher}{Pearson Education}, \bibinfo{year}{2002}).

\bibitem{Helmolzcond1}
\bibinfo{author}{Nigam, K.} \& \bibinfo{author}{Banerjee, K.}
\newblock \bibinfo{journal}{\bibinfo{title}{{A Brief Review of Helmholtz Conditions}}}.
\newblock {\emph{\JournalTitle{arXiv}}}  (\bibinfo{year}{2016}).

\bibitem{Sarlet1982}
\bibinfo{author}{Sarlet, W.}
\newblock \bibinfo{journal}{\bibinfo{title}{{The Helmholtz conditions revisited. A new approach to the inverse problem of Lagrangian dynamics}}}.
\newblock {\emph{\JournalTitle{Journal of Physics A: Mathematical and General}}} \textbf{\bibinfo{volume}{15}}, \bibinfo{pages}{1503}, \url{10.1088/0305-4470/15/5/013} (\bibinfo{year}{1982}).

\bibitem{Douglas1939}
\bibinfo{author}{Douglas, J.}
\newblock \bibinfo{journal}{\bibinfo{title}{{Solution of the Inverse Problem of the Calculus of Variations}}}.
\newblock {\emph{\JournalTitle{Proceedings of the National Academy of Sciences of the United States of America}}} \textbf{\bibinfo{volume}{25}}, \bibinfo{pages}{631--637}, \url{10.1073/PNAS.25.12.631} (\bibinfo{year}{1939}).

\bibitem{Craciun1996}
\bibinfo{author}{Craciun, D.} \& \bibinfo{author}{Opris, D.}
\newblock \bibinfo{journal}{\bibinfo{title}{{The Helmholtz conditions for the difference equations systems.}}}
\newblock {\emph{\JournalTitle{Balkan Journal of Geometry and its Applications (BJGA)}}} \textbf{\bibinfo{volume}{1}}, \bibinfo{pages}{21--30} (\bibinfo{year}{1996}).

\bibitem{Bourdin2013}
\bibinfo{author}{Bourdin, L.} \& \bibinfo{author}{Cresson, J.}
\newblock \bibinfo{journal}{\bibinfo{title}{{Helmholtz's inverse problem of the discrete calculus of variations}}}.
\newblock {\emph{\JournalTitle{Journal of Difference Equations and Applications}}} \textbf{\bibinfo{volume}{19}}, \bibinfo{pages}{1417--1436}, \url{10.1080/10236198.2012.754435} (\bibinfo{year}{2013}).

\bibitem{Gubbiotti1}
\bibinfo{author}{Gubbiotti, G.}
\newblock \bibinfo{journal}{\bibinfo{title}{{On the inverse problem of the discrete calculus of variations}}}.
\newblock {\emph{\JournalTitle{Journal of Physics A: Mathematical and Theoretical}}} \textbf{\bibinfo{volume}{52}}, \bibinfo{pages}{305203}, \url{10.1088/1751-8121/AB2919} (\bibinfo{year}{2019}).

\bibitem{Gubbiotti2}
\bibinfo{author}{Gubbiotti, G.}
\newblock \bibinfo{journal}{\bibinfo{title}{{Lagrangians and integrability for additive fourth-order difference equations}}}.
\newblock {\emph{\JournalTitle{The European Physical Journal Plus 2020 135:10}}} \textbf{\bibinfo{volume}{135}}, \bibinfo{pages}{1--30}, \url{10.1140/EPJP/S13360-020-00858-Y} (\bibinfo{year}{2020}).

\bibitem{Kistler2021}
\bibinfo{author}{Kistler, N.}
\newblock \bibinfo{journal}{\bibinfo{title}{{Solving spin systems: the Babylonian way}}}.
\newblock {\emph{\JournalTitle{arXiv}}}  (\bibinfo{year}{2021}).

\bibitem{WickRotation}
\bibinfo{author}{Wick, G.~C.}
\newblock \bibinfo{journal}{\bibinfo{title}{{Properties of Bethe-Salpeter Wave Functions}}}.
\newblock {\emph{\JournalTitle{Physical Review}}} \textbf{\bibinfo{volume}{96}}, \bibinfo{pages}{1124}, \url{10.1103/PhysRev.96.1124} (\bibinfo{year}{1954}).

\bibitem{Wickrotation1975}
\bibinfo{author}{O'Brien, D.}
\newblock \bibinfo{journal}{\bibinfo{title}{{The Wick rotation}}}.
\newblock {\emph{\JournalTitle{Australian Journal of Physics}}} \textbf{\bibinfo{volume}{28}}, \bibinfo{pages}{7--13} (\bibinfo{year}{1975}).

\bibitem{DiCastro1969}
\bibinfo{author}{Di~Castro, C.} \& \bibinfo{author}{Jona-Lasinio, G.}
\newblock \bibinfo{journal}{\bibinfo{title}{{On the microscopic foundation of scaling laws}}}.
\newblock {\emph{\JournalTitle{Physics Letters A}}} \textbf{\bibinfo{volume}{29}}, \bibinfo{pages}{322--323}, \url{10.1016/0375-9601(69)90148-0} (\bibinfo{year}{1969}).

\bibitem{Nguyen2017}
\bibinfo{author}{Nguyen, H.~C.}, \bibinfo{author}{Zecchina, R.} \& \bibinfo{author}{Berg, J.}
\newblock \bibinfo{journal}{\bibinfo{title}{{Inverse statistical problems: from the inverse Ising problem to data science}}}.
\newblock {\emph{\JournalTitle{http://dx.doi.org/10.1080/00018732.2017.1341604}}} \textbf{\bibinfo{volume}{66}}, \bibinfo{pages}{197--261}, \url{10.1080/00018732.2017.1341604} (\bibinfo{year}{2017}).

\bibitem{Carleo2019}
\bibinfo{author}{Carleo, G.} \emph{et~al.}
\newblock \bibinfo{journal}{\bibinfo{title}{{Machine learning and the physical sciences}}}.
\newblock {\emph{\JournalTitle{Reviews of Modern Physics}}} \textbf{\bibinfo{volume}{91}}, \bibinfo{pages}{045002}, \url{10.1103/REVMODPHYS.91.045002/FIGURES/8/MEDIUM} (\bibinfo{year}{2019}).

\bibitem{Zdeborova2016}
\bibinfo{author}{Zdeborov{\'{a}}, L.} \& \bibinfo{author}{Krzakala, F.}
\newblock \bibinfo{journal}{\bibinfo{title}{{Statistical physics of inference: thresholds and algorithms}}}.
\newblock {\emph{\JournalTitle{Advances in Physics}}} \textbf{\bibinfo{volume}{65}}, \bibinfo{pages}{453--552}, \url{10.1080/00018732.2016.1211393} (\bibinfo{year}{2016}).

\bibitem{Albert2014}
\bibinfo{author}{Albert, J.} \& \bibinfo{author}{Swendsen, R.~H.}
\newblock \bibinfo{journal}{\bibinfo{title}{{The Inverse Ising Problem}}}.
\newblock {\emph{\JournalTitle{Physics Procedia}}} \textbf{\bibinfo{volume}{57}}, \bibinfo{pages}{99--103}, \url{10.1016/J.PHPRO.2014.08.140} (\bibinfo{year}{2014}).

\bibitem{Swendsen1984}
\bibinfo{author}{Swendsen, R.~H.}
\newblock \bibinfo{journal}{\bibinfo{title}{{Monte Carlo Calculation of Renormalized Coupling Parameters}}}.
\newblock {\emph{\JournalTitle{Physical Review Letters}}} \textbf{\bibinfo{volume}{52}}, \bibinfo{pages}{1165}, \url{10.1103/PhysRevLett.52.1165} (\bibinfo{year}{1984}).

\bibitem{Aurell2012}
\bibinfo{author}{Aurell, E.} \& \bibinfo{author}{Ekeberg, M.}
\newblock \bibinfo{journal}{\bibinfo{title}{{Inverse ising inference using all the data}}}.
\newblock {\emph{\JournalTitle{Physical Review Letters}}} \textbf{\bibinfo{volume}{108}}, \bibinfo{pages}{090201}, \url{10.1103/PHYSREVLETT.108.090201/FIGURES/2/MEDIUM} (\bibinfo{year}{2012}).

\bibitem{SessakMonasson}
\bibinfo{author}{Sessak, V.} \& \bibinfo{author}{Monasson, R.}
\newblock \bibinfo{journal}{\bibinfo{title}{{Small-correlation expansions for the inverse Ising problem}}}.
\newblock {\emph{\JournalTitle{Journal of Physics A: Mathematical and Theoretical}}} \textbf{\bibinfo{volume}{42}}, \bibinfo{pages}{055001}, \url{10.1088/1751-8113/42/5/055001} (\bibinfo{year}{2009}).

\bibitem{Arous2009}
\bibinfo{author}{Arous, G.~B.} \& \bibinfo{author}{Kuptsov, A.}
\newblock \bibinfo{journal}{\bibinfo{title}{{REM Universality for Random Hamiltonians}}}.
\newblock {\emph{\JournalTitle{Progress in Probability}}} \textbf{\bibinfo{volume}{62}}, \bibinfo{pages}{45--84}, \url{10.1007/978-3-7643-9891-0{\_}2/COVER} (\bibinfo{year}{2009}).

\bibitem{Lovasz2012}
\bibinfo{author}{Lov{\'{a}}sz, L.}
\newblock \emph{\bibinfo{title}{{Large Networks and Graph Limits}}}, vol.~\bibinfo{volume}{60} of \emph{\bibinfo{series}{Colloquium Publications}} (\bibinfo{publisher}{American Mathematical Society}, \bibinfo{address}{Providence, Rhode Island}, \bibinfo{year}{2012}).

\bibitem{Breitenberg}
\bibinfo{author}{van Hemmen, J.~L.}, \bibinfo{author}{Sch{\"{u}}z, A.} \& \bibinfo{author}{Aertsen, A.}
\newblock \bibinfo{journal}{\bibinfo{title}{{Structural aspects of biological cybernetics: Valentino Braitenberg, neuroanatomy, and brain function}}}.
\newblock {\emph{\JournalTitle{Biological cybernetics}}} \textbf{\bibinfo{volume}{108}}, \bibinfo{pages}{517--525}, \url{10.1007/S00422-014-0630-6} (\bibinfo{year}{2014}).

\bibitem{Buzsaki2012}
\bibinfo{author}{Buzs{\'{a}}ki, G.}, \bibinfo{author}{Anastassiou, C.~A.} \& \bibinfo{author}{Koch, C.}
\newblock \bibinfo{journal}{\bibinfo{title}{{The origin of extracellular fields and currents--EEG, ECoG, LFP and spikes}}}.
\newblock {\emph{\JournalTitle{Nature reviews. Neuroscience}}} \textbf{\bibinfo{volume}{13}}, \bibinfo{pages}{407--420}, \url{10.1038/NRN3241} (\bibinfo{year}{2012}).

\bibitem{Segev2006}
\bibinfo{author}{Segev, R.}, \bibinfo{author}{Puchalla, J.} \& \bibinfo{author}{Berry, M.~J.}
\newblock \bibinfo{journal}{\bibinfo{title}{{Functional organization of ganglion cells in the salamander retina}}}.
\newblock {\emph{\JournalTitle{Journal of Neurophysiology}}} \textbf{\bibinfo{volume}{95}}, \bibinfo{pages}{2277--2292}, \url{10.1152/jn.00928.2005} (\bibinfo{year}{2006}).

\bibitem{Segev1998}
\bibinfo{author}{Segev, C.}
\newblock \emph{\bibinfo{title}{{CHAPTER 1 KINETIC MODELS OF SYNAPTIC TRANSMISSION}}} (\bibinfo{publisher}{MIT Press}, \bibinfo{year}{1998}).

\bibitem{Lubke2007}
\bibinfo{author}{L{\"{u}}bke, J.} \& \bibinfo{author}{Feldmeyer, D.}
\newblock \bibinfo{journal}{\bibinfo{title}{{Excitatory signal flow and connectivity in a cortical column: focus on barrel cortex}}}.
\newblock {\emph{\JournalTitle{Brain structure {\&} function}}} \textbf{\bibinfo{volume}{212}}, \bibinfo{pages}{3--17}, \url{10.1007/S00429-007-0144-2} (\bibinfo{year}{2007}).

\bibitem{Pachitariu2016}
\bibinfo{author}{Pachitariu, M.}, \bibinfo{author}{Steinmetz, N.~A.}, \bibinfo{author}{Kadir, S.~N.}, \bibinfo{author}{Carandini, M.} \& \bibinfo{author}{Harris, K.~D.}
\newblock \bibinfo{journal}{\bibinfo{title}{{Fast and accurate spike sorting of high-channel count probes with KiloSort}}}.
\newblock {\emph{\JournalTitle{Advances in Neural Information Processing Systems}}} \textbf{\bibinfo{volume}{29}} (\bibinfo{year}{2016}).

\bibitem{Churchland2012}
\bibinfo{author}{Churchland, M.~M.} \emph{et~al.}
\newblock \bibinfo{journal}{\bibinfo{title}{{Neural population dynamics during reaching}}}.
\newblock {\emph{\JournalTitle{Nature}}} \textbf{\bibinfo{volume}{487}}, \bibinfo{pages}{51--56}, \url{10.1038/nature11129} (\bibinfo{year}{2012}).

\bibitem{Mattia2013}
\bibinfo{author}{Mattia, M.} \emph{et~al.}
\newblock \bibinfo{journal}{\bibinfo{title}{{Heterogeneous attractor cell assemblies for motor planning in premotor cortex}}}.
\newblock {\emph{\JournalTitle{Journal of Neuroscience}}} \textbf{\bibinfo{volume}{33}}, \bibinfo{pages}{11155--11168}, \url{10.1523/JNEUROSCI.4664-12.2013} (\bibinfo{year}{2013}).

\bibitem{Kaufman2016}
\bibinfo{author}{Kaufman, M.~T.} \emph{et~al.}
\newblock \bibinfo{journal}{\bibinfo{title}{{The largest response component in the motor cortex reflects movement timing but not movement type}}}.
\newblock {\emph{\JournalTitle{eNeuro}}} \textbf{\bibinfo{volume}{3}}, \bibinfo{pages}{85--101}, \url{10.1523/ENEURO.0085-16.2016} (\bibinfo{year}{2016}).

\bibitem{Clawson}
\bibinfo{author}{Clawson, W.} \emph{et~al.}
\newblock \bibinfo{journal}{\bibinfo{title}{{Computing hubs in the hippocampus and cortex}}}.
\newblock {\emph{\JournalTitle{Science Advances}}} \textbf{\bibinfo{volume}{5}}, \bibinfo{pages}{eaax4843}, \url{10.1126/sciadv.aax4843} (\bibinfo{year}{2019}).

\bibitem{Weinrich1982}
\bibinfo{author}{Weinrich, M.} \& \bibinfo{author}{Wise, S.~P.}
\newblock \bibinfo{journal}{\bibinfo{title}{{The premotor cortex of the monkey}}}.
\newblock {\emph{\JournalTitle{The Journal of neuroscience : the official journal of the Society for Neuroscience}}} \textbf{\bibinfo{volume}{2}}, \bibinfo{pages}{1329--1345}, \url{10.1523/JNEUROSCI.02-09-01329.1982} (\bibinfo{year}{1982}).

\bibitem{Churchland2010}
\bibinfo{author}{Churchland, M.~M.}, \bibinfo{author}{Cunningham, J.~P.}, \bibinfo{author}{Kaufman, M.~T.}, \bibinfo{author}{Ryu, S.~I.} \& \bibinfo{author}{Shenoy, K.~V.}
\newblock \bibinfo{journal}{\bibinfo{title}{{Cortical preparatory activity: representation of movement or first cog in a dynamical machine?}}}
\newblock {\emph{\JournalTitle{Neuron}}} \textbf{\bibinfo{volume}{68}}, \bibinfo{pages}{387--400}, \url{10.1016/J.NEURON.2010.09.015} (\bibinfo{year}{2010}).

\bibitem{Bardella2024}
\bibinfo{author}{Bardella, G.} \emph{et~al.}
\newblock \bibinfo{journal}{\bibinfo{title}{{Response inhibition in premotor cortex corresponds to a complex reshuffle of the mesoscopic information network}}}.
\newblock {\emph{\JournalTitle{Network Neuroscience}}} \bibinfo{pages}{1--48}, \url{10.1162/NETN{\_}A{\_}00365} (\bibinfo{year}{2024}).

\bibitem{Kaufman2014}
\bibinfo{author}{Kaufman, M.~T.}, \bibinfo{author}{Churchland, M.~M.}, \bibinfo{author}{Ryu, S.~I.} \& \bibinfo{author}{Shenoy, K.~V.}
\newblock \bibinfo{journal}{\bibinfo{title}{{Cortical activity in the null space: permitting preparation without movement}}}.
\newblock {\emph{\JournalTitle{Nature neuroscience}}} \textbf{\bibinfo{volume}{17}}, \bibinfo{pages}{440}, \url{10.1038/NN.3643} (\bibinfo{year}{2014}).

\bibitem{Shenoy2013}
\bibinfo{author}{Shenoy, K.~V.}, \bibinfo{author}{Sahani, M.} \& \bibinfo{author}{Churchland, M.~M.}
\newblock \bibinfo{journal}{\bibinfo{title}{{Cortical Control of Arm Movements: A Dynamical Systems Perspective}}}.
\newblock {\emph{\JournalTitle{Annual Review of Neuroscience}}} \textbf{\bibinfo{volume}{36}}, \bibinfo{pages}{337--359}, \url{10.1146/annurev-neuro-062111-150509} (\bibinfo{year}{2013}).

\bibitem{Churchland2006}
\bibinfo{author}{Churchland, M.~M.}, \bibinfo{author}{Yu, B.~M.}, \bibinfo{author}{Ryu, S.~I.}, \bibinfo{author}{Santhanam, G.} \& \bibinfo{author}{Shenoy, K.~V.}
\newblock \bibinfo{journal}{\bibinfo{title}{{Neural variability in premotor cortex provides a signature of motor preparation}}}.
\newblock {\emph{\JournalTitle{The Journal of neuroscience : the official journal of the Society for Neuroscience}}} \textbf{\bibinfo{volume}{26}}, \bibinfo{pages}{3697--3712}, \url{10.1523/JNEUROSCI.3762-05.2006} (\bibinfo{year}{2006}).

\bibitem{Ames2014}
\bibinfo{author}{Ames, K.~C.}, \bibinfo{author}{Ryu, S.~I.} \& \bibinfo{author}{Shenoy, K.~V.}
\newblock \bibinfo{journal}{\bibinfo{title}{{Neural Dynamics of Reaching Following Incorrect or Absent Motor Preparation}}}.
\newblock {\emph{\JournalTitle{Neuron}}} \textbf{\bibinfo{volume}{81}}, \bibinfo{pages}{438}, \url{10.1016/J.NEURON.2013.11.003} (\bibinfo{year}{2014}).

\bibitem{Elsayed2016}
\bibinfo{author}{Elsayed, G.~F.}, \bibinfo{author}{Lara, A.~H.}, \bibinfo{author}{Kaufman, M.~T.}, \bibinfo{author}{Churchland, M.~M.} \& \bibinfo{author}{Cunningham, J.~P.}
\newblock \bibinfo{journal}{\bibinfo{title}{{Reorganization between preparatory and movement population responses in motor cortex}}}.
\newblock {\emph{\JournalTitle{Nature Communications 2016 7:1}}} \textbf{\bibinfo{volume}{7}}, \bibinfo{pages}{1--15}, \url{10.1038/ncomms13239} (\bibinfo{year}{2016}).

\bibitem{Mirabella2011}
\bibinfo{author}{Mirabella, G.}, \bibinfo{author}{Pani, P.} \& \bibinfo{author}{Ferraina, S.}
\newblock \bibinfo{journal}{\bibinfo{title}{{Neural correlates of cognitive control of reaching movements in the dorsal premotor cortex of rhesus monkeys}}}.
\newblock {\emph{\JournalTitle{Journal of Neurophysiology}}} \textbf{\bibinfo{volume}{106}}, \bibinfo{pages}{1454--1466}, \url{10.1152/jn.00995.2010} (\bibinfo{year}{2011}).

\bibitem{BattagliaMayer2014}
\bibinfo{author}{Battaglia-Mayer, A.} \emph{et~al.}
\newblock \bibinfo{title}{{Correction and suppression of reaching movements in the cerebral cortex: Physiological and neuropsychological aspects}}, \url{10.1016/j.neubiorev.2014.03.002} (\bibinfo{year}{2014}).

\bibitem{Caminiti1991}
\bibinfo{author}{Caminiti, R.}, \bibinfo{author}{Johnson, P.~B.}, \bibinfo{author}{Galli, C.}, \bibinfo{author}{Ferraina, S.} \& \bibinfo{author}{Burnod, Y.}
\newblock \bibinfo{journal}{\bibinfo{title}{{Making arm movements within different parts of space: the premotor and motor cortical representation of a coordinate system for reaching to visual targets}}}.
\newblock {\emph{\JournalTitle{Journal of Neuroscience}}} \textbf{\bibinfo{volume}{11}}, \bibinfo{pages}{1182--1197}, \url{10.1523/JNEUROSCI.11-05-01182.1991} (\bibinfo{year}{1991}).

\bibitem{Caminiti2017}
\bibinfo{author}{Caminiti, R.} \emph{et~al.}
\newblock \bibinfo{journal}{\bibinfo{title}{{Computational architecture of the parieto-frontal network underlying cognitive-motor control in monkeys}}}.
\newblock {\emph{\JournalTitle{eNeuro}}} \textbf{\bibinfo{volume}{4}}, \bibinfo{pages}{306--322}, \url{10.1523/ENEURO.0306-16.2017} (\bibinfo{year}{2017}).

\bibitem{Nambu2002}
\bibinfo{author}{Nambu, A.}, \bibinfo{author}{Tokuno, H.} \& \bibinfo{author}{Takada, M.}
\newblock \bibinfo{journal}{\bibinfo{title}{{Functional significance of the cortico-subthalamo-pallidal 'hyperdirect' pathway}}}.
\newblock {\emph{\JournalTitle{Neuroscience Research}}} \textbf{\bibinfo{volume}{43}}, \bibinfo{pages}{111--117}, \url{10.1016/S0168-0102(02)00027-5} (\bibinfo{year}{2002}).

\bibitem{Middleton2001}
\bibinfo{author}{Middleton, F.~A.} \& \bibinfo{author}{Strick, P.~L.}
\newblock \bibinfo{journal}{\bibinfo{title}{{Cerebellar Projections to the Prefrontal Cortex of the Primate}}}.
\newblock {\emph{\JournalTitle{The Journal of Neuroscience}}} \textbf{\bibinfo{volume}{21}}, \bibinfo{pages}{700}, \url{10.1523/JNEUROSCI.21-02-00700.2001} (\bibinfo{year}{2001}).

\bibitem{Marconi2001}
\bibinfo{author}{Marconi, B.} \emph{et~al.}
\newblock \bibinfo{journal}{\bibinfo{title}{{Eye–Hand Coordination during Reaching. I. Anatomical Relationships between Parietal and Frontal Cortex}}}.
\newblock {\emph{\JournalTitle{Cerebral Cortex}}} \textbf{\bibinfo{volume}{11}}, \bibinfo{pages}{513--527}, \url{10.1093/CERCOR/11.6.513} (\bibinfo{year}{2001}).

\bibitem{Johnson1996}
\bibinfo{author}{Johnson, P.~B.}, \bibinfo{author}{Ferraina, S.}, \bibinfo{author}{Bianchi, L.} \& \bibinfo{author}{Caminiti, R.}
\newblock \bibinfo{journal}{\bibinfo{title}{{Cortical Networks for Visual Reaching: Physiological and Anatomical Organization of Frontal and Parietal Lobe Arm Regions}}}.
\newblock {\emph{\JournalTitle{Cerebral Cortex}}} \textbf{\bibinfo{volume}{6}}, \bibinfo{pages}{102--119}, \url{10.1093/CERCOR/6.2.102} (\bibinfo{year}{1996}).

\bibitem{Gallego2017}
\bibinfo{author}{Gallego, J.~A.}, \bibinfo{author}{Perich, M.~G.}, \bibinfo{author}{Miller, L.~E.} \& \bibinfo{author}{Solla, S.~A.}
\newblock \bibinfo{journal}{\bibinfo{title}{{Neural Manifolds for the Control of Movement}}}.
\newblock {\emph{\JournalTitle{Neuron}}} \textbf{\bibinfo{volume}{94}}, \bibinfo{pages}{978--984}, \url{10.1016/J.NEURON.2017.05.025} (\bibinfo{year}{2017}).

\bibitem{Yu2009}
\bibinfo{author}{Yu, B.~M.} \emph{et~al.}
\newblock \bibinfo{journal}{\bibinfo{title}{{Gaussian-process factor analysis for low-dimensional single-trial analysis of neural population activity}}}.
\newblock {\emph{\JournalTitle{Journal of neurophysiology}}} \textbf{\bibinfo{volume}{102}}, \bibinfo{pages}{614--635}, \url{10.1152/JN.90941.2008} (\bibinfo{year}{2009}).

\bibitem{Langdon2023}
\bibinfo{author}{Langdon, C.}, \bibinfo{author}{Genkin, M.} \& \bibinfo{author}{Engel, T.~A.}
\newblock \bibinfo{journal}{\bibinfo{title}{{A unifying perspective on neural manifolds and circuits for cognition}}}.
\newblock {\emph{\JournalTitle{Nature Reviews Neuroscience 2023 24:6}}} \textbf{\bibinfo{volume}{24}}, \bibinfo{pages}{363--377}, \url{10.1038/s41583-023-00693-x} (\bibinfo{year}{2023}).

\bibitem{Genkin2023}
\bibinfo{author}{Genkin, M.}, \bibinfo{author}{Shenoy, K.~V.}, \bibinfo{author}{Chandrasekaran, C.} \& \bibinfo{author}{Engel, T.~A.}
\newblock \bibinfo{journal}{\bibinfo{title}{{The dynamics and geometry of choice in premotor cortex}}}.
\newblock {\emph{\JournalTitle{bioRxiv}}} \bibinfo{pages}{2023.07.22.550183}, \url{10.1101/2023.07.22.550183} (\bibinfo{year}{2023}).

\bibitem{Tersenghi2014}
\bibinfo{author}{Decelle, A.} \& \bibinfo{author}{Ricci-Tersenghi, F.}
\newblock \bibinfo{journal}{\bibinfo{title}{{Solving the inverse Ising problem by mean-field methods in a clustered phase space with many states}}}.
\newblock {\emph{\JournalTitle{Physical Review E}}} \textbf{\bibinfo{volume}{94}}, \bibinfo{pages}{012112}, \url{10.1103/PHYSREVE.94.012112/FIGURES/3/MEDIUM} (\bibinfo{year}{2016}).

\bibitem{Robert1984}
\bibinfo{author}{Robert, M.} \& \bibinfo{author}{Widom, B.}
\newblock \bibinfo{journal}{\bibinfo{title}{{Field-induced phase separation in one dimension}}}.
\newblock {\emph{\JournalTitle{Journal of Statistical Physics}}} \textbf{\bibinfo{volume}{37}}, \bibinfo{pages}{419--437}, \url{10.1007/BF01011842/METRICS} (\bibinfo{year}{1984}).

\bibitem{Percus1977}
\bibinfo{author}{Percus, J.~K.}
\newblock \bibinfo{journal}{\bibinfo{title}{{One-dimensional Ising model in arbitrary external field}}}.
\newblock {\emph{\JournalTitle{Journal of Statistical Physics}}} \textbf{\bibinfo{volume}{16}}, \bibinfo{pages}{299--309}, \url{10.1007/BF01020384/METRICS} (\bibinfo{year}{1977}).

\bibitem{Tejero1987}
\bibinfo{author}{Tejero, C.~F.}
\newblock \bibinfo{journal}{\bibinfo{title}{{One-dimensional inhomogeneous Ising model: A new approach}}}.
\newblock {\emph{\JournalTitle{Journal of Statistical Physics}}} \textbf{\bibinfo{volume}{48}}, \bibinfo{pages}{531--538}, \url{10.1007/BF01019686/METRICS} (\bibinfo{year}{1987}).

\bibitem{Derrida1986}
\bibinfo{author}{Derrida, B.}, \bibinfo{author}{Mend{\`{e}}s~France, M.} \& \bibinfo{author}{Peyri{\`{e}}re, J.}
\newblock \bibinfo{journal}{\bibinfo{title}{{Exactly solvable one-dimensional inhomogeneous models}}}.
\newblock {\emph{\JournalTitle{Journal of Statistical Physics}}} \textbf{\bibinfo{volume}{45}}, \bibinfo{pages}{439--449}, \url{10.1007/BF01021080/METRICS} (\bibinfo{year}{1986}).

\bibitem{Pani2018DY}
\bibinfo{author}{Pani, P.} \emph{et~al.}
\newblock \bibinfo{journal}{\bibinfo{title}{{Persistence of cortical neuronal activity in the dying brain}}}.
\newblock {\emph{\JournalTitle{Resuscitation}}} \textbf{\bibinfo{volume}{130}}, \bibinfo{pages}{e5--e7}, \url{10.1016/J.RESUSCITATION.2018.07.001} (\bibinfo{year}{2018}).

\bibitem{Zheng2022}
\bibinfo{author}{Zheng, H.}, \bibinfo{author}{Feng, Y.}, \bibinfo{author}{Tang, J.} \& \bibinfo{author}{Ma, S.}
\newblock \bibinfo{journal}{\bibinfo{title}{{Interfacing brain organoids with precision medicine and machine learning}}}.
\newblock {\emph{\JournalTitle{Cell Reports Physical Science}}} \textbf{\bibinfo{volume}{3}}, \bibinfo{pages}{100974}, \url{10.1016/J.XCRP.2022.100974} (\bibinfo{year}{2022}).

\bibitem{Sharf2022}
\bibinfo{author}{Sharf, T.} \emph{et~al.}
\newblock \bibinfo{journal}{\bibinfo{title}{{Functional neuronal circuitry and oscillatory dynamics in human brain organoids}}}.
\newblock {\emph{\JournalTitle{Nature Communications 2022 13:1}}} \textbf{\bibinfo{volume}{13}}, \bibinfo{pages}{1--20}, \url{10.1038/s41467-022-32115-4} (\bibinfo{year}{2022}).

\end{thebibliography}

\end{document}